\documentclass[12pt,preprint]{aastex}

\shorttitle{Chemical Composition of TrES-1}
\shortauthors{Sozzetti et al.}

\begin{document}


\title{Chemical Composition of the Planet-Harboring Star TrES-1}


\author{Alessandro Sozzetti\altaffilmark{1,2}, David Yong\altaffilmark{3}, 
Bruce W. Carney\altaffilmark{3}, John B. Laird\altaffilmark{4}, 
David W. Latham\altaffilmark{1}, and 
Guillermo Torres\altaffilmark{1}} 
\altaffiltext{1}{Harvard-Smithsonian Center for Astrophysics, 60
Garden Street, Cambridge, MA 02138 USA}
\altaffiltext{2}{INAF - Osservatorio Astronomico di Torino, 
10025 Pino Torinese, Italy}
\altaffiltext{3}{Department of Physics \& Astronomy,
University of North Carolina at Chapel Hill, Chapel Hill, NC 27599 USA}
\altaffiltext{4}{Department of Physics \& Astronomy,
Bowling Green State University, Bowling Green, OH 43403 USA}
\email{asozzett@cfa.harvard.edu}
\email{yong@physics.unc.edu}
\email{bruce@physics.unc.edu}
\email{laird@tycho.bgsu.edu}
\email{dlatham@cfa.harvard.edu}
\email{gtorres@cfa.harvard.edu}

\begin{abstract}

We present a detailed chemical abundance analysis of the parent star of the
transiting extrasolar planet TrES-1. Based on high-resolution Keck/HIRES and 
HET/HRS spectra, we have determined abundances relative to the Sun for 16 elements 
(Na, Mg, Al, Si, Ca, Sc, Ti, V, Cr, Mn, Co, Ni, Cu, Zn, Y, and Ba). The resulting 
average abundance of $<[$$X$/H$]>$ $= -0.02\pm0.06$ is in good agreement with 
initial estimates of solar metallicity based on iron. We compare the elemental 
abundances of TrES-1 with those of the sample of stars with planets, searching 
for possible chemical abundance anomalies. TrES-1 appears not to be chemically
peculiar in any measurable way.  We investigate possible signs of selective 
accretion of refractory elements in TrES-1 and other stars with planets, and 
find no 
statistically significant trends of metallicity [$X$/H] with condensation 
temperature $T_c$. We use published abundances and kinematic information for 
the sample of planet-hosting stars (including TrES-1) and several 
statistical indicators to provide an updated classification in terms 
of their likelihood to belong to either the thin disk or the thick disk of 
the Milky Way Galaxy. TrES-1 is found to be a very likely member of the 
thin disk population. By comparing $\alpha$-element abundances of 
planet hosts and a large control sample of field stars, we also find that 
metal-rich ([Fe/H]$\gtrsim 0.0$) stars with planets appear to be systematically 
underabundant in [$\alpha$/Fe] by $\approx 0.1$ dex with respect to comparison 
field stars. The reason for this signature is unclear, but systematic 
differences in the analysis procedures adopted by different groups cannot 
be ruled out.

\end{abstract}




\section{Introduction}

The possibility that super-solar metallicity could imply 
a higher likelihood of a given star to harbor a planet was
investigated since the first detections by precision
radial-velocity surveys (\citeauthor{gonzalez97}~\citeyear{gonzalez97},
~\citeyear{gonzalez98a},~\citeyear{gonzalez98b};
\citeauthor{fuhrmann97}~\citeyear{fuhrmann97},~\citeyear{fuhrmann98};
\citeauthor{laughlin97}~\citeyear{laughlin97}).
A number of studies have been performed throughout these years, with 
increasingly larger sample sizes, employing both spectroscopic and
photometric techniques for metallicity determination (using iron as the
primary reference element), and
adopting control samples of field stars without detected
planets
(\citeauthor{santos00}~\citeyear{santos00},~\citeyear{santos01},
~\citeyear{santos03},~\citeyear{santos04a},~\citeyear{santos05};
\citeauthor{reid02}~\citeyear{reid02};
\citeauthor{laughlin00}~\citeyear{laughlin00};
\citeauthor{gonzalez00}~\citeyear{gonzalez00};
\citeauthor{gonzalez01}~\citeyear{gonzalez01};
\citeauthor{israelian01}~\citeyear{israelian01};
\citeauthor{queloz00a}~\citeyear{queloz00a};
\citeauthor{smith01}~\citeyear{smith01};
\citeauthor{gimenez00}~\citeyear{gimenez00};
\citeauthor{martell02}~\citeyear{martell02};
\citeauthor{heiter03}~\citeyear{heiter03};
\citeauthor{sadakane02}~\citeyear{sadakane02};
\citeauthor{pinsonneault01}~\citeyear{pinsonneault01};
\citeauthor{murray02}~\citeyear{murray02};
\citeauthor{laws03}~\citeyear{laws03};
\citeauthor{fischer05}~\citeyear{fischer05}).

The global trend is that planet-harboring stars are indeed more metal rich than
stars without known planets. Based on observationally unbiased stellar samples,
the strong dependence of planetary frequency on the host star metallicity
has been clearly demonstrated by e.g., Santos et al. (\citeyear{santos01}, 
\citeyear{santos04a}),
and Fischer \& Valenti (\citeyear{fischer05}). Furthermore, the metallicity
enhancement is likely to be ``primordial'' in nature, i.e. due to the
intrinsically high metal content of the protoplanetary cloud from which the 
planetary systems formed, as opposed to the 
possibility of ``self-enrichment'', caused by
accretion of rocky planetesimal material onto the parent star (see Gonzalez 
\citeyear{gonzalez03} 
for a review of the subject). This conclusion is primarily based upon the 
evidence of no dependence of the iron-abundance enhancement on the stellar 
effective temperature, as theoretical calculations would predict 
(e.g., ~\citealp{dotter03}; \citealp{cody05}, and references therein, but see also 
\citealp{vauclair04} for somewhat different arguments), and it 
bears important consequences for the proposed models of giant planet 
formation by core accretion (e.g., \citealp{ida05}; \citealp{kornet05}) 
and disk instability (\citealp{boss02}).

Based on detailed chemical abundance analyses of metals other than iron, 
several attempts have been made in the recent past to confirm the observed trend
and to put on firmer grounds (or refute) the idea that stars with 
planets are primordially
metal-rich, and have not been polluted. Many authors have determined the abundances
of over a dozen other elements for planet hosts, including light elements such
as Li and the isotopic ratio $^6$Li/$^7$Li 
(\citeauthor{gonzalez00}~\citeyear{gonzalez00};
\citeauthor{ryan00}~\citeyear{ryan00};
\citeauthor{israelian01}~\citeyear{israelian01},~\citeyear{israelian03},
~\citeyear{israelian04}; \citeauthor{reddy02}~\citeyear{reddy02};
\citeauthor{mandell04}~\citeyear{mandell04}) and Be 
(\citeauthor{garcia98}~\citeyear{garcia98}; 
\citeauthor{deliyannis00}~\citeyear{deliyannis00};
\citeauthor{santos02}~\citeyear{santos02},~\citeyear{santos04b}), 
refractories such as the
$\alpha$-elements Si, Mg, Ca, Ti, and the iron-group elements Cr, Ni, and Co,
and volatiles such as C, N, O, S, and Zn 
(\citeauthor{santos00}~\citeyear{santos00};
\citeauthor{gonzalez01}~\citeyear{gonzalez01};
\citeauthor{smith01}~\citeyear{smith01};
\citeauthor{takeda01}~\citeyear{takeda01};
\citeauthor{sadakane02}~\citeyear{sadakane02};
\citeauthor{zhao02}~\citeyear{zhao02}; 
\citeauthor{bodaghee03}~\citeyear{bodaghee03};
\citeauthor{ecuvillon04a}~\citeyear{ecuvillon04a},~\citeyear{ecuvillon04b},
~\citeyear{ecuvillon05a}; \citeauthor{beirao05}~\citeyear{beirao05}; 
\citeauthor{gilli05}~\citeyear{gilli05}).

For instance, detection of anomalous light-element abundances in the
atmosphere of a star could be indicative of recent planetary accretion events. 
While evidence for Li excesses in some planet-harboring
stars has been reported in the literature
(\citeauthor{israelian01}~\citeyear{israelian01},~\citeyear{israelian03};
\citeauthor{laws01}~\citeyear{laws01}), clearly suggesting that 
accretion of planetary
material can actually take place in some stars, as implied by theoretical arguments
(\citeauthor{montalban02}~\citeyear{montalban02}; 
\citeauthor{boesgaard02}~\citeyear{boesgaard02}; \citealp{sandquist02}),
in general stars with planets have normal light-element abundances, 
typical of field stars (e.g., \citeauthor{ryan00}~\citeyear{ryan00}; 
\citeauthor{israelian04}~\citeyear{israelian04}).

Arguments in favor of the ``self-enrichment'' hypothesis 
could also be substantiated if volatile elements were to exhibit 
different abundance trends with respect to refractory elements. One way 
to approach the problem is to make use of the condensation temperatures 
$T_c$ of the elements, a typical diagnostic employed for investigating chemical 
fractionation patterns in many areas of planetary science and astronomy 
(e.g., \citeauthor{lodders03}~\citeyear{lodders03}, 
and references therein). In this particular case, volatiles, having 
low $T_c$-values, are expected to show a deficiency in accreted material 
with respect to refractories. However, the most recent evidence 
(e.g., \citeauthor{bodaghee03}~\citeyear{bodaghee03}; 
\citeauthor{ecuvillon04a}~\citeyear{ecuvillon04a},~\citeyear{ecuvillon04b},
~\citeyear{ecuvillon05a}; \citeauthor{gilli05}~\citeyear{gilli05}) 
is that the abundance distributions of other elements
in stars with planets are simply the extension of the observed behavior for [Fe/H],
a result quantified by trends of decreasing [$X$/Fe] with increasing [Fe/H], for 
both refractories and volatiles. 
It thus seems unlikely that pollution effects can be responsible for the overall
metallicity enhancement of the planet host stellar sample. 

The primary goal of this work is to present a detailed study of 
the chemical composition of the parent star of the recently discovered 
transiting extrasolar planet TrES-1 (GSC 02652-01324; \citealp{alonso04}). 
We have done so by undertaking a detailed chemical abundance 
analysis using our Keck and Hobby Eberly Telescope (HET)
spectra of TrES-1. Secondly, we have compared the elemental abundances of TrES-1 
with those of the sample of stars with planets, in order to search for possible
chemical abundance anomalies in the former. To this end, we have utilized 
results from uniform studies of
elemental abundances of large sets of planet hosts available in the literature.
Third, in an attempt to find circumstantial evidence of possible 
selective accretion of 
planetary material, we have further investigated the sample 
of planet hosts and TrES-1, 
searching for statistically significant trends of [$X$/H] 
with condensation temperature. 
Finally, we have utilized the chemical composition information for TrES-1 and
a large sample of planet-hosting stars along with their kinematic properties
in order to classify them, based on a number of diagnostic indicators,
in terms of their likelihood of being members of the thin or 
thick disk populations of the Milky Way Galaxy (e.g., \citealp{gilmore83}; 
\citealp{carney89}. See Majewski \citeyear{majewski93}, and references therein, 
for a comprehensive review and discussion of formation scenarios).
This analysis has the purpose of revisiting and updating the results
of a few past studies (\citealp{gonzalez99}; 
\citeauthor{reid02}~\citeyear{reid02}; 
\citeauthor{barbgratt02} \citeyear{barbgratt02}; \citeauthor{santos03} 
\citeyear{santos03}) 
which, using limited sample sizes, confirmed the strong similarity
between the kinematic properties of stars with planets and that of control samples
of stars without known planets.

This paper is organized as follows. In Section 2 we present our chemical
abundance analysis for the planet-hosting star TrES-1. All elemental
abundances are compared in Section 3 with those of selected, uniformly studied
samples of planet hosts. Section 4 is dedicated to an updated classification of
planet-harboring stars in terms of different stellar populations in our Galaxy.
Finally, Section 5 contains a summary of the main results and concluding remarks.

\section{Observations and Abundance Analysis}

The Keck/HIRES and HET/HRS spectra analyzed in this paper have been studied 
by Sozzetti et al. (\citeyear{sozzetti04}) for an improved determination 
of the stellar and planetary parameters of the system. We refer the reader 
to that paper for a description of the data. 

\subsection{Abundances}

The abundance analysis of TrES-1 in the spectral region 3820-7840 \AA\, 
covered by our data was carried out using a modified version of the 
local thermodynamic equilibrium (LTE) spectral synthesis code MOOG 
(\citealp{sneden73}) and a grid of Kurucz (\citeyear{kurucz93}) LTE 
model stellar atmospheres. Overall, we present here results for 16 
additional elements 
(Na, Mg, Al, Si, Ca, Sc, Ti, V, Cr, Mn, Co, Ni, Cu, Zn, Y, and Ba), 
plus Fe and Li which had already been the subject of 
study in the Sozzetti et al. (\citeyear{sozzetti04}) paper. Our set 
of elements spans a range of condensation temperatures of about 
1000 K, the element with the lowest $T_c$ being Zn. 
Zinc is notably interesting for 
at least two reasons. For instance, this element is commonly used to 
investigate abundance patterns between different stellar populations 
resulting from chemical evolution processes in our Milky Way Galaxy 
(\citealp{sneden91}; \citeauthor{pn00}~\citeyear{pn00}; 
\citeauthor{bensby03}~\citeyear{bensby03}). The 
accurate determination of the abundance ratio [Zn/Fe] is also 
of great importance in studies addressing questions on the chemical evolution 
of the early universe, which employ quasar absorption line abundance analyses, 
in particular damped Ly$\alpha$ systems, believed to be the progenitors of 
modern galaxies (e.g., \citealp{prowolfe00}, and references therein). 
Recent studies (e.g., \citeauthor{mishenina02}~\citeyear{mishenina02}; 
\citeauthor{bensby03}~\citeyear{bensby03}; 
\citeauthor{ecuvillon04b}~\citeyear{ecuvillon04b}; \citealp{nissenetal04}. 
See Chen et al. (\citeyear{chen04}) for a review of the subject of Zn 
abundances determination) have provided indications 
that Zn might not be an exact tracer of Fe, as it is often assumed.
Unfortunately, attempts to measure abundances for other important volatiles, 
such as C, N, O, and S, with even lower $T_c$-values, were not 
successful, the limiting factors being the fact that 
TrES-1 is cool and not metal-poor, thus lines of these elements are too 
strong and lie in regions too crowded to be analyzed, or they are 
outside our wavelength coverage.

For each element in the spectrum for which at least one relatively weak, 
unblended line could be found, we determined equivalent widths (EW) 
using the SPLOT task in IRAF\footnote{IRAF is distributed by the 
National Optical Astronomy Observatories, operated by the 
Association of Universities for Research in Astronomy, Inc., under 
contract with the National Science Foundation, USA.}. Abundances
were computed using the ABFIND driver in MOOG, and by imposing excitation and 
ionization equilibrium (e.g., \citeauthor{sozzetti04}~\citeyear{sozzetti04}; 
\citeauthor{santos04a}~\citeyear{santos04a}, and references therein). 
The solar abundances of reference were taken from Grevesse \& Sauval 
(\citeyear{grevesse98}). Hyperfine and isotopic splitting was 
taken into account for Sc, V, Mn, Co, Cu, and Ba. 
In our analyses, with the exception of Cu, for which a line list from 
Simmerer et al. (\citeyear{ss03}) was utilized, we 
adopted the hyperfine line lists from Prochaska et al. 
(\citeyear{pn00}), and solar isotopic ratios 
(\citealp{anders89}) . In four cases (Mg, Al, Cu, and Zn), 
all the lines in the spectral domain of our data were slightly blended and/or in 
regions where the continuum was difficult to determine such that an EW analysis
would not give reliable results. For these four elements, abundances 
were obtained by fitting synthetic spectra to the data. In the 
panels of Figure~\ref{synth} we show two examples of spectral synthesis 
for \ion{Al}{1} and \ion{Zn}{1} lines, respectively.

For each element 
analyzed, we summarize in Table~\ref{tb1} the final list of lines adopted, the 
lower excitation potentials, the oscillator strength values and the 
literature sources from which they were taken, and the relative EWs 
(where applicable). 
The abundance ratios [$X$/Fe] (and [Fe/H]) for each element, 
averaged over all useful 
lines, are presented in Table~\ref{tb2}, along with the actual number of 
lines used in 
each case. The quoted errors correspond to the dispersion around the mean. 
Finally, in Figure~\ref{abund0} we plot the elemental abundances of TrES-1, 
expressed 
as [$X$/H], as a function of element number. Iron is included, but not lithium, 
for which only an upper limit had been obtained by Sozzetti et al. (2004). 
As one can see, the mean abundance ratio for this star 
($<$[$X$/H]$> =  -0.02$, indicated by the 
horizontal solid line, with a dispersion of $\pm0.06$ dex) 
is very similar to solar, confirming the first 
estimates by Sozzetti et al. (2004), who used iron as a proxy. 

\subsection{Sources of Uncertainty}

Abundance determinations can be subject to a significant number of 
uncertainties, which can be random or systematic in nature. For 
example, EWs can be measured incorrectly due to unrecognized blends or 
poor location of the continuum, a problem that can become severe 
when only a few lines are available for a given element. When a large 
set of lines can be found, then uncertainties in the determination of 
stellar atmospheric parameters (effective temperature $T_\mathrm{eff}$, 
surface gravity $\log g$, and microturbulent velocity $\xi_t$) are 
more likely to constitute the more significant sources of error in the 
abundance determination for a given species. 
Sozzetti et al. (\citeyear{sozzetti04}) 
have derived (following the prescriptions of Neuforge \& Magain 
(\citeyear{neuforge97}) 
and Gonzalez \& Vanture (\citeyear{gonzvant98})) errors in $T_\mathrm{eff}$, 
$\log g$, and $\xi_t$ of $\pm 75$ K, $\pm 0.2$ dex, and $\pm 0.10$ km s$^{-1}$, 
respectively. 
In Table~\ref{tb3} we show the sensitivity of the abundances for all elements 
measured in TrES-1 to changes of the above amounts in the atmospheric parameters, 
with respect to the nominal values of $T_\mathrm{eff} = 5250$ K, 
$\log g = 4.6$, and $\xi_t = 0.95$ km s$^{-1}$ (\citeauthor{sozzetti04} 
\citeyear{sozzetti04}). In most cases, variations are comparable to, 
or smaller than, the quoted uncertainties in Table~\ref{tb2}. 

Finally, non-LTE effects have not been taken into account in our 
analysis. In principle, systematic uncertainties can arise due to 
the use of plane-parallel, LTE model atmospheres. This issue has been 
a matter of debate for quite some time. Recent studies argue that non-LTE 
effects are particularly strong in very cool and metal-poor stars 
(e.g, \citeauthor{edvard93} 
\citeyear{edvard93}; \citeauthor{feltzing98}~\citeyear{feltzing98}; 
\citealp{thevenin99}; \citealp{chen00}; \citealp{yong04}). 
For stars with the temperature (or higher) and metallicity 
(or higher) of TrES-1, such corrections are typically of the 
same order (or smaller) 
than the quoted uncertainties from our abundance analyses. 
We thus believe that ignoring non-LTE effects in the analysis has not introduced 
any major source of error.

\section{TrES-1 vs. Other Planet Hosts}

We attempt here to put TrES-1 in context with the other planet-bearing stars known 
today. 
In particular, we have searched for 1) possible chemical abundance anomalies, 
by comparing the abundances in TrES-1 with those determined for other stars with 
planets, and 2) possible signs of selective accretion of refractory elements and 
chemical evolutionary effects, by analyzing the dependence of abundances 
on condensation temperature $T_c$ for both TrES-1 and the sample of planet hosts.

\subsection{Comparison Between Abundance Ratios}

The Jupiter-sized transiting planet found orbiting TrES-1 (\citealp{alonso04}) 
is the first success of wide-field, ground-based photometric surveys which target 
relatively bright ($9\lesssim V\lesssim 13$) stars, typically lying at a few 
hundred pc from the Sun (for a review see 
for example \citealp{hor04} and \citealp{charbonneau04}). These samples have 
little or no distance overlap with that of solar-neighborhood stars 
($D < 50-100$ pc) targeted by Doppler surveys. On the one hand, the 
evidence for a mild radial metallicity 
gradient of $\sim -0.09$ dex/kpc in the disk of the Milky Way 
presented by Nordstr\"om et al. (\citeyear{nordstrom04}) suggests that, 
when attempting to undertake statistical studies of possible correlations between 
planet properties and the chemical abundances of the host stars 
(e.g., \citealp{sozz04}, and references therein) using stellar 
samples covering a distance range of hundreds of pc, large-scale 
metallicity trends should also be taken into account, in order to cope with this 
potential bias. On the other hand, 
based on the photometric distance estimate of $\sim 150$ pc for TrES-1 inferred 
by Sozzetti et al. (\citeyear{sozzetti04}) and 
Laughlin et al. (\citeyear{laughlin05}), 
one could {\it a priori} expect that no significant anomalies should be 
found for this object in particular. One way or the other, 
it is thus beneficial to carry out the experiment to 
compare the chemical composition 
of planet-hosts discovered by radial-velocity surveys and that of TrES-1 presented 
in this work. Our study is illustrative of those that will be undertaken to better
characterize the parent stars of all transiting planets that will
likely be found in the coming years by the large number of ongoing and
planned wide-field, ground-based photometric surveys.

We have collected chemical composition information for stars with planets 
from a variety of literature sources. The results we have utilized in our 
comparison have all been obtained in the context of systematic, uniform, and 
detailed spectroscopic studies of large sets of planet-harboring stars. We present 
in Figures~\ref{abund1} and~\ref{abund2}, in the plane [$X$/H]-[Fe/H], 
the comparison between the abundances for 16 elements measured in TrES-1 and those 
obtained for other planet hosts. Except for one case, every element measured in 
TrES-1 has been compared with the same element measured in a large subset of 
planet hosts in the context of a single study 
(see captions of Figures~\ref{abund1} 
and~\ref{abund2}). In this way, we have attempted to minimize the possibility 
that unknown systematics 
may have been introduced by considering for each element a set 
of results from a variety of studies which used different spectral 
line lists and model atmospheres to derive stellar parameters and abundances. 

As a general remark, TrES-1 does not appear to be anomalous in any 
conspicuous way. 
All abundance estimates for the elements appear to be rather `normal'. The only 
possible exception is Chromium, which appears to be slightly 
overabundant. However, 
in the TrES-1 spectra we could locate only one measurable line for \ion{Cr}{1}, 
and none for \ion{Cr}{2}. The possibility of an unrecognized 
blend in the measurement of the EW for the 5238.96 \AA\, Chromium line cannot be 
ruled out. Alternatively, although the general features of the 
spectroscopic abundance determination are similar, systematic 
differences in the details of the analysis carried 
out in this work and that of Bodaghee et al. (\citeyear{bodaghee03}) could also 
contribute to explain the observed discrepancy. 
The first possible systematic difference to look for between
the Bodaghee et al. (\citeyear{bodaghee03}) and our analyses is the $\log gf$ 
value. Our \ion{Cr}{1} line was not used by Bodaghee 
et al. (\citeyear{bodaghee03}), nor were any of Bodaghee's 
\ion{Cr}{1} lines in the Thor\'en \& Feltzing (\citeyear{tf00}) paper
from which we took our $\log gf$ value. Therefore, we suspect an offset 
in $\log gf$ values but cannot track it down since there are no lines in 
common between the studies. We note, however, that for all 
other elemental abundances in TrES-1 for which we have used the Bodaghee et al. 
(\citeyear{bodaghee03}) sample for comparison no anomaly is apparent. 

\subsection{Signatures of Accretion}

In order to find possible circumstantial evidence of accretion of metal-enriched 
material onto the parent star, Sozzetti et al. (\citeyear{sozzetti04}) 
attempted to measure the Li abundance in TrES-1, but could place only 
a low upper limit of $\log\epsilon(\mathrm{Li}) < 0.1$. A comparison of the 
result for TrES-1 with other Li abundance estimates for planet hosts obtained 
by Israelian et al. (\citeyear{israelian04}) is presented in 
Figure~\ref{lithium}. As one can see, TrES-1 appears not to be peculiar in any 
measurable way, further corroborating the conclusion that no recent accretion 
events have occurred in this star. 

Another test that has been proposed to investigate further 
(in a statistical sense) the idea of planetesimal 
accretion is to search for evidence of a dependence between elemental abundances 
[$X$/H] and condensation temperature $T_c$. As any accretion events would 
occur in a high temperature environment (very close to the star), refractories 
which condense at high $T_c$ (e.g., $\alpha$-group and iron-peak elements) 
might be added in larger quantities with respect to volatiles with much 
lower $T_c$ (e.g., C, N, O, and Zn). A further constraint in this case is 
for enrichment not to occur  rapidly, so as to give sufficient time to volatile 
elements to evaporate, otherwise any possible chemical differentiation might not 
be detectable. 
We show in Figure~\ref{t_c_abund} abundances [$X$/H] as a function of $T_c$ 
for TrES-1. Any trend of [$X$/H] with $T_c$ can be quantified in terms of a 
significant positive slope in a linear least-squares fit. In our case 
(solid line in Figure~\ref{t_c_abund}) the derived slope of 
$2.74(\pm 6.80)\times 10^{-5}$ dex K$^{-1}$ is statistically insignificant, 
thus no measurable correlation between chemical abundances and $T_c$ is found. 
We note, however, that the element with the lowest abundance (Zn) is 
also the one with the lowest $T_c$. We are fully aware of the importance 
of extending the range of condensation temperatures to put our results 
on more solid grounds. Unfortunately, as discussed before, we have not 
been able to measure reliable abundances for C, N, O, and S within 
the Keck and HET spectra. Additional very high resolution, very high $S/N$ 
spectroscopic observations of TrES-1 (possibly extending to the infrared 
to study molecular line features of CO, CN, and OH), aimed at 
studying volatiles and refractories with a very wide range of 
condensation temperatures, are clearly encouraged. 

Based on relatively small sample sizes of stars with planets, 
the possibility of a trend of [$X$/H] with $T_c$ has been investigated 
in the past by several 
authors (\citeauthor{gonzalez97}~\citeyear{gonzalez97}; 
\citeauthor{smith01}~\citeyear{smith01}; 
\citeauthor{takeda01}~\citeyear{takeda01}; 
\citeauthor{sadakane02}~\citeyear{sadakane02}). In particular, 
Smith et al. (\citeyear{smith01}), by comparing a set of planet 
hosts studied by Gonzalez et al. (\citeyear{gonzalez01}) and of field dwarfs 
without known planets from the Edvardsson et al. (\citeyear{edvard93}) and 
Feltzing \& Gustafsson (\citeyear{feltzing98}) surveys, showed a trend of 
decreasing $T_c$-slope with [Fe/H], explained in terms of Galactic chemical 
evolution. They then highlighted a small number of high-metallicity 
stars with planets exhibiting 
positive $T_c$-slopes above the established trend, and suggested these might be 
good candidates to have undergone selective accretion of planetary material. 
However, Smith et al. (\citeyear{smith01}) pointed out that the 
heterogeneity of the data utilized for the analysis prevented them 
from drawing more than tentative 
conclusions. 
Takeda et al. (\citeyear{takeda01}) and Sadakane et al. (\citeyear{sadakane02}), 
on the other hand, conducted independent abundance analyses of a couple dozen 
planet hosts and a handful of planetless field stars, including some from the 
Gonzalez et al. (\citeyear{gonzalez01}) sample which, according to 
Smith et al. (\citeyear{smith01}), showed signs of a correlation between chemical 
abundances and $T_c$, and concluded that no statistically convincing trend 
could be found. Due to the very small number of comparison field stars observed, 
the conclusions they draw can only be considered suggestive. 

In an attempt to improve on the above results, we have revisited this issue by 
determining the $T_c$-slopes for a set of $\sim 100$ 
planet-harboring stars and $\sim 40$ stars not known to have planets, 
utilizing the abundances presented in the large, 
uniform studies of Bodaghee et al. (\citeyear{bodaghee03}), Ecuvillon et al. 
(\citeyear{ecuvillon04a}, \citeyear{ecuvillon04b}, \citeyear{ecuvillon05a}), 
and Beir\~ao et al. (\citeyear{beirao05}). While the sample of comparison field 
stars is smaller by a factor $\sim 2$ with respect to those utilized by 
Smith et al. (\citeyear{smith01}), and it does not extend to quite as low 
metallicities, it has the crucial advantage of having been analyzed by 
the same group, thus possible systematics arising from the comparison between 
different analysis methods should be avoided. However, in their study 
Bodaghee et al. (\citeyear{bodaghee03}) utilized the atmospheric parameters and iron 
abundances from Santos et al. (\citeyear{santos00}, \citeyear{santos01}, 
\citeyear{santos03}), while the works of Ecuvillon et al. 
(\citeyear{ecuvillon04a}, \citeyear{ecuvillon04b}, \citeyear{ecuvillon05a}), 
and Beir\~ao et al. (\citeyear{beirao05}) used updated values from 
Santos et al. (\citeyear{santos04a}) and Pepe et al. (\citeyear{pepe04}). 
In order to assess the impact of possible residual systematic differences, 
we compared the $T_\mathrm{eff}$ and [Fe/H] values for stars with planets 
in common between the two studies. The effective temperature is the most critical 
parameter, and in this case the different values adopted by Bodaghee et al. and
Ecuvillon et al. and Beirao et al., respectively, are in excellent
agreement with each other, with a mean difference $\Delta T_\mathrm{eff}=-3$ K 
and a standard deviation of 34 K. Such a difference is not expected to
affect abundances. Indeed, the average difference $\Delta$[Fe/H] in the
two studies is null, with a dispersion of 0.03 dex. No trends 
of $\Delta$[Fe/H] vs. [Fe/H] are found, as shown in Figure~\ref{delta_feh}. 
Thus, the datasets from Bodaghee et al. and Ecuvillon et al. and Beirao et al.
may be safely combined.

In the abovementioned works, abundances were determined for elements
covering a large range of condensation temperatures (low values of $T_c$ as 
for C, N, and O, intermediate values as for S, Cu, Zn, and Na, and high values 
as for Fe, Mg, Ti, and Ca). 
We show the derived values of the $T_c$-slopes (including the one for TrES-1 
obtained in this work) in Figure~\ref{t_c_slope}, as a function of [Fe/H]. 
First, as shown by the the straight line through the points, only 
a weak trend of decreasing $T_c$-slope with decreasing [Fe/H] can be detected 
(the linear least-squares fit has 
a slope of $4.65(\pm3.01)\times 10^{-5}$ K$^{-1}$). The lack of a 
measurable $T_c$-[Fe/H] trend, to be interpreted as a signature of 
Galactic disk chemical evolution, is likely due to the metallicity range encompassed 
by the data utilized in the analysis. In fact, the vast majority of the field stars 
from the Edvardsson et al. (\citeyear{edvard93}) survey with [Fe/H]$\lesssim -0.6$ 
have negative $T_c$-slopes in Figure 10 of Smith et al. (\citeyear{smith01}), 
while the planet hosts and control samples used here do not extend 
below [Fe/H]$\simeq -0.5$.

Second, the average and dispersion of the $T_c$-slopes of the combined 
sample shown in 
Figure~\ref{t_c_slope} are $4.79(\pm8.23)\times 10^{-5}$ dex K$^{-1}$. 
Taking into account uncertainties on the $T_c$-slopes, 
three planet hosts (HD 40979, HD 162020 and HD 222404) and three comparison field 
stars (HD 23356, HD 50281, and HD 191408) deviate by $2-3\sigma$ from the average. 
This is, however, far from being a statistically firm result, and these objects 
do not appear to cluster at large [Fe/H], but rather span a range of metallicities 
$-0.5\lesssim$[Fe/H]$\lesssim 0.2$. 
In addition, none of the stars suggested by Smith et al. (\citeyear{smith01}) 
to be candidates to have undergone selective accretion of refractory elements 
shows significant positive slopes. The magnitude of the effect shown by 
Smith et al. (\citeyear{smith01}) is relatively small, and the presence of 
systematics between the methods for abundance determinations adopted 
by different authors in the Smith et al. (\citeyear{smith01}) sample 
and the one utilized in this work 
could easily be invoked to explain this difference. Furthermore, 
as suggested by Gonzalez (\citeyear{gonzalez03}), evidence for self-enrichment 
resulting from large $T_c$-slopes should also translate into trends with 
$T_\mathrm{eff}$, the hottest stars displaying the largest slope values. 
We show in Figure~\ref{t_c_teff} the $T_c$-slopes as a function of 
$T_\mathrm{eff}$ 
for the combined sample utilized in the analysis. Indeed, a weak trend 
with a {\it negative} slope of $-0.005(\pm0.002)\times 10^{-5}$ dex K$^{-2}$ 
(of statistically low significance) appears 
to be present in the opposite direction (the objects with largest slopes being 
the coolest). However, we interpret this feature as 
mostly due to the intrinsic difficulty to very accurately determine abundances 
for a large set of elements in cool stars ($T_\mathrm{eff}\lesssim 5000$ K) 
without 
the danger of introducing greater uncertainties in the abundance results and 
systematic errors primarily caused by departure from 
LTE, granulation convective motions, and crowdedness of the spectra. 

In conclusion, the absence of any statistically convincing 
evidence for differences in the 
relative abundances of volatiles compared to refractories is one more
piece of circumstantial evidence which indicates that
pollution by accreted planetary material is not
likely to play a significant role in the observed metallicity
enhancement of stars with detected planets. 
Our findings are generally in agreement with the results presented in 
similar studies recently undertaken by Ecuvillon et al. (\citeyear{ecuvillon05b}, 
\citeyear{ecuvillon05c}), in which a larger comparison sample is being used, with
revised abundances (\citeauthor{gilli05}~\citeyear{gilli05}) for a significant 
number of elements for both stars with and without known planets. Condensation 
temperature trends among stars with planets have also been recently investigated 
by Gonzalez (\citeyear{gonzalez05}) who, similarly to this work, also used a 
homogeneous set of published abundance data. In that paper, 
Gonzalez (\citeyear{gonzalez05}) comes essentially to our same conclusions. 
In their series of papers on elemental abundance determinations for a large subset of
planet-harboring stars and a control sample of stars without planets,
Bodaghee et al. (\citeyear{bodaghee03}), 
Santos et al. (\citeyear{santos04a}, \citeyear{santos04b}, \citeyear{santos05}), 
Ecuvillon et al. (\citeyear{ecuvillon04a}, \citeyear{ecuvillon04b}, 
\citeyear{ecuvillon05a}), Beir\~ao et al. (\citeyear{beirao05}, and Gilli et al. 
(\citeyear{gilli05})) drew similar 
conclusions, studying separately abundance trends for various elements and 
concluding that each abundance distribution for planet hosts is 
indistinguishable from that of the comparison sample, the former 
simply being the extension of the latter at high metallicities. 
Our result is also in line 
with the recent findings by Fischer \& Valenti (\citeyear{fischer05}) 
and Valenti \& Fischer (\citeyear{valenti05}), 
who could not detect any abundance variations for Na, Si, Ti, and Ni as a 
function of condensation temperature in a sample of 1040 FGK-type stars with 
and without planets from the Keck, Lick, and Anglo-Australian Telescope 
planet search programs. This conclusion is further 
corroborated by the observed trend from studies of the Li and Be abundances in 
planet hosts (as discussed in the introduction and at the beginning of this 
Section), and by the evidence for no dependence of metallicity on the stellar 
convective envelope mass, as extensively discussed by e.g. Pinsonneault et 
al. (\citeyear{pinsonneault01}) and Fischer \& 
Valenti (\citeyear{fischer05}). 

\section{TrES-1, Other Planet Hosts, and Stellar Populations}

Elemental abundances and galactic kinematics are often used to assign, 
on an observational basis, individual objects to different stellar populations 
in the Milky Way Galaxy. A few studies have concentrated on comparing the 
kinematics of planet hosts with that of comparison samples of field stars 
(\citealp{gonzalez99}; \citeauthor{barbgratt02} \citeyear{barbgratt02}; 
\citeauthor{santos03} \citeyear{santos03}). No statistically significant 
kinematic peculiarity was uncovered between stars with and without known 
planets, the former simply being more metal-rich on average than the latter 
at any given distance from the galactic center. 
We focus here instead on assigning, 
in a statistical sense, TrES-1 to be either a thin- or thick-disk object, 
by comparison with other planet-bearing stars and a large sample 
of field stars. Whether a star belongs to one or the other galactic disk 
population has been determined on the 
basis of the agreement between a few indicators, which are purely
kinematic in nature, a combination of kinematics and chemistry, or are
solely chemistry-based. 

\subsection{Kinematic Indicators}

While TrES-1 lacks a parallax estimate, photometric distances have been 
derived by Sozzetti et al. (\citeyear{sozzetti04}) and
Laughlin et al. (\citeyear{laughlin05}), which place the star+planet
system at $d \approx 150$ pc. The combination of distance, proper motion,
and radial-velocity allows one to calculate the galactic velocity vector 
 ($U$, $V$, $W$, with $U$ positive toward the galactic anti-center) 
with respect to the Local Standard of Rest (LSR), 
adjusting for the standard solar motion ($U_\odot$, $V_\odot$, $W_\odot$) = 
(-9.0,+12.0,+7.0) km s$^{-1}$ 
(following Mihalas \& Binney (\citeyear{mihalas81})). 
The basic data are summarized in Table~\ref{tb4} (columns 1 through 10), 
together with the same information for a sample of 120 stars with planets. 

One way to classify TrES-1 as either a thin or thick disk object is to 
use statistical indicators purely based on kinematics. We have compared the 
results obtained for TrES-1 with those for the other planet hosts listed 
in Table~\ref{tb4} and with a large comparison sample of 639 field stars 
taken from the catalog compiled by Soubiran \& Girard (\citeyear{soubiran05}). 
In order to calculate the likelihood of any given object 
to belong to either of the two populations on the basis of its galactic 
kinematics, a number of approaches can be adopted. We elect to carry out 
population assignments using the classifications by 
Mishenina et al. (\citeyear{mishenina04}), 
Bensby et al. (\citeyear{bensby03}, \citeyear{bensby05}), 
Venn et al. (\citeyear{venn04}), and Brewer \& Carney (\citeyear{brewer05}). 
Mishenina et al. (\citeyear{mishenina04}) adopt a classification scheme based 
on the assumption of a Gaussian velocity ellipsoid for the thin and 
thick disk populations, with kinematical parameters (velocity dispersion and 
asymmetric drift) taken from Soubiran et al. (\citeyear{soubiran03}), 
and relative densities of the two populations of 75\% and 25\%, respectively.
The approach of Bensby et al. (\citeyear{bensby03}, \citeyear{bensby05}) 
is similar in principle, although the authors used the 
thick-disk-to-thin-disk probability ratio $TD/D$, a slightly different 
velocity ellipsoid, and the observed fractions of each population in the 
solar neighborhood (4\% and 96\%, respectively). 
Finally, Venn et al. (\citeyear{venn04}) and 
Brewer \& Carney (\citeyear{brewer05}) 
employ a standard Bayesian classification scheme assuming a Gaussian velocity 
ellipsoid with components from Dehnen \& Binney (\citeyear{dehnen98}) and 
Soubiran et al. (\citeyear{soubiran03}), and uniform prior probability 
distributions for both populations. In columns 11 through 17 of Table~\ref{tb4} 
we report the membership probabilities for TrES-1 and the sample 
of stars with planets computed with the methods 
described above, plus a population assignment based on Venn's scheme 
but using the Bensby et al. (\citeyear{bensby03}) values for 
the prior probability distributions. 

As a consistency check between the various methods, let us first 
consider objects classified as very likely thick disk members by the different 
methods. According to the Mishenina et al. (\citeyear{mishenina04}) approach, 
five planet-bearing stars (HD 13445, HD 47536, HD 111232, HD 114762, and 
HD 195019A) have $P^\mathrm{thick} \ge 0.90$. Bensby et al. (\citeyear{bensby03}) 
classify as thick disk members objects with $TD/D \ge 10$ (i.e., objects 
which are ten times more likely of being thick disk rather than thin 
disk members). With this 
prescription, the second classification scheme assigns to the thick 
disk only HD 47536 and HD 114762, a K0III giant and the lowest-metallicity 
object known to-date to harbor a planetary mass object and a 
well-known member of the thick-disk population, respectively. 
The Bayesian approach of Venn et al. (\citeyear{venn04}) with the 
Bensby et al. (\citeyear{bensby03}) prior probability distributions 
classifies the same two stars as thick disk members with 
$P_2^\mathrm{thick} \ge 0.90$. According to these last two schemes, 
HD 13445, HD 111232, and HD 195019A all have intermediate kinematics, 
with $TD/D\simeq 2-3$ and $P_2^\mathrm{thick} \simeq 0.60$, 
although the thick disk membership is still more probable, albeit 
without high confidence. Finally, if uniform priors are 
used, $P_1^\mathrm{thick} \ge 0.90$ for 10 objects, a sample 
including all the five stars with $P^\mathrm{thick} \ge 0.90$, 
plus HD 27894, HD 88133, HD 114729, HD 190360, and HD 330075. 
These objects all have $P^\mathrm{thick} \ge 0.85$, $TD/D > 1$, 
and $P_2^\mathrm{thick} = 0.50$, with the exception of HD 114729, for 
which $TD/D = 0.65$, and HD 190360, for which $P^\mathrm{thick} = 0.63$ 
and $P_2^\mathrm{thick} = 0.40$. Then, we can state as a general 
conclusion that all the purely kinematic indicators agree to a 
significant extent, and the very likely thick disk members, with 
extreme kinematics, are 
all readily identified, regardless of the classification scheme. 
In this context, TrES-1 appears to be a very likely member of the 
thin disk population, with $P^\mathrm{thin} = 0.92$, $TD/D = 0.02$, 
$P_1^\mathrm{thin} = 0.70$, and $P_2^\mathrm{thin} = 1.0$. 

Finally, in Figure~\ref{kinem} we show the Toomre diagram 
$UW-V$ ($UW = \sqrt{U^2+W^2}$) for the Soubiran \& Girard (\citeyear{soubiran05}) 
and the planet host samples (left and right panel, respectively). 
TrES-1 is indicated with a large filled black dot. 
In the two panels of Figure~\ref{kinem}, 
symbols of different shapes indicate objects 
with $P^\mathrm{thin}\ge 0.90$ and $P^\mathrm{thick}\ge 0.90$. For the 
planet hosts sample we also plot objects with intermediate kinematics. 
In addition, the solid lines identify regions of constant peculiar 
space velocities $v_\mathrm{p} = \sqrt{U^2+V^2+W^2}$, 
with $v_\mathrm{p} = 85$ km s$^{-1}$ and $v_\mathrm{p} = 180$ km s$^{-1}$, 
respectively, which is the simple recipe proposed by Feltzing et al. 
(\citeyear{feltzing03}) and Nissen (\citeyear{nissen04}) 
to operationally distinguish between thick and thin disk populations. 
By inspection of Figure~\ref{kinem}, we see how, in both cases, 
the two classification schemes select basically the same clean samples of
stars in the two kinematic populations. Also, many of the stars with 
planets assigned to either the thick or thin disk populations with lower 
confidence appear to be borderline cases, according to 
the simple criterion based on $v_\mathrm{p}$. Again, a comparison 
between the two different classification schemes results in very broad 
agreement. And again, simply based on its value of $v_\mathrm{p}$, 
TrES-1 is confidently assigned to the thin disk population. 

\subsection{Hybrid Indicators}

In Figure~\ref{hybrid} the distribution of the $X$ parameter defined by 
Schuster et al. (\citeyear{schuster93}) 
is shown, for the comparison sample of field stars (solid histogram), the 
planet hosts (dashed-dotted histogram), and TrES-1 (solid arrow). 
The $X$ parameter is a linear combination of $V_\mathrm{rot}$ and [Fe/H], 
where $V_\mathrm{rot} = V + 220$ km s$^{-1}$ (corrected for the rotation
velocity of the LSR). All the $X$ values are also listed for the 
planet hosts sample in column 18 of Table~\ref{tb4}. 

According to Schuster et al. (\citeyear{schuster93}) 
and Karata\c{s} et al. (\citeyear{karatas05}), values of $-21 < X < -6$ 
identify a clean sample of thick disk stars. Within the 
context of this scheme, 
94 stars are assigned to the thick disk with minimal contamination. 
On the other hand, 47 of the 70 field stars with $P^\mathrm{thick} \ge 0.90$, 
and 46 out of 66 objects with both $P^\mathrm{thick} \ge 0.90$ {\em and} 
$85\leq v_\mathrm{p}\leq 180$ km s$^{-1}$ fall in this range. 
As for what concerns the sample of stars with planets, five objects 
(HD 6434, HD 47536, HD 111232, HD 114729, and HD 114762) have 
$-21 < X < -6$. These include two assigned to the thick disk by all other 
criteria (HD 47536 and HD 114762), two low-confidence thick 
disk members (HD 111232, HD 114729), and one star (HD 6434) for 
which the other criteria give mixed results. 
The hybrid method thus appears to perform slightly more poorly with 
respect to both those based on the determination of membership 
probabilities and on the values of peculiar velocities, but its 
accuracy appears nonetheless comparable. Finally, objects 
with $X \lesssim -33$ are assigned with high confidence to the thin disk, 
according to Schuster et al. (\citeyear{schuster93})
and Karata\c{s} et al. (\citeyear{karatas05}). The arrow at $X \simeq -33$
identifies TrES-1, which is thus confirmed as a likely member of the
thin disk also on the basis of the $X$ indicator.

\subsection{Chemical Indicators}

Recent studies (e.g., Bensby et al. \citeyear{bensby03}, \citeyear{bensby05};
Brewer \& Carney \citeyear{brewer05}. See Nissen \citeyear{nissen04}, and references therein,
for a review of the subject) have highlighted how the thin and thick disk of the
Milky Way Galaxy appear to overlap significantly in metallicity, when [Fe/H]
is utilized as a reference, while they are separated in [$\alpha$/Fe]
abundances. The variations in the abundances in $\alpha$-elements can
then be used to not only explain the history of star-formation processes
in the Galaxy (as it is usually done), but also to identify and understand
systematic differences in the chemical composition of the thin and thick disk.

It is customary to use averages of $\alpha$-element abundances for
a variety of studies of chemical abundance trends in the Milky Way Galaxy
and beyond (e.g., Venn et al. \citeyear{venn04}, and references therein).
Usually, elements with similar trends are considered.
For the purpose of this analysis, we have utilized [$\alpha$/Fe] defined as
$\frac{1}{4}$([Mg/Fe]+[Si/Fe]+[Ca/Fe]+[Ti/Fe]). One could in
principle adopt other combinations, and include other
$\alpha$-elements such as O and S. We have chosen to not consider
these elements in our discussion for a number of reasons. First, as already 
mentioned in Section 2.1 and Section 3.2, the sulfur lines and the UV OH 
lines are outside the wavelength domain of our Keck/HIRES and HET/HRS spectra. 
Second, the forbidden [OI] lines at 6300 \AA\, and 6363 \AA\, are too weak to 
be measured reliably in dwarf stars such as TrES-1. Third, the oxygen triplet 
lines near 7770 \AA\, have very high excitation levels (9.15 eV), thus the
sensitivity to non-LTE effects is significantly greater than in the case of
the much lower excitation potential lines of Mg, Si, Ca, and Ti.
Finally, there is significant disagreement in the literature
on abundance trends of both oxygen and sulfur (e.g,
Nissen et al. \citeyear{nissen02}; Jonsell et al. \citeyear{jonsell05};
Caffau et al. \citeyear{caffau05}). In the case of oxygen, for 
example, typical discrepancies of 0.1-0.2 dex are found when 
[OI] and oxygen triplet abundances are compared (e.g., 
Ecuvillon et al. \citeyear{ecuvillon05a}).

We show in Figure~\ref{chemical1} the [$\alpha$/Fe] vs. [Fe/H] diagram for 425 
stars in the catalog of Soubiran \& Girard (\citeyear{soubiran05}) and for 
78 stars with planets and 41 comparison field stars 
for which a value of [$\alpha$/Fe] could be derived 
from the Bodaghee et al. (\citeyear{bodaghee03}) and 
Beir\~ao et al. (\citeyear{beirao05}) samples 
(left and right panels, respectively). TrES-1 is indicated by the 
large filled dot, as before. In the left panel, crosses are thick disk stars and
open circles are thin disk objects, based on the purely kinematic criterion 
of Mishenina et al. (\citeyear{mishenina04}) described above. 
While there is a significant overlap in metallicity, the separation
between the two populations is clear when considering metal-poor objects
([Fe/H] $< -0.5$) with large values ($> 0.2$) of [$\alpha$/Fe] abundances. 
Again, based on a purely chemistry-based indicator, 
TrES-1 appears to belong clearly to the thin disk. In the right panel, 
the same plot for the planet-harboring stars and the control sample 
closely follows the same trend. One of the two objects assigned to the 
thick disk by all the classification schemes discussed before 
(HD 114762) is also among those with the largest values of [$\alpha$/Fe] 
(as it can be seen by looking at the last column of Table~\ref{tb4}, 
in which all the [$\alpha$/Fe] abundances for planet hosts are reported), while 
no measurement is available for the other thick disk star (HD 47536). Two 
other objects (HD 6434 and HD 37124) with [$\alpha$/Fe]$\sim 0.15-0.20$ 
have intermediate kinematics. As a further confirmation, the three planet 
hosts are the most metal-poor objects in the sample studied here. 

However, a visual comparison between the two panels in Figure~\ref{chemical1} 
shows evidence of an interesting feature. The metal-rich sample ([Fe/H]$> 0.0$) 
of planet hosts appears to be systematically less abundant in [$\alpha$/Fe] than the 
large sample of field stars from Soubiran \& Girard 
(\citeyear{soubiran05}), by $\sim 0.1$ dex. In order to make a more meaningful 
statement on the reality of this difference, we first show in Figure~\ref{chemical2} 
a comparison between the [$\alpha$/Fe] values for all stars in the 
Soubiran \& Girard (\citeyear{soubiran05}) sample, the planet hosts 
and comparison field stars from Bodaghee et al. (\citeyear{bodaghee03}) 
and Beir\~ao et al. (\citeyear{beirao05}), and TrES-1, expressed as a function 
of [Fe/H] and $T_\mathrm{eff}$, respectively. No clear trend with temperature 
is apparent: Stars with planets have lower [$\alpha$/Fe]-values than 
planetless field stars for a broad range of $T_\mathrm{eff}$, although 
slightly less so at the cooler side. On the other hand, the trend with 
metallicity does appear significant. 
We summarize in Table~\ref{tb5} the results of a series of Kolmogorov-Smirnov
(K-S) tests we have run to measure to what extent the [$\alpha$/Fe] distributions 
for the three samples might differ one from the other, and 
expressed in terms of the probability $Pr(D)$ of the null hypothesis 
(i.e. that the distributions are the same). We have done so 
by considering the [$\alpha$/Fe] abundances for all values of [Fe/H], 
and by restricting the comparison to [$\alpha$/Fe] values in the two 
regimes [Fe/H]$\leq 0.0$ and [Fe/H]$> 0.0$. Indeed, the [$\alpha$/Fe] 
distributions for the Soubiran \& Girard (\citeyear{soubiran05}) sample 
and for the control sample of field stars of Bodaghee et al. 
(\citeyear{bodaghee03}) and Beir\~ao et al. (\citeyear{beirao05}) 
appear globally indistinguishable, and this holds true when the two 
samples are compared in different metallicity bins. However, the [$\alpha$/Fe] 
distribution for the planet hosts sample appears to differ 
significantly from both the other cases, when no restriction on 
[Fe/H] is imposed. The distribution of [$\alpha$/Fe] values 
for the planet hosts is significantly different from that of the 
stars in the large catalog of Soubiran \& Girard (\citeyear{soubiran05}) 
when different [Fe/H] ranges are considered, but less so for 
the [Fe/H]$\leq 0.0$ bin. This confirms the conclusions drawn from the comparison 
presented in Figure~\ref{chemical2}. Stars with planets and the smaller 
control sample appear instead to have the same distribution 
in the metal-poor regime, while they exhibit somewhat significant 
differences in the metal-rich regime. 

Assuming the difference is real, what is a likely cause for its existence? 
One possible reason could be systematics. For instance, inhomogeneous 
$T_\mathrm{eff}$ scales determinations using different methods of 
spectroscopic analyses are a likely outcome, and systematic differences 
in this critical parameter translate in systematic differences in 
[Fe/H] and thus in any elemental abundance [$X$/Fe]. 
Some of the planet hosts and the stars in the control sample of Bodaghee et al. 
(\citeyear{bodaghee03}) and Beir\~ao et al. (\citeyear{beirao05}), 22 and 25 
respectively, are in common with the Soubiran \& Girard (\citeyear{soubiran05}) 
sample, thus the values of [$\alpha$/Fe] obtained in the two cases can be directly 
compared. In the two upper panels of Figure~\ref{systematic} we show the 
comparison between the $T_\mathrm{eff}$ and [Fe/H] determinations for 
stars common to both studies. Average systematic differences in the two cases amount 
to $\Delta T_\mathrm{eff} \approx 50-60$ K and $\Delta$[Fe/H]$\approx 0.03$ dex. 
A mild dependence of the differences between the $\alpha$-element 
abundances $\Delta$[$\alpha$/Fe] on $T_\mathrm{eff}$ 
is found, but less so on [Fe/H], as shown in the two bottom panels 
of Figure~\ref{systematic}. In all cases, the trend for planet hosts is 
closely traced by the control sample, and systematic differences of the same 
magnitude appear to be present whether one compares stars with planets or 
control sample field stars in common to both studies. On average, 
stars with planets and comparison field stars in common with the two 
studies appear to differ by $\sim 0.06$ dex and $\sim  0.04$ dex in 
[$\alpha$/Fe], respectively. 

In light of this, intriguing as it might be, 
one could simply interpret the observed signature in Figure~\ref{chemical2} 
in terms of unknown systematics in the abundance determination procedures 
adopted by the different authors. However, if systematics were to be the dominant 
effect, then one would not expect to find significant differences in 
the two samples analyzed in a uniform manner by the same authors, as instead
the K-S test analysis seems to indicate. While on the one hand 
the discrepancies between the two samples of Bodaghee et al. (\citeyear{bodaghee03}) 
and Beir\~ao et al. (\citeyear{beirao05}) could still be in part explained 
invoking small-number statistics and selection effects 
(e.g., too few metal-rich stars in their control sample), 
on the other hand the possibility that the observed feature is 
not an artifact due to systematics might not be discarded completely. 

To put this point under further scrutiny, 
we have investigated whether age might be a factor in the equation. 
For example, if the sub-sample of metal-rich planet hosts is systematically 
composed of young objects, which are more likely to have their chemical 
composition dominated by heavy-element materials from Type Ia supernovae, 
this might in turn contribute to explain the underabundance in [$\alpha$/Fe].  
We present in the two panels of Figure~\ref{ages} the age-metallicity diagram 
and a plot of [$\alpha$/Fe] vs. age for stars in the  
Soubiran \& Girard (\citeyear{soubiran05}) catalog, for TrES-1, and for the stars with 
planets and the control sample of field stars from Bodaghee et al. 
(\citeyear{bodaghee03}) and Beir\~ao et al. (\citeyear{beirao05}). 
Stellar age estimates were obtained from the
Nordstr\"om et al. (\citeyear{nordstrom04}) catalog, except for TrES-1,
for which the preliminary estimate by Sozzetti et al. (\citeyear{sozzetti04})
was used. As one can see, stars with planets do not appear to be particularly younger
than planetless field stars at any given metallicity, and the former
sample seems underabundant in [$\alpha$/Fe] at all ages.
Thus, one would tend to exclude age as primary responsible for
the feature observed in Figure~\ref{chemical2}.

A large number of spectroscopic studies of stars with planets
and control samples have been undertaken in the past
(see the Introduction Section), yet none have noticed the apparent
deficiency of [$\alpha$/Fe] among the former relative to the latter.
There are three possible reasons why we have discerned the effect while
others have not. First, the magnitude of the effect is small, and few workers
have averaged the results from several elements. As long as the elements
have similar sensitivities to systematic effects such as temperature and gravity,
using four elements rather than one has an obvious advantage. Second, a
number of studies have not employed large enough numbers of the two classes
of stars. Finally, some studies may have been vulnerable to systematic 
differences in the analysis procedures between differing datasets. An
extensive review of all such prior comparisons is beyond the scope of
this paper, but we hope that future work will indeed consider these
four $\alpha$-elements both individually and jointly.

While on the basis of the evidence presented here no definite
conclusion can be drawn, the possible existence of
an underabundance in [$\alpha$/Fe] in stars with planets
with respect to non-planet hosts
is nevertheless intriguing and worthy of further investigation,
to ascertain its reality or to firmly rule it out on the basis
of e.g. the presence of identifiable systematics.

\section{Concluding Remarks}

In this work we have carried out an abundance analysis of 16 elements for the
parent star of the transiting extrasolar planet TrES-1. The resulting
average abundance of $<[X$/H$]>= -0.02\pm0.06$ is in good agreement with
initial estimates of solar metallicity based on iron
(\citeauthor{sozzetti04} \citeyear{sozzetti04}; 
\citeauthor{laughlin05} \citeyear{laughlin05}). TrES-1 appears not to be 
chemically peculiar in any measurable way when compared to a large sample 
of known stars with planets. No convincing evidence for statistically 
significant trends of metallicity [$X$/H] with condensation temperature $T_c$ 
can be found for TrES-1 or other planet hosts, a further indication that 
selective accretion of planetary material is not likely to be responsible 
for the observed high metal content of stars with detected planets, 
a conclusion similar to those drawn by others (e.g., 
\citeauthor{santos01} \citeyear{santos01},~\citeyear{santos04a}; 
\citeauthor{fischer05} \citeyear{fischer05}). Using its abundance and 
kinematic information, we have classified TrES-1 as a likely member of 
the thin disk population, and provided updated membership probabilities 
for a large set of planet hosts, based on the relative agreement between 
different statistical indicators (purely kinematic in nature, solely 
based on chemistry, and a combination of the two). Finally, we have 
highlighted an apparent systematic underabundance in [$\alpha$/Fe] 
of stars with planets compared to a large comparison sample of field stars.  
The more likely cause for this signature resides in unknown systematics 
in the details of the abundance analysis procedures adopted by different 
authors. However, we have found hints for differences between the 
[$\alpha$/Fe] abundances of planet hosts and control samples analyzed 
in exactly the same way. In this respect, we stress the importance of continuously 
updating and expanding uniform, detailed studies of the chemical composition 
of planet hosts (including both refractory and volatile elements 
spanning a wide range of condensation temperatures) as new objects 
are added to the list, as well as statistically significant control 
samples of stars without any detected planets, following those 
recently undertaken by e.g., Bodaghee et al. (\citeyear{bodaghee03}), 
Santos et al. (\citeyear{santos04a}, \citeyear{santos04b}, \citeyear{santos05}), 
Israelian et al. (\citeyear{israelian04}), 
Ecuvillon et al. (\citeyear{ecuvillon04a}, \citeyear{ecuvillon04b}, 
\citeyear{ecuvillon05a}, \citeyear{ecuvillon05b},  \citeyear{ecuvillon05c}), 
Beir\~ao et al. (\citeyear{beirao05}), Gilli et al. (\citeyear{gilli05}), 
Fischer \& Valenti (\citeyear{fischer05}), Valenti \& Fischer (\citeyear{valenti05}), 
and Gonzalez (\citeyear{gonzalez05}). Such investigations 
would also help to disentangle possible signatures induced by the 
presence of planets from effects related instead to Galactic chemical 
evolution.

\acknowledgments

A.S. acknowledges support from the Keck PI Data Analysis Fund (JPL 1262605). 
G.T. acknowledges partial support for this work from NASA Origins grant 
NNG04LG89G. J.L. is partially supported by an NSF grant AST-0307340. 
It is a pleasure to thank C. Allende Prieto for stimulating discussions. 
An anonymous referee provided very useful suggestions and comments. 
Some of the data presented herein were obtained at the Hobby-Eberly Telescope, 
which is operated by the McDonald Observatory on behalf of the
University of Texas at Austin, the Pennsylvania State University,
Stanford University, Ludwig Maximillians Universit\"at M\"unchen,
and Georg August Universit\"at G\"ottingen. The other data were 
obtained at the W.M. Keck Observatory, which
is operated as a scientific partnership among the California
Institute of Technology, the University of California and the
National Aeronautics and Space Administration. The Observatory was
made possible by the generous financial support of the W.M. Keck
Foundation. The authors wish to recognize and acknowledge the very
significant cultural role and reverence that the summit of Mauna
Kea has always had within the indigenous Hawaiian community. 
Without their generous hospitality, the Keck observations presented 
here would not have been possible. 
This research has made use of NASA's Astrophysics Data System Abstract Service
and of the SIMBAD database, operated at CDS, Strasbourg, France.

\clearpage

\figcaption[]{Top panel: a 5\AA\, region in the observed spectrum (dots) 
of TrES-1 containing two \ion{Al}{1} lines (6696.0\AA\, and 6698.7\AA) 
used in the analysis. Superposed are three spectral syntheses for 
different values of [Al/Fe]. Bottom panel: same as top panel, but for a 
3\AA\, region containing the \ion{Zn}{1} line at 4722.2\AA.\label{synth}}

\figcaption[]{Chemical abundances [$X$/H] measured in TrES-1 as a 
function of element 
number Z. Error bars correspond to the error of the mean $\sigma/\sqrt{n}$. 
The average solar abundance ratio $<$[$X$/H]$> = 0.0$ is indicated 
by a horizontal dashed line. The solid line is the average abundance ratio 
determined for TrES-1 ($<$[$X$/H]$>\simeq -0.02$).\label{abund0}}

\figcaption[]{[$X$/H] versus [Fe/H] for 8 elements measured in TrES-1 (filled 
black circle) and in comparison samples of planet hosts (open circles). 
The literature sources for the comparison samples are the 
following. [V/H], [Cr/H], 
[Mn/H], [Co/H], [Ni/H], [Si/H], [Ca/H], and 
[Sc/H]: \citeauthor{bodaghee03}~\citeyear{bodaghee03}.\label{abund1}}

\figcaption[]{Same as Figure~\ref{abund1}, but for the remaining 8 
elements that could be measured in TrES-1. The literature sources 
for the comparison samples are the following. 
[Ti/H]: \citeauthor{bodaghee03}~\citeyear{bodaghee03}; 
[Na/H], [Mg/H], and [Al/H]: \citeauthor{beirao05}~\citeyear{beirao05}; 
[Cu/H] and [Zn/H]: \citeauthor{ecuvillon04b}~\citeyear{ecuvillon04b}; 
[Ba/H]: \citeauthor{sadakane02}~\citeyear{sadakane02} and \citealp{zhao02}; 
[Y/H]: \citeauthor{sadakane02}~\citeyear{sadakane02}. \label{abund2}}

\figcaption[]{Lithium abundance $\log\epsilon$(Li) as a function of effective 
temperature $T_\mathrm{eff}$ for planet hosts. 
The black filled circle corresponds to TrES-1, while open circles are 
Lithium measurements for the sample of stars with planets analyzed by Israelian et 
al. (\citeyear{israelian04}). Arrows indicate that only upper limits on 
$\log\epsilon$(Li) are available. \label{lithium}}

\figcaption[]{Chemical abundances [$X$/H] measured in TrES-1 versus 
condensation temperature $T_c$ (taken from Lodders (\citeyear{lodders03})). 
A linear least square fit to the data in the form [$X$/H] = a + b$T_c$ 
(solid line) provides no evidence of a measurable trend of abundance with 
condensation temperature.\label{t_c_abund}}

\figcaption[]{Differences $\Delta$[Fe/H] between the iron abundances derived by 
Santos et al. (\citeyear{santos04a}) (S04) and those obtained by 
Santos et al. (\citeyear{santos00}, \citeyear{santos01}, 
\citeyear{santos03}). The average difference (dotted line) is null, 
with a standard deviation of $\sim 0.03$ dex. No trend is visible as a function [Fe/H] 
(a rank-correlation test gave a probability of no correlation $p=0.84$). 
\label{delta_feh}}

\figcaption[]{Slopes of the [$X$/H]-$T_c$ relation as a function of 
[Fe/H] for TrES-1 (this work, filled black circle) 
and for a sample of planet hosts (filled black 
circles) and comparison field stars (open triangles) uniformly studied 
by Bodaghee et al. (\citeyear{bodaghee03}), 
Ecuvillon et al. (\citeyear{ecuvillon04a}, 
\citeyear{ecuvillon04b}, \citeyear{ecuvillon05a}), 
and Beir\~ao et al. (\citeyear{beirao05}).\label{t_c_slope}}

\clearpage

\figcaption[]{Slopes of the [$X$/H]-$T_c$ relation as a function of 
$T_\mathrm{eff}$ for TrES-1 (this work, 
filled black circle) and for a sample of planet hosts 
(filled black circles) and comparison field stars (open squares) uniformly 
studied by Bodaghee et al. (\citeyear{bodaghee03}), 
Ecuvillon et al. (\citeyear{ecuvillon04a}, 
\citeyear{ecuvillon04b},\citeyear{ecuvillon05a}), and 
Beir\~ao et al. (\citeyear{beirao05}).\label{t_c_teff}}

\figcaption[]{Left: Toomre diagram for the Soubiran \& Girard 
(\citeyear{soubiran05}) stellar 
sample. Solid lines identify regions of constant peculiar 
velocity $v_\mathrm{p,1} = 85$ km s$^{-1}$ and 
$v_\mathrm{p,2} = 180$ km s$^{-1}$, respectively. 
Objects with $v_\mathrm{p} < v_\mathrm{p,1}$ are thin disk stars, 
while those with $v_\mathrm{p,1} < v_\mathrm{p} < v_\mathrm{p,2}$ 
are assigned to the thick disk population. Asterisks and triangles 
are likely members of the thin and thick disk, respectively, 
according to the Mishenina et al. (\citeyear{mishenina04}) 
kinematic classification. Crosses identify stars with 
intermediate kinematics. The large filled dot represents TrES-1. 
Right: The same diagram, but for TrES-1 and a large sample of 
planet hosts. Squares are objects with $P^\mathrm{thin}\ge 0.90$, 
asterisks are stars with $P^\mathrm{thick}\ge 0.90$, and crosses identify 
objects with intermediate kinematics. \label{kinem}}

\figcaption[]{Distributions of the $X$ indicator, defined by 
Schuster et al. (\citeyear{schuster93}), for the Soubiran \& 
Girard (\citeyear{soubiran05}) sample (solid histogram) and 
for a sample of planet hosts (dashed-dotted histogram). The solid 
arrow corresponds to the $X$ value for TrES-1. \label{hybrid}}

\figcaption[]{Left: [$\alpha$/Fe] vs. [Fe/H] for the Soubiran \& 
Girard (\citeyear{soubiran05}) sample. Open circles have 
$P^\mathrm{thin}\ge 0.90$, while crosses are stars with 
$P^\mathrm{thick}\ge 0.90$. 
Stars with intermediate kinematics are not shown. 
TrES-1 is identified by the filled black dot. Right: the same 
diagram for a sample of planet hosts (open circles) and 
comparison field stars (crosses) studied by 
Bodaghee et al. (\citeyear{bodaghee03}) 
and Beir\~ao et al. (\citeyear{beirao05}). \label{chemical1}}

\figcaption[]{Left: same as Figure~\ref{chemical1}, but now 
comparing all stars in the Soubiran \& Girard (\citeyear{soubiran05}) catalog 
(filled black circles) with TrES-1 (this work, green circle), 
planet hosts (green circles) and comparison field 
stars (red circles) from Bodaghee et al. (\citeyear{bodaghee03}) 
and Beir\~ao et al. (\citeyear{beirao05}). Right: the same comparison, in the 
[$\alpha$/Fe]-$T_\mathrm{eff}$ plane. \label{chemical2}}

\figcaption[]{Top panels: differences in $T_\mathrm{eff}$ and [Fe/H] values 
for a sample of planet hosts (filled dots) and stars without 
known planets (open squares) in common with the 
Soubiran \& Girard (\citeyear{soubiran05}) 
and the Bodaghee et al. (\citeyear{bodaghee03}) 
and Beir\~ao et al. (\citeyear{beirao05}) samples. Bottom panels: 
differences $[\alpha/\mathrm{Fe}]_1-[\alpha/\mathrm{Fe}]_2$ between 
the former and the latter $\alpha$-element abundances as a function of 
$T_\mathrm{eff}$ and [Fe/H]. Solid lines indicate a linear fit to the data. 
\label{systematic}}

\figcaption[]{Left: a comparison, in the age-metallicity diagram, 
between stars in the Soubiran \& Girard (\citeyear{soubiran05}) catalog, 
TrES-1 (this work), and planet hosts and comparison field 
stars from Bodaghee et al. (\citeyear{bodaghee03}) 
and Beir\~ao et al. (\citeyear{beirao05}). The legend for the symbols 
is the same as in Figure~\ref{chemical2}. Right: the same comparion 
in the age-[$\alpha$/Fe] plane. \label{ages}}

\clearpage

\begin{deluxetable}{cccccc}
\tablecaption{TrES-1 equivalent widths\label{tb1}}
\tablewidth{0pt}
\tablehead{
\colhead{$\lambda$ (\AA)} & \colhead{Species} & \colhead{$\chi_l$} &
\colhead{$\log gf$} & \colhead{EW$_\lambda$ (m\AA)} & \colhead{References}
}
\startdata
6154.23 &   \ion{Na}{1} (Z= 11; $\log\epsilon_\odot = 6.33$)     &2.10   &$-1.57$  &60.6   &1\\
6318.72 &   \ion{Mg}{1} (Z= 12; $\log\epsilon_\odot = 7.58$)& 5.11& $-1.97$& Synth\tablenotemark{a} &2\\
6319.24 &   \ion{Mg}{1}    &5.11   &$-2.22$  &Synth\tablenotemark{a} &2\\
6965.41 &   \ion{Mg}{1}     &5.75   &$-1.51$  &Synth\tablenotemark{a} &3\\
6696.02 &   \ion{Al}{1} (Z= 13; $\log\epsilon_\odot = 6.47$) &3.14   &$-1.34$  &Synth\tablenotemark{a} &2\\
6698.67 &   \ion{Al}{1}     &3.14   &$-1.64$  &Synth\tablenotemark{a} &2\\
5665.56 &   \ion{Si}{1} (Z= 14; $\log\epsilon_\odot = 7.55$)    &4.92   &$-2.04$  &47.0   &4\\
5690.43 &   \ion{Si}{1}     &4.93   &$-1.87$  &49.6   &4\\
6145.02 &   \ion{Si}{1}     &5.61   &$-1.48$  &38.7   &1\\
5260.39 &   \ion{Ca}{1} (Z= 20; $\log\epsilon_\odot = 6.36$)    &2.52   &$-1.72$  &51.7   &5\\
5867.56 &   \ion{Ca}{1}     &2.93   &$-1.57$  &48.3   &6\\
6166.44 &   \ion{Ca}{1}     &2.52   &$-1.14$  &98.7   &5\\
6455.60 &   \ion{Ca}{1}     &2.52   &$-1.29$  &81.7   &5\\
5657.88 &   \ion{Sc}{2} (Z= 21; $\log\epsilon_\odot = 3.17$)   &1.51   &$-0.50$  &63.1   &7\\
5669.04 &   \ion{Sc}{2}    &1.50   &$-1.10$  &31.2   &7\\
6245.64 &   \ion{Sc}{2}    &1.51   &$-1.13$  &32.6   &7\\
4562.64 &   \ion{Ti}{1} (Z= 22; $\log\epsilon_\odot = 5.02$)    &0.02   &$-2.60$  &40.3   &8\\
4820.42 &   \ion{Ti}{1}     &1.50   &$-0.39$  &68.7   &9\\
4926.16 &   \ion{Ti}{1}     &0.82   &$-2.11$  &25.9   &10\\
5219.70 &   \ion{Ti}{1}     &0.02   &$-2.24$  &64.7   &8\\
5426.26 &   \ion{Ti}{1}     &0.02   &$-2.95$  &30.3   &8\\
5880.31 &   \ion{Ti}{1}     &1.05   &$-1.99$  &25.8   &10\\
5903.33 &   \ion{Ti}{1}     &1.05   &$-2.09$  &20.9   &10\\
5922.12 &   \ion{Ti}{1}     &1.05   &$-1.41$  &54.8   &10\\
5941.75 &   \ion{Ti}{1}     &1.05   &$-1.45$  &48.2   &1\\
6064.63 &   \ion{Ti}{1}     &1.05   &$-1.89$  &34.6   &1\\
6126.22 &   \ion{Ti}{1}     &1.05   &$-1.37$  &57.0   &10\\
6312.24 &   \ion{Ti}{1}     &1.46   &$-1.50$  &27.5   &9\\
6554.22 &   \ion{Ti}{1}     &1.46   &$-1.02$  &50.0   &9\\
6556.06 &   \ion{Ti}{1}     &1.44   &$-1.07$  &51.4   &9\\
6599.11 &   \ion{Ti}{1}     &0.90   &$-2.03$  &35.2   &10\\
4568.31 &   \ion{Ti}{2}    &1.22   &$-2.52$  &36.9   &11\\
4583.44 &   \ion{Ti}{2}    &1.17   &$-2.77$ &28.7   &11\\
4589.96 &   \ion{Ti}{2}    &1.24   &$-1.75$  &82.3   &11\\
5336.81 &   \ion{Ti}{2}    &1.58   &$-1.70$  &66.6   &11\\
5418.80 &   \ion{Ti}{2}    &1.58   &$-1.86$  &48.6   &11\\
6090.22 &   \ion{V}{1}  (Z= 23; $\log\epsilon_\odot = 4.00$)    &1.08   &$-0.06$  &78.4   &7\\
6216.34 &   \ion{V}{1}      &0.28   &$-1.29$  &74.2   &7\\
6251.82 &   \ion{V}{1}      &0.29   &$-1.34$  &60.7   &7\\
6274.64 &   \ion{V}{1}      &0.27   &$-1.67$  &39.7   &7\\
5238.96 &   \ion{Cr}{1} (Z= 24; $\log\epsilon_\odot = 5.67$)    &2.71   &$-1.51$  &33.3   &12\\
5537.74 &   \ion{Mn}{1} (Z= 25; $\log\epsilon_\odot = 5.39$)    &2.19   &$-2.02$  &79.5   &7\\
4602.00 &   \ion{Fe}{1} (Z= 26; $\log\epsilon_\odot = 7.48$)    &1.61   &$-3.15$  &98.5   &13 \\
4745.80 &   \ion{Fe}{1}     &3.65   &$-1.29$  &108.9  &13\\
4788.75 &   \ion{Fe}{1}     &3.23   &$-1.78$  &86.5   &13\\
4802.52 &   \ion{Fe}{1}     &4.60   &$-1.69$  &27.6   &13\\
4802.88 &   \ion{Fe}{1}     &3.69   &$-1.53$  &78.1   &13\\
4809.94 &   \ion{Fe}{1}     &3.57   &$-2.57$  &32.6   &13\\
4961.91 &   \ion{Fe}{1}     &3.63   &$-2.33$  &41.3   &13\\
5016.48 &   \ion{Fe}{1}     &4.25   &$-1.54$ &43.3   &13\\
5242.49 &   \ion{Fe}{1}     &3.63   &$-1.03$  &110.3  &13\\
5373.70 &   \ion{Fe}{1}     &4.47   &$-0.80$  &83.4   &13\\
5379.57 &   \ion{Fe}{1}     &3.69   &$-1.54$  &81.5   &13\\
5386.34 &   \ion{Fe}{1}     &4.15   &$-1.72$  &49.7   &13\\
5417.04 &   \ion{Fe}{1}     &4.41   &$-1.45$  &49.9   &13\\
5472.71 &   \ion{Fe}{1}     &4.21   &$-1.56$  &64.5   &13\\
5538.52 &   \ion{Fe}{1}     &4.21   &$-1.55$  &52.2   &13\\
5560.21 &   \ion{Fe}{1}     &4.43   &$-1.06$  &64.2   &13\\
5577.03 &   \ion{Fe}{1}     &5.03   &$-1.47$  &17.1   &13\\
5638.26 &   \ion{Fe}{1}     &4.22   &$-0.74$  &110.1  &13\\
5679.03 &   \ion{Fe}{1}     &4.65   &$-0.70$  &86.8   &13\\
5696.10 &   \ion{Fe}{1}     &4.55   &$-1.79$  &20.7   &13\\
5741.85 &   \ion{Fe}{1}     &4.25   &$-1.65$  &45.5   &13\\
5778.45 &   \ion{Fe}{1}     &2.59   &$-3.46$  &40.2   &13\\
5809.22 &   \ion{Fe}{1}     &3.88   &$-1.55$  &78.5   &13\\
5811.92 &   \ion{Fe}{1}     &4.14   &$-2.27$  &18.2   &13\\
5855.09 &   \ion{Fe}{1}     &4.60   &$-1.55$  &30.0   &13\\
5909.97 &   \ion{Fe}{1}     &3.21   &$-2.63$  &55.3   &13\\
5956.69 &   \ion{Fe}{1}     &0.86   &$-4.61$  &80.2   &13\\
6027.05 &   \ion{Fe}{1}     &4.07   &$-1.14$  &83.6   &13\\
6151.62 &   \ion{Fe}{1}     &2.17   &$-3.30$  &69.3   &13\\
6165.36 &   \ion{Fe}{1}     &4.14   &$-1.51$  &57.2   &13\\
6265.13 &   \ion{Fe}{1}     &2.17   &$-2.55$  &114.4  &13\\
6322.69 &   \ion{Fe}{1}     &2.59   &$-2.43$  &99.2   &13\\
6481.87 &   \ion{Fe}{1}     &2.28   &$-2.98$  &84.3   &13\\
6609.11 &   \ion{Fe}{1}     &2.56   &$-2.69$  &85.5   &13\\
6739.52 &   \ion{Fe}{1}     &1.56   &$-4.82$  &30.8   &13\\
6750.15 &   \ion{Fe}{1}     &2.42   &$-2.62$  &103.4  &13\\
4656.98 &   \ion{Fe}{2}    &2.89   &$-3.55$  &27.8   &14\\
4670.18 &   \ion{Fe}{2}    &2.58   &$-3.90$  &22.0   &14 \\
4993.36 &   \ion{Fe}{2}    &2.80   &$-3.49$  &36.2   &14 \\
5197.58 &   \ion{Fe}{2}    &3.23   &$-2.23$  &70.0   &14 \\
5425.26 &   \ion{Fe}{2}    &3.20   &$-3.37$  &31.5   &14 \\
6149.26 &   \ion{Fe}{2}    &3.89   &$-2.72$  &23.4   &14 \\
6247.56 &   \ion{Fe}{2}    &3.89   &$-2.33$  &35.7   &14 \\
6456.38 &   \ion{Fe}{2}    &3.90   &$-2.08$  &48.8   &14 \\
5342.71 &   \ion{Co}{1}  (Z= 27; $\log\epsilon_\odot = 4.91$)   &4.02   & $0.54$  &40.2   &7\\
5352.05 &   \ion{Co}{1}     &3.58   & $0.06$  &39.0   &7\\
5647.23 &   \ion{Co}{1}     &2.28   &$-1.56$  &29.6   &7\\
5578.73 &   \ion{Ni}{1}  (Z= 28; $\log\epsilon_\odot = 6.30$)   &1.68   &$-2.79$  &72.1   &3\\
5748.35 &   \ion{Ni}{1}     &1.68   &$-3.26$  &51.7   &2\\
6007.31 &   \ion{Ni}{1}     &1.68   &$-3.34$  &42.2   &2\\
6111.07 &   \ion{Ni}{1}     &4.09   &$-0.87$  &46.2   &12\\
6128.96 &   \ion{Ni}{1}     &1.68   &$-3.43$  &45.0   &12\\
6133.96 &   \ion{Ni}{1}     &4.09   &$-1.83$   &7.8   &12\\
6176.81 &   \ion{Ni}{1}     &4.09   &$-0.35$  &81.2   &12\\
6177.24 &   \ion{Ni}{1}     &1.83   &$-3.60$  &28.0   &12\\
5105.00 &   \ion{Cu}{1}  (Z= 29; $\log\epsilon_\odot = 4.21$)&1.39   &$-1.52$  &Synth\tablenotemark{a} &15\\
4722.15 &   \ion{Zn}{1}  (Z= 30; $\log\epsilon_\odot = 4.60$)&4.03   &$-0.34$  &Synth\tablenotemark{a} &3\\
4810.53 &   \ion{Zn}{1}     &4.08   &$-0.14$  &Synth\tablenotemark{a} &3\\
5087.43 &   \ion{Y}{2}   (Z= 39; $\log\epsilon_\odot = 2.12$)  &1.08   &$-0.16$  &44.1   &16\\
5402.78 &   \ion{Y}{2}     &1.84   &$-0.44$  &12.1   &16\\
5853.67 &   \ion{Ba}{2}  (Z= 56; $\log\epsilon_\odot = 2.50$)  &0.60   &$-1.01$  &66.6   &7\\
\enddata
\tablenotetext{a}{The abundance was determined from spectrum synthesis
as no reliable EW measurement was available}
\tablerefs{
(1)~\citealp{ps03}; (2)~\citealp{rc02}; (3)~\citealp{kb95}; 
(4)~\citealp{g73}; (5)~\citealp{sr81}; (6)~\citealp{smith88}; 
(7)~\citeauthor{pn00}~\citeyear{pn00}; (8)~\citealp{bl82}; (9)~\citealp{bl86}; 
(10)~\citealp{bl83}; (11)~\citealp{sava90}; (12)~\citeauthor{tf00}~\citeyear{tf00};
(13)~\citealp{lee02}; (14)~\citealp{b91}; (15)~\citeauthor{ss03}~\citeyear{ss03}; 
(16)~\citealp{reddy03}
}
\tablecomments{Spectral lines used in the elemental abundance analysis of
the planet-host star TrES-1. Columns 1 through 6 report wavelength $\lambda$ 
(in \AA), species name, nominal solar abundance $\log\epsilon_\odot$ and Z number, 
lower excitation potential $\xi_l$ (in eV), oscillator strengths 
$\log gf$, equivalent widths EW$_\lambda$ (in m\AA), and 
the reference number from which the $\log gf$ values were taken, respectively}
\end{deluxetable}

\clearpage

\begin{deluxetable}{cccc}
\tablecaption{TrES-1: average abundance ratios\label{tb2}}
\tablewidth{0pt}
\tablehead{
\colhead{Species} & \colhead{Mean} & \colhead{$\sigma$} & \colhead{$n$}
}
\startdata
$[\mathrm{Na/Fe}]$   &  $-0.06$   &  \dots    &   1  \\
$[\mathrm{Mg/Fe}]$   &  $-0.04$   &   0.05    &   3\\
$[\mathrm{Al/Fe}]$   &  $-0.07$   &   0.03    &   2\\
$[\mathrm{Si/Fe}]$   &   0.07   &   0.06    &   3\\
$[\mathrm{Ca/Fe}]$   &  $-0.06$   &   0.09    &   4\\
$[\mathrm{Sc/Fe}]$   &   $-0.08$   &   0.03    &   3\\
$[\mathrm{Ti/Fe}]$  &  $-0.03$   &   0.05    &  15\\
$[\mathrm{V/Fe}]$    &  0.00    &   0.03    &   4\\
$[\mathrm{Cr/Fe}]$  &  0.10    &  \dots    &   1\\
$[\mathrm{Mn/Fe}]$   &  0.09    &   \dots    &   1\\
$[\mathrm{Fe/H}]$   &  0.00    &   0.09    &  36\\
$[\mathrm{Co/Fe}]$   & $-0.05$    &   0.04    &   3\\
$[\mathrm{Ni/Fe}]$   &  0.07    &   0.05    &   8\\
$[\mathrm{Cu/Fe}]$   & $-0.05$    &  \dots    &   1\\
$[\mathrm{Zn/Fe}]$   & $-0.13$    &   0.04    &   2\\
$[\mathrm{Y/Fe}]$    &  0.01    &   0.10    &   2\\
$[\mathrm{Ba/Fe}]$   & $-0.05$    &  \dots    &   1  \\
\enddata
\end{deluxetable}

\clearpage

\begin{deluxetable}{cccc}
\tablecaption{TrES-1: sensitivities to atmospheric parameters\label{tb3}}
\tablewidth{0pt}
\tablehead{
\colhead{Species} & \colhead{$\Delta T_\mathrm{eff} = +75$ K} &
\colhead{$\Delta\log g = +0.2$ dex} & \colhead{$\Delta\xi_t = +0.10$ km s$^{-1}$}
}
\startdata
$[\mathrm{Na/Fe}]$  & 0.05  &$-0.06$  & 0.01 \\
$[\mathrm{Mg/Fe}]$  & 0.02  &$-0.04$  & 0.02 \\
$[\mathrm{Al/Fe}]$  & 0.05  &$-0.06$  & 0.01 \\
$[\mathrm{Si/Fe}]$  &$-0.01$  &$-0.01$  & 0.01  \\
$[\mathrm{Ca/Fe}]$  & 0.07  &$-0.08$  & 0.01  \\
$[\mathrm{Sc/Fe}]$  &-0.01  & $0.05$  & 0.00 \\
$[\mathrm{Ti\,I/Fe}]$ & 0.10  &$-0.04$  & 0.00 \\
$[\mathrm{Ti\,II/Fe}]$    & 0.01  & $0.03$  & 0.00 \\
$[\mathrm{V/Fe}]$   & 0.10  &$-0.04$  & 0.01 \\
$[\mathrm{Cr/Fe}]$ & 0.07  &$-0.05$  & 0.02 \\
$[\mathrm{Mn/Fe}]$  & 0.07  &$-0.05$  &$-0.01$ \\
$[\mathrm{Co/Fe}]$  & 0.02  &$-0.01$  & 0.01 \\
$[\mathrm{Ni/Fe}]$  & 0.04  &$-0.02$  & 0.01 \\
$[\mathrm{Cu/Fe}]$  & 0.05  &$-0.05$  &$-0.03$ \\
$[\mathrm{Zn/Fe}]$  &$-0.01$  &$-0.02$  &$-0.01$ \\
$[\mathrm{Y/Fe}]$   & 0.01  &$ 0.04$  & 0.00 \\
$[\mathrm{Ba/Fe}]$  & 0.04  &$ 0.00$  &$-0.04$ \\
\enddata
\end{deluxetable}


\topmargin      0mm \headheight      0cm
\headsep         0cm \textheight    245mm \footskip        -1.cm
\textwidth     175mm
\oddsidemargin   -0.2cm  
\evensidemargin  -0.2cm  %

{
\begin{deluxetable}{lcccccccccccccccccc}
\centering
\tabletypesize{\tiny}
\rotate
\tablenum{4}
\tablewidth{1.2\textheight}
\tablecaption{Kinematic data of stars with planets and TrES-1\label{tb4}}
\tablewidth{0pt}
\tablehead{
\colhead{Name} & \colhead{$\pi$} &
\colhead{HRV} & \colhead{$\mu_\alpha$} & \colhead{$\mu_\delta$} & 
\colhead{$\alpha$ (J2000)} 
& \colhead{$\delta$ (J2000)} & \colhead{$U$} & \colhead{$V$} & \colhead{$W$} 
& \colhead{$P^{\mathrm{thin}}$} & \colhead{$P^{\mathrm{thick}}$} 
& \colhead{$TD/D$} & \colhead{$P_1^{\mathrm{thin}}$} 
& \colhead{$P_1^{\mathrm{thick}}$} & \colhead{$P_2^{\mathrm{thin}}$} 
& \colhead{$P_2^{\mathrm{thick}}$} & \colhead{$X$} & \colhead{[$\alpha$/Fe]} \\
\colhead{} & \colhead{(mas)} & \colhead{km s$^{-1}$} & 
\colhead{mas yr$^{-1}$} & \colhead{mas yr$^{-1}$} & \colhead{h:m:s} & 
\colhead{d:m:s}&  \colhead{km s$^{-1}$} & \colhead{km s$^{-1}$} & \colhead{km s$^{-1}$} 
&\colhead{(M04)} &\colhead{(M04)} & \colhead{(B03)} & \colhead{(V04)} 
&\colhead{(V04)} &\colhead{(V04)} & \colhead{(V04)} & \colhead{(S93)} &
}
\startdata
  TrES-1   &  6.66   & $-20.520$ &  $ -42.00$ &  $ -22.00$ &  19 04  09.80 & $+36$ 37 57.5 &    $-24.5$ &  $-21.4$ &   $ 23.1$ &  0.92 &   0.08 &   0.02 &   0.70 &   0.30 &   1.00 &   0.00 & $-32.37$ &  $-0.00$ \\
  HD142    &  39.00  & $  2.600$ &  $ 575.21$ &  $ -39.92$ &  00 06 19.18 &  $-49$ 04 30.7 &    $ 49.2$ &  $-24.8$ &   $ -5.8$ &  0.92 &   0.08 &   0.02 &   0.70 &   0.30 &   1.00 &   0.00 & $-34.50$ &  $-0.01$ \\
  HD1237   &  56.76  & $ -5.808$ &  $ 433.88$ &  $ -57.91$ &  00 16 12.68 &  $-79$ 51 04.3 &    $ 24.1$ &  $ -4.3$ &   $  9.3$ &  0.97 &   0.03 &   0.01 &   0.80 &   0.20 &   1.00 &   0.00 & $-36.76$ &  $-0.04$ \\
  HD2039   &  11.13  & $  8.400$ &  $  78.53$ &  $  15.23$ &  00 24 20.28 &  $-56$ 39 00.2 &    $ 19.6$ &  $ -2.6$ &   $ -6.9$ &  0.97 &   0.03 &   0.01 &   0.90 &   0.10 &   1.00 &   0.00 & $-40.72$ &  $-0.01$ \\
  HD2638   &  18.62  & $  9.550$ &  $-107.08$ &  $-224.06$ &  00 29 59.87 &  $-05$ 45 50.4 &    $-59.7$ &  $-15.7$ &   $-20.3$ &  0.90 &   0.10 &   0.03 &   0.60 &   0.30 &   1.00 &   0.00 & $-36.05$ &  \nodata \\
  HD3651   &  90.03  & $-34.200$ &  $-461.09$ &  $-370.90$ &  00 39 21.81 &  $+21$ 15 01.7 &    $-49.5$ &  $ -8.3$ &   $ 16.2$ &  0.94 &   0.06 &   0.02 &   0.80 &   0.20 &   1.00 &   0.00 & $-36.25$ &  \nodata \\
  HD4203   &  12.85  & $-14.140$ &  $ 125.25$ &  $-123.99$ &  00 44 41.20 &  $+20$ 26 56.1 &    $  7.5$ &  $-47.2$ &   $-18.4$ &  0.74 &   0.26 &   0.05 &   0.30 &   0.60 &   0.90 &   0.10 & $-36.47$ &  $ 0.01$ \\
  HD4208   &  30.58  & $ 55.400$ &  $ 313.51$ &  $ 150.00$ &  00 44 26.65 &  $-26$ 30 56.4 &    $ 43.5$ &  $  7.1$ &   $-49.4$ &  0.72 &   0.28 &   0.43 &   0.40 &   0.60 &   0.90 &   0.10 & $-31.52$ &  $ 0.06$ \\
  HD6434   &  24.80  & $ 22.962$ &  $-168.97$ &  $-527.70$ &  01 04 40.15 &  $-39$ 29 17.6 &    $-94.0$ &  $-55.0$ &   $  4.4$ &  0.23 &   0.77 &   0.83 &   0.10 &   0.80 &   0.70 &   0.30 & $-18.26$ &  $ 0.22$ \\
  HD8574   &  22.65  & $ 18.864$ &  $ 252.59$ &  $-158.59$ &  01 25 15.52 &  $+28$ 34 00.1 &    $ 35.1$ &  $-24.7$ &   $-23.8$ &  0.90 &   0.10 &   0.03 &   0.60 &   0.40 &   1.00 &   0.00 & $-33.01$ &  $-0.04$ \\
  HD9826   &  74.25  & $-27.700$ &  $-172.57$ &  $-381.03$ &  01 36 47.84 &  $+41$ 24 19.7 &    $-37.1$ &  $ -9.8$ &   $ -7.5$ &  0.96 &   0.04 &   0.01 &   0.80 &   0.20 &   1.00 &   0.00 & $-36.24$ &  $ 0.04$ \\
  HD10647  &  57.63  & $ 12.900$ &  $ 166.97$ &  $-106.71$ &  01 42 29.32 &  $-53$ 44 27.0 &    $ -5.8$ &  $ -7.9$ &   $  2.1$ &  0.97 &   0.03 &   0.00 &   0.90 &   0.10 &   1.00 &   0.00 & $-33.50$ &  \nodata \\
  HD10697  &  30.71  & $-44.800$ &  $ -45.05$ &  $-105.39$ &  01 44 55.82 &  $+20$ 04 59.3 &    $-44.9$ &  $-15.6$ &   $ 23.1$ &  0.92 &   0.08 &   0.02 &   0.70 &   0.30 &   1.00 &   0.00 & $-35.69$ &  $-0.01$ \\
  HD12661  &  26.91  & $-52.200$ &  $-107.81$ &  $-175.26$ &  02 04 34.29 &  $+25$ 24 51.5 &    $-63.8$ &  $-20.0$ &   $  7.4$ &  0.91 &   0.09 &   0.02 &   0.70 &   0.30 &   1.00 &   0.00 & $-39.23$ &  $-0.02$ \\
  HD13445  &  91.63  & $ 56.570$ &  $2092.59$ &  $ 654.49$ &  02 10 25.93 &  $-50$ 49 25.4 &    $ 88.6$ &  $-63.2$ &   $-22.6$ &  0.08 &   0.92 &   3.00 &   0.00 &   0.90 &   0.40 &   0.60 & $-22.44$ &  $ 0.12$ \\
  HD16141  &  27.85  & $-51.500$ &  $-156.89$ &  $-437.07$ &  02 35 19.93 &  $-03$ 33 38.2 &    $-93.7$ &  $-29.4$ &   $  9.0$ &  0.70 &   0.30 &   0.15 &   0.40 &   0.60 &   0.90 &   0.10 & $-34.08$ &  $-0.03$ \\
  HD17051  &  58.00  & $ 15.500$ &  $ 333.72$ &  $ 219.21$ &  02 42 33.47 &  $-50$ 48 01.1 &    $ 22.2$ &  $ -4.4$ &   $ -0.6$ &  0.97 &   0.03 &   0.01 &   0.90 &   0.10 &   1.00 &   0.00 & $-39.37$ &  $-0.05$ \\
  HD19994  &  44.69  & $ 19.331$ &  $ 193.43$ &  $ -69.23$ &  03 12 46.44 &  $-01$ 11 46.0 &    $ 11.5$ &  $ -7.4$ &   $  0.1$ &  0.97 &   0.03 &   0.00 &   0.90 &   0.10 &   1.00 &   0.00 & $-38.61$ &  $-0.02$ \\
  HD20367  &  36.86  & $  5.300$ &  $-103.09$ &  $ -56.65$ &  03 17 40.05 &  $+31$ 07 37.4 &    $-11.8$ &  $ 18.0$ &   $ -7.5$ &  0.96 &   0.04 &   0.01 &   0.90 &   0.10 &   1.00 &   0.00 & $-40.59$ &  $-0.08$ \\
  HD22049  & 310.74  & $ 16.300$ &  $-976.36$ &  $  17.98$ &  03 32 55.84 &  $-09$ 27 29.7 &    $ -5.4$ &  $ 19.0$ &   $-13.6$ &  0.96 &   0.04 &   0.01 &   0.90 &   0.10 &   1.00 &   0.00 & $-35.12$ &  $ 0.03$ \\
  HD23079  &  28.90  & $  0.500$ &  $-193.62$ &  $ -91.92$ &  03 39 43.10 &  $-52$ 54 57.0 &    $-36.8$ &  $ 26.8$ &   $ -8.5$ &  0.94 &   0.06 &   0.02 &   0.90 &   0.10 &   1.00 &   0.00 & $-36.49$ &  $ 0.02$ \\
  HD23596  &  19.24  & $-10.200$ &  $  53.56$ &  $  21.06$ &  03 48 00.37 &  $+40$ 31 50.3 &    $-13.1$ &  $  2.4$ &   $ 21.0$ &  0.96 &   0.04 &   0.01 &   0.80 &   0.20 &   1.00 &   0.00 & $-41.20$ &  $-0.02$ \\
  HD27442  &  54.84  & $ 29.300$ &  $ -47.99$ &  $-167.81$ &  04 16 29.03 &  $-59$ 18 07.8 &    $-24.1$ &  $-10.1$ &   $-12.1$ &  0.96 &   0.04 &   0.01 &   0.80 &   0.20 &   1.00 &   0.00 & $-41.06$ &  $-0.03$ \\
  HD27894  &  23.60  & $ 82.900$ &  $ 182.25$ &  $ 272.33$ &  04 20 47.05 &  $-59$ 24 39.0 &    $ 54.3$ &  $-61.3$ &   $-35.6$ &  0.13 &   0.87 &   1.80 &   0.00 &   0.90 &   0.50 &   0.50 & $-32.78$ &  \nodata \\
  HD28185  &  25.28  & $ 50.246$ &  $  80.85$ &  $ -60.29$ &  04 26 26.32 &  $-10$ 33 02.9 &    $ 24.6$ &  $-22.7$ &   $-16.2$ &  0.93 &   0.07 &   0.01 &   0.70 &   0.30 &   1.00 &   0.00 & $-36.26$ &  $-0.04$ \\
  HD33636  &  34.85  & $  5.300$ &  $ 180.83$ &  $-137.32$ &  05 11 46.45 &  $+04$ 24 12.7 &    $ -8.8$ &  $-18.0$ &   $ 16.0$ &  0.95 &   0.05 &   0.01 &   0.80 &   0.20 &   1.00 &   0.00 & $-31.27$ &  $ 0.01$ \\
  HD37124  &  30.08  & $-19.000$ &  $ -79.75$ &  $-419.96$ &  05 37 02.49 &  $+20$ 43 50.8 &    $-37.6$ &  $-34.7$ &   $-36.4$ &  0.70 &   0.30 &   0.15 &   0.30 &   0.70 &   0.90 &   0.10 & $-23.50$ &  $ 0.17$ \\
  HD37605  &  23.32  & $-22.050$ &  $  54.70$ &  $-245.76$ &  05 40 01.73 &  $+06$ 03 38.1 &    $-48.4$ &  $-26.0$ &   $ -2.7$ &  0.92 &   0.08 &   0.02 &   0.70 &   0.30 &   1.00 &   0.00 & $-37.51$ &  \nodata \\
  HD38529  &  23.57  & $ 30.000$ &  $ -80.05$ &  $-141.79$ &  05 46 34.91 &  $+01$ 10 05.5 &    $  4.3$ &  $-13.1$ &   $-27.0$ &  0.94 &   0.06 &   0.02 &   0.70 &   0.30 &   1.00 &   0.00 & $-40.86$ &  $-0.00$ \\
  HD39091  &  54.92  & $  9.400$ &  $ 311.88$ &  $1050.20$ &  05 37 09.89 &  $-80$ 28 08.8 &    $ 74.1$ &  $-33.9$ &   $  6.5$ &  0.78 &   0.22 &   0.06 &   0.40 &   0.50 &   0.90 &   0.10 & $-32.58$ &  $-0.01$ \\
  HD40979  &  30.00  & $ 32.800$ &  $  95.05$ &  $-152.23$ &  06 04 29.94 &  $+44$ 15 37.6 &    $ 27.7$ &  $ -9.3$ &   $ 15.3$ &  0.96 &   0.04 &   0.01 &   0.80 &   0.20 &   1.00 &   0.00 & $-37.80$ &  \nodata \\
  HD41004A &  23.24  & $ 42.200$ &  $ -42.27$ &  $  65.16$ &  05 59 49.65 &  $-48$ 14 22.9 &    $ 13.7$ &  $-17.5$ &   $-18.3$ &  0.95 &   0.05 &   0.01 &   0.70 &   0.30 &   1.00 &   0.00 & $-35.81$ &  \nodata \\
  HD46375  &  29.93  & $  4.000$ &  $ 114.24$ &  $ -96.79$ &  06 33 12.62 &  $+05$ 27 46.5 &    $-14.8$ &  $ -9.5$ &   $ 15.8$ &  0.96 &   0.04 &   0.01 &   0.80 &   0.20 &   1.00 &   0.00 & $-37.59$ &  $ 0.05$ \\
  HD47536  &   8.24  & $ 78.800$ &  $ 108.96$ &  $  64.13$ &  06 37 47.62 &  $-32$ 20 23.0 &    $ 46.8$ &  $-67.0$ &   $ 53.1$ &  0.02 &   0.98 &  31.00 &   0.00 &   0.90 &   0.10 &   0.90 & $-16.34$ &  \nodata \\
  HD49674  &  24.55  & $  11.80$ &  $  34.96$ &  $-122.85$ &  06 51 30.52 &  $+40$ 52 03.9 &    $  4.8$ &  $-11.5$ &   $  8.3$ &  0.97 &   0.03 &   0.01 &   0.80 &   0.20 &   1.00 &   0.00 & $-39.77$ &  \nodata \\
  HD50554  &  32.23  & $ -3.861$ &  $ -37.29$ &  $ -96.36$ &  06 54 42.83 &  $+24$ 14 44.0 &    $-12.6$ &  $  1.9$ &   $ -4.4$ &  0.97 &   0.03 &   0.01 &   0.90 &   0.10 &   1.00 &   0.00 & $-35.52$ &  $ 0.00$ \\
  HD52265  &  35.63  & $ 53.600$ &  $-115.76$ &  $  80.35$ &  07 00 18.04 &  $-05$ 22 01.8 &    $ 43.1$ &  $ -8.5$ &   $ -2.2$ &  0.96 &   0.04 &   0.01 &   0.80 &   0.20 &   1.00 &   0.00 & $-38.28$ &  $-0.04$ \\
  HD59686  &  10.81  & $-40.200$ &  $  42.65$ &  $ -75.39$ &  07 31 48.40 &  $+17$ 05 09.8 &    $-59.7$ &  $ -8.4$ &   $ -0.8$ &  0.94 &   0.06 &   0.01 &   0.80 &   0.20 &   1.00 &   0.00 & $-39.23$ &  \nodata \\
  HD63454  &  27.93  & $ 33.840$ &  $ -20.65$ &  $ -39.69$ &  07 39 21.85 &  $-78$ 16 44.3 &    $-24.0$ &  $-13.0$ &   $-11.7$ &  0.96 &   0.04 &   0.01 &   0.80 &   0.20 &   1.00 &   0.00 & $-35.46$ &  \nodata \\
  HD65216  &  28.10  & $ 42.300$ &  $-122.12$ &  $ 145.90$ &  07 53 41.32 &  $-63$ 38 50.4 &    $ 17.8$ &  $-29.1$ &   $-13.2$ &  0.92 &   0.08 &   0.01 &   0.60 &   0.30 &   1.00 &   0.00 & $-29.08$ &  \nodata \\
  HD68988  &  17.00  & $-69.700$ &  $ 128.33$ &  $  31.73$ &  08 18 22.17 &  $+61$ 27 38.6 &    $-84.1$ &  $ -9.5$ &   $ -2.9$ &  0.88 &   0.12 &   0.04 &   0.70 &   0.30 &   1.00 &   0.00 & $-40.58$ &  \nodata \\
  HD70642  &  34.77  & $ 48.100$ &  $-202.07$ &  $ 225.59$ &  08 21 28.14 &  $-39$ 42 19.5 &    $ 41.2$ &  $-26.0$ &   $  0.4$ &  0.93 &   0.07 &   0.01 &   0.70 &   0.30 &   1.00 &   0.00 & $-35.09$ &  \nodata \\
  HD72659  &  19.47  & $-18.400$ &  $-113.75$ &  $ -98.30$ &  08 34 03.19 &  $-01$ 34 05.6 &    $-16.4$ &  $ 10.0$ &   $-33.2$ &  0.92 &   0.08 &   0.04 &   0.80 &   0.20 &   1.00 &   0.00 & $-36.94$ &  \nodata \\
  HD73256  &  27.38  & $ 29.500$ &  $-180.58$ &  $  65.71$ &  08 36 23.02 &  $-30$ 02 15.5 &    $ 27.2$ &  $ -8.9$ &   $ -7.9$ &  0.96 &   0.04 &   0.01 &   0.80 &   0.20 &   1.00 &   0.00 & $-38.79$ &  \nodata \\
  HD73526  &  10.57  & $ 26.100$ &  $ -60.31$ &  $ 161.70$ &  08 37 16.48 &  $-41$ 19 08.8 &    $ 68.3$ &  $ -1.8$ &   $ 29.1$ &  0.86 &   0.14 &   0.08 &   0.60 &   0.40 &   1.00 &   0.00 & $-39.90$ &  \nodata \\
  HD74156  &  15.49  & $  3.813$ &  $  24.96$ &  $-200.48$ &  08 42 55.12 &  $+04$ 34 41.2 &    $-37.6$ &  $-39.8$ &   $-11.1$ &  0.83 &   0.17 &   0.03 &   0.50 &   0.50 &   0.90 &   0.10 & $-32.94$ &  $-0.02$ \\
  HD75289  &  34.55  & $  9.258$ &  $ -20.50$ &  $-227.68$ &  08 47 40.39 &  $-41$ 44 12.4 &    $-30.0$ &  $ -0.6$ &   $-14.6$ &  0.96 &   0.04 &   0.01 &   0.80 &   0.20 &   1.00 &   0.00 & $-40.24$ &  $-0.04$ \\
  HD75732  &  79.80  & $ 27.800$ &  $-485.46$ &  $-234.40$ &  08 52 35.81 &  $+28$ 19 50.9 &    $ 28.4$ &  $ -6.2$ &   $ -0.7$ &  0.97 &   0.03 &   0.01 &   0.80 &   0.20 &   1.00 &   0.00 & $-40.44$ &  $ 0.01$ \\
  HD76700  &  16.75  & $ 36.700$ &  $-283.05$ &  $ 121.22$ &  08 53 55.52 &  $-66$ 48 03.6 &    $ 59.9$ &  $-29.3$ &   $-42.7$ &  0.57 &   0.43 &   0.46 &   0.20 &   0.80 &   0.80 &   0.20 & $-38.96$ &  \nodata \\
  HD80606  &  17.13  & $  3.768$ &  $  46.98$ &  $   6.92$ &  09 22 37.57 &  $+50$ 36 13.4 &    $-15.7$ &  $ 14.8$ &   $ 18.5$ &  0.96 &   0.04 &   0.01 &   0.90 &   0.10 &   1.00 &   0.00 & $-42.98$ &  $ 0.00$ \\
  HD82943  &  36.42  & $  8.060$ &  $   2.38$ &  $-174.05$ &  09 34 50.74 &  $-12$ 07 46.4 &    $-19.2$ &  $ -7.8$ &   $ -1.8$ &  0.97 &   0.03 &   0.01 &   0.80 &   0.20 &   1.00 &   0.00 & $-39.68$ &  $-0.06$ \\
  HD83443  &  22.97  & $ 28.917$ &  $  22.35$ &  $-120.76$ &  09 37 11.83 &  $-43$ 16 19.9 &    $-28.9$ &  $-18.6$ &   $ -4.8$ &  0.95 &   0.05 &   0.01 &   0.80 &   0.20 &   1.00 &   0.00 & $-39.22$ &  $ 0.01$ \\
  HD88133  &  13.43  & $  -3.53$ &  $ -12.87$ &  $-263.91$ &  10 10 07.68 &  $+18$ 11 12.7 &    $-39.6$ &  $-72.2$ &   $-19.0$ &  0.11 &   0.89 &   1.10 &   0.10 &   0.90 &   0.50 &   0.50 & $-31.93$ &  \nodata \\
  HD89744  &  25.65  & $  6.500$ &  $-120.17$ &  $-138.60$ &  10 22 10.56 &  $+41$ 13 46.3 &    $  8.6$ &  $-17.4$ &   $  3.5$ &  0.96 &   0.04 &   0.01 &   0.80 &   0.20 &   1.00 &   0.00 & $-36.95$ &  \nodata \\
  HD92788  &  30.94  & $ -4.000$ &  $ -12.63$ &  $-222.75$ &  10 42 48.53 &  $-02$ 11 01.5 &    $-25.0$ &  $-10.6$ &   $-13.3$ &  0.96 &   0.04 &   0.01 &   0.80 &   0.20 &   1.00 &   0.00 & $-39.70$ &  $-0.02$ \\
  HD93083  &  34.60  & $  41.70$ &  $ -92.84$ &  $-151.12$ &  10 44 20.91 &  $-33$ 34 37.3 &    $-12.3$ &  $-35.6$ &   $  0.2$ &  0.91 &   0.09 &   0.01 &   0.60 &   0.40 &   1.00 &   0.00 & $-33.29$ &  \nodata \\
  HD95128  &  71.04  & $ 12.000$ &  $-315.92$ &  $  55.15$ &  10 59 27.97 &  $+40$ 25 48.9 &    $ 15.3$ &  $  9.5$ &   $  8.3$ &  0.97 &   0.03 &   0.01 &   0.90 &   0.10 &   1.00 &   0.00 & $-37.44$ &  $-0.01$ \\
  HD99492  &  55.59  & $  1.700$ &  $-730.15$ &  $ 191.17$ &  11 26 46.28 &  $+03$ 00 22.8 &    $ 53.2$ &  $  0.7$ &   $ -4.8$ &  0.95 &   0.05 &   0.01 &   0.80 &   0.20 &   1.00 &   0.00 & $-40.05$ &  \nodata \\
  HD101930 &  32.79  & $ 18.360$ &  $  15.00$ &  $ 347.49$ &  11 43 30.11 &  $-58$ 00 24.8 &    $ -5.1$ &  $ -6.3$ &   $ 57.1$ &  0.63 &   0.37 &   0.80 &   0.20 &   0.70 &   0.80 &   0.20 & $-37.44$ &  \nodata \\
  HD102117 &  23.81  & $  48.90$ &  $ -63.05$ &  $ -69.87$ &  11 44 50.46 &  $-58$ 42 13.4 &    $-21.6$ &  $-36.8$ &   $ -7.0$ &  0.89 &   0.11 &   0.02 &   0.60 &   0.40 &   1.00 &   0.00 & $-36.31$ &  \nodata \\
  HD104985 &   9.80  & $-19.800$ &  $ 147.22$ &  $ -92.36$ &  12 05 15.12 &  $+76$ 54 20.6 &    $-88.2$ &  $  3.6$ &   $ 40.3$ &  0.63 &   0.37 &   0.65 &   0.40 &   0.60 &   0.90 &   0.10 & $-30.32$ &  \nodata \\
  HD106252 &  26.71  & $ 15.481$ &  $  23.77$ &  $-279.41$ &  12 13 29.51 &  $+10$ 02 29.9 &    $-37.5$ &  $-31.6$ &   $  7.5$ &  0.90 &   0.10 &   0.02 &   0.60 &   0.40 &   1.00 &   0.00 & $-30.82$ &  $-0.03$ \\
  HD108147 &  25.93  & $ -5.065$ &  $-181.60$ &  $ -60.80$ &  12 25 46.27 &  $-64$ 01 19.5 &    $ 21.3$ &  $  0.6$ &   $ -7.2$ &  0.97 &   0.03 &   0.01 &   0.90 &   0.10 &   1.00 &   0.00 & $-38.90$ &  $-0.05$ \\
  HD108874 &  14.59  & $-30.700$ &  $ 129.16$ &  $ -89.40$ &  12 30 26.88 &  $+22$ 52 47.4 &    $-60.8$ &  $ 11.6$ &   $-22.2$ &  0.91 &   0.09 &   0.04 &   0.80 &   0.20 &   1.00 &   0.00 & $-40.89$ &  $-0.04$ \\
  HD111232 &  34.63  & $102.200$ &  $  27.82$ &  $ 112.81$ &  12 48 51.75 &  $-68$ 25 30.5 &    $-67.8$ &  $-72.8$ &   $ 12.5$ &  0.07 &   0.93 &   2.10 &   0.00 &   0.90 &   0.40 &   0.60 & $-18.95$ &  \nodata \\
  HD114386 &  35.66  & $ 33.370$ &  $-138.23$ &  $-325.10$ &  13 10 39.82 &  $-35$ 03 17.2 &    $-21.1$ &  $-40.1$ &   $-14.2$ &  0.84 &   0.16 &   0.03 &   0.50 &   0.50 &   0.90 &   0.10 & $-28.41$ &  $ 0.07$ \\
  HD114729 &  28.57  & $ 64.700$ &  $-202.11$ &  $-308.49$ &  13 12 44.26 &  $-31$ 52 24.1 &    $-28.3$ &  $-74.5$ &   $ -0.8$ &  0.12 &   0.88 &   0.65 &   0.10 &   0.90 &   0.50 &   0.50 & $-20.79$ &  $ 0.07$ \\
  HD114762 &  24.65  & $ 49.300$ &  $-582.68$ &  $  -1.98$ &  13 12 19.74 &  $+17$ 31 01.6 &    $ 73.9$ &  $-57.3$ &   $ 64.4$ &  0.01 &   0.99 & 220.00 &   0.00 &   0.90 &   0.10 &   0.90 & $-14.60$ &  $ 0.19$ \\
  HD114783 &  48.95  & $-12.800$ &  $-138.13$ &  $   9.62$ &  13 12 43.79 &  $-02$ 15 54.1 &    $  6.7$ &  $  9.5$ &   $ -2.5$ &  0.97 &   0.03 &   0.01 &   0.90 &   0.10 &   1.00 &   0.00 & $-38.00$ &  $-0.01$ \\
  HD117176 &  55.22   &$  5.200$  & $-234.81$  & $-576.19$  & 13 28 25.81  & $+13$ 46 43.6  &   $-22.1$ &  $-39.9$ &   $  3.3$ &  0.87 &   0.13 &   0.02 &   0.50 &   0.50 &   0.90 &   0.10 & $-28.81$ &  $ 0.02$ \\
  HD117207 &  30.29   &$-17.900$  & $-204.99$  & $ -71.73$  & 13 29 21.11  & $-35$ 34 15.6  &   $ 24.9$ &  $ -0.1$ &   $ -6.2$ &  0.97 &   0.03 &   0.01 &   0.90 &   0.10 &   1.00 &   0.00 & $-39.36$ &  \nodata \\
  HD117618 &  26.30   &$  0.900$  & $  25.04$  & $-124.63$  & 13 32 25.56  & $-47$ 16 16.9  &   $-13.9$ &  $  7.2$ &   $-14.8$ &  0.97 &   0.03 &   0.01 &   0.90 &   0.10 &   1.00 &   0.00 & $-37.15$ &  \nodata \\
  HD120136 &  64.12   &$-15.800$  & $-480.34$  & $  54.18$  & 13 47 15.74  & $+17$ 27 24.9  &   $ 24.5$ &  $ -6.7$ &   $  0.1$ &  0.97 &   0.03 &   0.01 &   0.80 &   0.20 &   1.00 &   0.00 & $-38.51$ &  \nodata \\
  HD121504 &  22.54   &$ 19.548$  & $-250.55$  & $ -84.02$  & 13 57 17.24  & $-56$ 02 24.2  &   $ 18.8$ &  $-39.8$ &   $  5.2$ &  0.87 &   0.13 &   0.02 &   0.50 &   0.50 &   0.90 &   0.10 & $-32.93$ &  $-0.05$ \\
  HD128311 &  60.35   &$ -9.600$  & $ 205.46$  & $-249.68$  & 14 36 00.56  & $+09$ 44 47.5  &   $-25.8$ &  $  7.5$ &   $-13.7$ &  0.96 &   0.04 &   0.01 &   0.90 &   0.10 &   1.00 &   0.00 & $-36.62$ &  $-0.01$ \\
  HD130322 &  33.60   &$-12.504$  & $-129.60$  & $-140.79$  & 14 47 32.73  & $-00$ 16 53.3  &   $  0.3$ &  $-14.0$ &   $ -3.8$ &  0.97 &   0.03 &   0.01 &   0.80 &   0.20 &   1.00 &   0.00 & $-33.83$ &  $-0.00$ \\
  HD134987 &  38.98   &$  5.200$  & $-399.01$  & $ -75.10$  & 15 13 28.67  & $-25$ 18 33.6  &   $ 11.5$ &  $-28.0$ &   $ 27.8$ &  0.88 &   0.12 &   0.03 &   0.50 &   0.50 &   0.90 &   0.10 & $-37.07$ &  $-0.00$ \\
  HD136118 &  19.13   &$ -3.600$  & $-124.05$  & $  23.50$  & 15 18 55.47  & $-01$ 35 32.6  &   $ 12.3$ &  $ -4.0$ &   $ 23.6$ &  0.95 &   0.05 &   0.01 &   0.80 &   0.20 &   1.00 &   0.00 & $-33.81$ &  $ 0.03$ \\
  HD137759 &  31.92  & $-10.700$ &  $  -8.27$ &  $  17.30$ &  15 24 55.77 &  $+58$ 57 57.8 &    $ -6.6$ &  $  5.0$ &   $ -1.2$ &  0.97 &   0.03 &   0.01 &   0.90 &   0.10 &   1.00 &   0.00 & $-38.16$ &  $-0.03$ \\
  HD141937 &  29.89   &$ -2.994$  & $  97.12$  & $  24.00$  & 15 52 17.55  & $-18$ 26 09.8  &   $-11.9$ &  $ 25.2$ &   $ -1.6$ &  0.96 &   0.04 &   0.01 &   0.90 &   0.10 &   1.00 &   0.00 & $-40.22$ &  $ 0.04$ \\
  HD142022 &  27.88   &$-10.500$  & $-337.60$  & $ -31.10$  & 16 10 15.02  & $-84$ 13 53.8  &   $ 21.1$ &  $-18.5$ &   $ 46.8$ &  0.75 &   0.25 &   0.21 &   0.30 &   0.70 &   0.90 &   0.10 & $-36.24$ &  \nodata \\
  HD142415 &  28.93   &$-12.000$  & $-113.96$  & $-102.35$  & 15 57 40.79  & $-60$ 12 00.9  &   $ 15.4$ &  $ -1.3$ &   $  7.4$ &  0.97 &   0.03 &   0.01 &   0.90 &   0.10 &   1.00 &   0.00 & $-38.84$ &  \nodata \\
  HD143761 &  57.38   &$ 18.000$  & $-196.88$  & $-773.00$  & 16 01 02.66  & $+33$ 18 12.6  &   $-63.1$ &  $-23.9$ &   $ 28.0$ &  0.81 &   0.19 &   0.08 &   0.50 &   0.50 &   0.90 &   0.10 & $-28.06$ &  $ 0.09$ \\
  HD145675 &  55.11   &$-13.842$  & $ 132.52$  & $-298.38$  & 16 10 24.31  & $+43$ 49 03.5  &   $-32.8$ &  $ -0.3$ &   $ -9.0$ &  0.96 &   0.04 &   0.01 &   0.90 &   0.10 &   1.00 &   0.00 & $-43.08$ &  $-0.06$ \\ 
  HD147513 &  77.69   &$ 10.100$  & $  72.64$  & $   3.41$  & 16 24 01.29  & $-39$ 11 34.7  &   $-19.8$ &  $ 11.7$ &   $  5.2$ &  0.97 &   0.03 &   0.01 &   0.90 &   0.10 &   1.00 &   0.00 & $-37.73$ &  $-0.02$ \\
  HD150706 &  36.73   &$-14.000$  & $  95.83$  & $ -87.97$  & 16 31 17.59  & $+79$ 47 23.2  &   $-27.2$ &  $  9.8$ &   $ -4.7$ &  0.97 &   0.03 &   0.01 &   0.90 &   0.10 &   1.00 &   0.00 & $-36.17$ &  $-0.04$ \\
  HD154857 &  14.59   &$ 27.900$  & $  87.19$  & $ -55.37$  & 17 11 15.72  & $-56$ 40 50.9  &   $-29.0$ &  $  3.6$ &   $-30.8$ &  0.93 &   0.07 &   0.03 &   0.70 &   0.20 &   1.00 &   0.00 & $-31.25$ &  \nodata \\
  HD160691 &  65.46   &$ -9.000$  & $ -15.06$  & $-191.17$  & 17 44 08.70  & $-51$ 50 02.6  &   $  4.6$ &  $  3.5$ &   $  3.0$ &  0.97 &   0.03 &   0.00 &   0.90 &   0.10 &   1.00 &   0.00 & $-41.53$ &  $-0.03$ \\
  HD162020 &  31.99   &$-27.600$  & $  20.99$  & $ -25.20$  & 17 50 38.36  & $-40$ 19 06.1  &   $ 18.8$ &  $ 14.8$ &   $  5.6$ &  0.97 &   0.03 &   0.01 &   0.90 &   0.10 &   1.00 &   0.00 & $-36.25$ &  $-0.02$ \\
  HD168443 &  26.40   &$-49.000$  & $ -92.15$  & $-224.16$  & 18 20 03.93  & $-09$ 35 44.6  &   $ 21.1$ &  $-45.8$ &   $  0.5$ &  0.80 &   0.20 &   0.03 &   0.40 &   0.60 &   0.90 &   0.10 & $-30.30$ &  $ 0.04$ \\
  HD168746 &  23.19   &$-25.645$  & $ -22.13$  & $ -69.23$  & 18 21 49.78  & $-11$ 55 21.7  &   $ 10.4$ &  $-10.1$ &   $  3.8$ &  0.97 &   0.03 &   0.01 &   0.80 &   0.20 &   1.00 &   0.00 & $-32.28$ &  $ 0.12$ \\
  HD169830 &  27.53   &$-17.215$  & $  -0.84$  & $ -15.16$  & 18 27 49.48  & $-29$ 49 00.7  &   $  7.9$ &  $  8.4$ &   $  8.5$ &  0.97 &   0.03 &   0.01 &   0.90 &   0.10 &   1.00 &   0.00 & $-40.11$ &  $-0.07$ \\
  HD177830 &  16.94   &$-72.300$  & $ -40.68$  & $ -51.84$  & 19 05 20.77  & $+25$ 55 14.4  &   $ 14.2$ &  $-58.5$ &   $ -0.0$ &  0.55 &   0.45 &   0.09 &   0.20 &   0.80 &   0.80 &   0.20 & $-33.70$ &  $ 0.03$ \\
  HD178911 &  20.42   &$-40.432$  & $  47.12$  & $ 194.51$  & 19 09 04.38  & $+34$ 36 01.6  &   $ 49.4$ &  $ -7.3$ &   $  7.8$ &  0.95 &   0.05 &   0.01 &   0.80 &   0.20 &   1.00 &   0.00 & $-39.18$ &  \nodata \\
  HD179949 &  36.97   &$-25.500$  & $ 114.78$  & $-101.81$  & 19 15 33.23  & $-24$ 10 45.7  &   $ 18.3$ &  $ -0.9$ &   $ -4.0$ &  0.97 &   0.03 &   0.01 &   0.90 &   0.10 &   1.00 &   0.00 & $-39.07$ &  $-0.04$ \\
  HD183263 &  18.93   &$-50.700$  & $ -17.78$  & $ -33.14$  & 19 28 24.57  & $+08$ 21 29.0  &   $ 20.3$ &  $-30.2$ &   $ 10.8$ &  0.92 &   0.08 &   0.01 &   0.60 &   0.30 &   1.00 &   0.00 & $-37.54$ &  \nodata \\
  HD186427 &  46.70   &$-27.500$  & $-135.15$  & $-163.53$  & 19 41 51.97  & $+50$ 31 03.1  &   $-26.6$ &  $-18.1$ &   $  5.3$ &  0.95 &   0.05 &   0.01 &   0.80 &   0.20 &   1.00 &   0.00 & $-34.24$ &  $-0.02$ \\
  HD187123 &  20.87   &$-17.500$  & $ 143.13$  & $-123.23$  & 19 46 58.11  & $+34$ 25 10.3  &   $-11.6$ &  $ -3.9$ &   $-36.4$ &  0.91 &   0.09 &   0.04 &   0.70 &   0.30 &   1.00 &   0.00 & $-37.01$ &  $-0.05$ \\
  HD188015 &  19.00   &$  2.600$  & $  53.89$  & $ -91.03$  & 19 52 04.54  & $+28$ 06 01.4  &   $-21.6$ &  $  9.1$ &   $-16.1$ &  0.96 &   0.04 &   0.01 &   0.90 &   0.10 &   1.00 &   0.00 & $-41.87$ &  \nodata \\
  HD190228 &  16.10   &$-50.218$  & $ 104.91$  & $ -69.85$  & 20 03 00.77  & $+28$ 18 24.7  &   $ 10.8$ &  $-35.1$ &   $-28.8$ &  0.82 &   0.18 &   0.05 &   0.40 &   0.60 &   0.90 &   0.10 & $-25.70$ &  $ 0.05$ \\
  HD190360 &  62.92   &$-45.300$  & $ 683.35$  & $-524.10$  & 20 03 37.41  & $+29$ 53 48.5  &   $  3.1$ &  $-32.8$ &   $-57.0$ &  0.37 &   0.63 &   1.70 &   0.10 &   0.90 &   0.60 &   0.40 & $-35.34$ &  $ 0.00$ \\
  HD192263 &  50.27   &$-10.817$  & $ -63.37$  & $ 262.26$  & 20 13 59.85  & $-00$ 52 00.8  &   $  7.5$ &  $ 22.1$ &   $ 26.6$ &  0.93 &   0.07 &   0.03 &   0.80 &   0.20 &   1.00 &   0.00 & $-37.57$ &  $ 0.01$ \\
  HD195019 &  26.77   &$-93.100$  & $ 349.49$  & $ -56.85$  & 20 28 18.64  & $+18$ 46 10.2  &   $ 63.5$ &  $-65.1$ &   $-30.2$ &  0.10 &   0.90 &   2.20 &   0.00 &   0.90 &   0.40 &   0.60 & $-28.18$ &  $-0.00$ \\
  HD196050 &  21.31   &$ 60.900$  & $-190.97$  & $ -64.27$  & 20 37 51.71  & $-60$ 38 04.1  &   $-74.5$ &  $-25.6$ &   $  6.4$ &  0.85 &   0.15 &   0.05 &   0.60 &   0.40 &   1.00 &   0.00 & $-35.89$ &  $-0.00$ \\
  HD202206 &  21.58   &$ 14.720$  & $ -38.23$  & $-119.77$  & 21 14 57.77  & $-20$ 47 21.2  &   $-31.6$ &  $ -7.2$ &   $ -2.8$ &  0.96 &   0.04 &   0.01 &   0.80 &   0.20 &   1.00 &   0.00 & $-40.69$ &  $-0.06$ \\
  HD208487 &  22.73   &$  5.300$  & $ 101.45$  & $-117.99$  & 21 57 19.85  & $-37$ 45 49.0  &   $  0.8$ &  $-14.8$ &   $ -9.2$ &  0.96 &   0.04 &   0.01 &   0.80 &   0.20 &   1.00 &   0.00 & $-34.29$ &  \nodata \\
  HD209458 &  21.24   &$-14.765$  & $  28.90$  & $ -18.37$  & 22 03 10.77  & $+18$ 53 03.5  &   $ -3.3$ &  $ -3.6$ &   $  7.5$ &  0.97 &   0.03 &   0.01 &   0.90 &   0.10 &   1.00 &   0.00 & $-34.99$ &  $-0.02$ \\
  HD210277 &  46.97   &$-21.100$  & $  85.48$  & $-449.83$  & 22 09 29.87  & $-07$ 32 55.2  &   $-12.9$ &  $-38.2$ &   $  0.9$ &  0.89 &   0.11 &   0.02 &   0.60 &   0.40 &   1.00 &   0.00 & $-33.70$ &  $ 0.04$ \\
  HD213240 &  24.54   &$ -0.458$  & $-135.16$  & $-194.06$  & 22 31 00.37  & $-49$ 25 59.8  &   $-34.2$ &  $-18.0$ &   $ 30.3$ &  0.89 &   0.11 &   0.04 &   0.60 &   0.40 &   1.00 &   0.00 & $-35.93$ &  $-0.03$ \\
  HD216435 &  30.04   &$ -1.100$  & $ 216.70$  & $ -81.49$  & 22 53 37.93  & $-48$ 35 53.8  &   $ 18.4$ &  $ -9.7$ &   $ -3.4$ &  0.97 &   0.03 &   0.01 &   0.80 &   0.20 &   1.00 &   0.00 & $-38.32$ &  $-0.04$ \\
  HD216437 &  37.71   &$ -3.000$  & $ -43.19$  & $  73.20$  & 22 54 39.48  & $-70$ 04 25.4  &   $-12.2$ &  $ 22.5$ &   $  6.1$ &  0.96 &   0.04 &   0.01 &   0.90 &   0.10 &   1.00 &   0.00 & $-42.67$ &  $-0.04$ \\
  HD216770 &  26.39   &$ 30.700$  & $ 228.60$  & $-178.18$  & 22 55 53.71  & $-26$ 39 31.5  &   $  3.1$ &  $-23.7$ &   $-40.1$ &  0.82 &   0.18 &   0.09 &   0.40 &   0.60 &   0.90 &   0.10 & $-36.87$ &  \nodata \\
  HD217014 &  65.10   &$-33.600$  & $ 208.07$  & $  60.96$  & 22 57 27.98  & $+20$ 46 07.8  &   $  6.2$ &  $-17.9$ &   $ 22.8$ &  0.94 &   0.06 &   0.01 &   0.70 &   0.30 &   1.00 &   0.00 & $-36.51$ &  $-0.00$ \\
  HD217107 &  50.71   &$-14.000$  & $  -6.05$  & $ -16.03$  & 22 58 15.54  & $-02$ 23 43.4  &   $ -7.4$ &  $  3.1$ &   $ 17.8$ &  0.96 &   0.04 &   0.01 &   0.90 &   0.10 &   1.00 &   0.00 & $-42.40$ &  $-0.04$ \\
  HD219449 &  21.97   &$-26.400$  & $ 368.56$  & $ -17.02$  & 23 15 53.49  & $-09$ 05 15.9  &   $ 63.2$ &  $-29.9$ &   $  0.2$ &  0.86 &   0.14 &   0.03 &   0.60 &   0.40 &   1.00 &   0.00 & $-32.15$ &  \nodata \\
  HD222404 &  72.50   &$-42.400$  & $ -48.85$  & $ 127.19$  & 23 39 20.85  & $+77$ 37 56.2  &   $-30.5$ &  $-25.4$ &   $  4.3$ &  0.94 &   0.06 &   0.01 &   0.70 &   0.30 &   1.00 &   0.00 & $-34.79$ &  \nodata \\
  HD222582 &  23.84   &$ 12.067$  & $-145.41$  & $-111.10$  & 23 41 51.53  & $-05$ 59 08.7  &   $-45.6$ &  $ 11.4$ &   $ -4.1$ &  0.96 &   0.04 &   0.01 &   0.90 &   0.10 &   1.00 &   0.00 & $-37.50$ &  $-0.05$ \\
  HD330075 &  19.92   &$ 61.280$  & $-235.58$  & $ -94.14$  & 15 49 37.69  & $-49$ 57 48.7  &   $-31.9$ &  $-68.1$ &   $ 28.1$ &  0.13 &   0.87 &   1.10 &   0.10 &   0.90 &   0.50 &   0.50 & $-27.79$ &  \nodata \\
  BD-103166 &  10.00  &$ 26.400$  & $-183.00$  & $  -4.80$  & 10 58 28.78  & $+10$ 46 13.4  &   $ 71.3$ &  $-27.4$ &   $ -7.6$ &  0.85 &   0.15 &   0.04 &   0.50 &   0.40 &   0.90 &   0.10 & $-38.08$ &  \nodata \\
  GJ876    & 212.69  & $ -1.902$ &  $ 960.31$ &  $-675.61$ &  22 53 16.73 &  $-14$ 15 49.3  &   $  3.5$ &  $ -7.9$ &   $ -4.5$ &  0.97 &   0.03 &   0.00 &   0.80 &   0.10 &   1.00 &   0.00 & $ \dots$ &  \nodata \\
  GJ436    &  97.73   &$ 10.000$  & $ 896.34$  & $-813.70$  & 11 42 11.09  & $+26$ 42 23.7  &   $-61.5$ &  $ -7.4$ &   $ 27.1$ &  0.89 &   0.11 &   0.05 &   0.60 &   0.40 &   1.00 &   0.00 & $ \dots$ &  \nodata \\
\enddata
\tablecomments{Spatial properties and population assignments for stars 
with planets and TrES-1. For each entry in the table, the 19 columns report 
star name, parallax ($\pi$), heliocentric radial-velocity (HRV), proper motion 
in right ascension and declination ($\mu_\alpha$ and $\mu_\delta$), 
equatorial coordinates ($\alpha$ and $\delta$), galactic velocity vector 
($U$, $V$, and $W$), thin and thick disk membership probabilities 
$P^\mathrm{thin}$ and $P^\mathrm{thick}$ according 
to Mishenina et al. (\citeyear{mishenina04}) (M04), the thick-to-thin disk 
probability ratio values $TD/D$ according to Bensby et al. 
(\citeyear{bensby03}) (B03)), 
thin and thick disk membership probabilities according to 
Venn et al. (\citeyear{venn04}) (V04), assuming both uniform and Bensby's 
prior probability distributions ($P_1^\mathrm{thin}$, $P_1^\mathrm{thick}$ 
and $P_2^\mathrm{thin}$, $P_2^\mathrm{thick}$, respectively), the $X$ 
stellar population criterion according to Schuster et al. 
(\citeyear{schuster93}) (S93), and the abundance ratio [$\alpha$/Fe] 
from Bodaghee et al. (\citeyear{bodaghee03}) and 
Beir\~ao et al. (\citeyear{beirao05}). 
All values of positions, proper motions, and parallax are taken from 
$Hipparcos$ (ESA 1997) and the SIMBAD database, except for TrES-1, 
for which RA, DEC, $\mu_\alpha$ and $\mu_\delta$ were taken from 
the USNO-B1.0 catalog (\citealp{monet03}), while the photometric parallax estimate 
was taken from Sozzetti et al. (\citeyear{sozzetti04}). 
The values of heliocentric radial-velocity are taken from SIMBAD and 
from the corresponding planet discovery papers.}
\end{deluxetable}
\thispagestyle{empty} 
}


\clearpage

\begin{deluxetable}{cccc}
\tablecaption{Results of the K-S tests\label{tb5}}
\tablewidth{0pt}
\tablenum{5}
\tablehead{
\colhead{Samples} & \colhead{$Pr(D)$} & \colhead{$Pr(D)$} &\colhead{$Pr(D)$} \\
 & \colhead{(All [Fe/H])} & \colhead{([Fe/H]$\leq 0.0$)} & \colhead{([Fe/H]$> 0.0$)}
}
\startdata
$[\alpha/\mathrm{Fe}]_\mathrm{pl}$ vs. $[\alpha/\mathrm{Fe}]_\mathrm{cs}$ & $5.34\times 10^{-10}$ & $0.24$ & $2.41\times 10^{-3}$  \\
$[\alpha/\mathrm{Fe}]_\mathrm{pl}$ vs. $[\alpha/\mathrm{Fe}]_\mathrm{SG}$ & $2.94\times 10^{-30}$ & $2.26\times 10^{-3}$ & $1.78\times 10^{-23}$  \\
$[\alpha/\mathrm{Fe}]_\mathrm{cs}$ vs. $[\alpha/\mathrm{Fe}]_\mathrm{SG}$ &$0.06$ & $0.21$ & $0.07$  \\
\enddata
\tablecomments{Results of the K-S test on different 
stellar samples: $[\alpha/\mathrm{Fe}]_\mathrm{pl}$ and 
$[\alpha/\mathrm{Fe}]_\mathrm{cs}$ 
are the distributions of [$\alpha$/Fe] abundances for the samples of 
planet hosts and comparison field stars from Bodaghee et al. (\citeyear{bodaghee03}) 
and Beir\~ao et al. (\citeyear{beirao05}), while $[\alpha/\mathrm{Fe}]_\mathrm{SG}$ 
is the analogous distribution from the Soubiran \& Girard (\citeyear{soubiran05}) 
sample. The significance level $Pr(D)$ is calculated when comparing the 
[$\alpha$/Fe] distributions in three cases: a) for all values of [Fe/H], 
b) for [Fe/H]$\leq 0.0$, and c) for [Fe/H]$> 0.0$.}
\end{deluxetable}

\clearpage



\begin{figure}
\centering
$\begin{array}{c}
\includegraphics[width=0.80\textwidth]{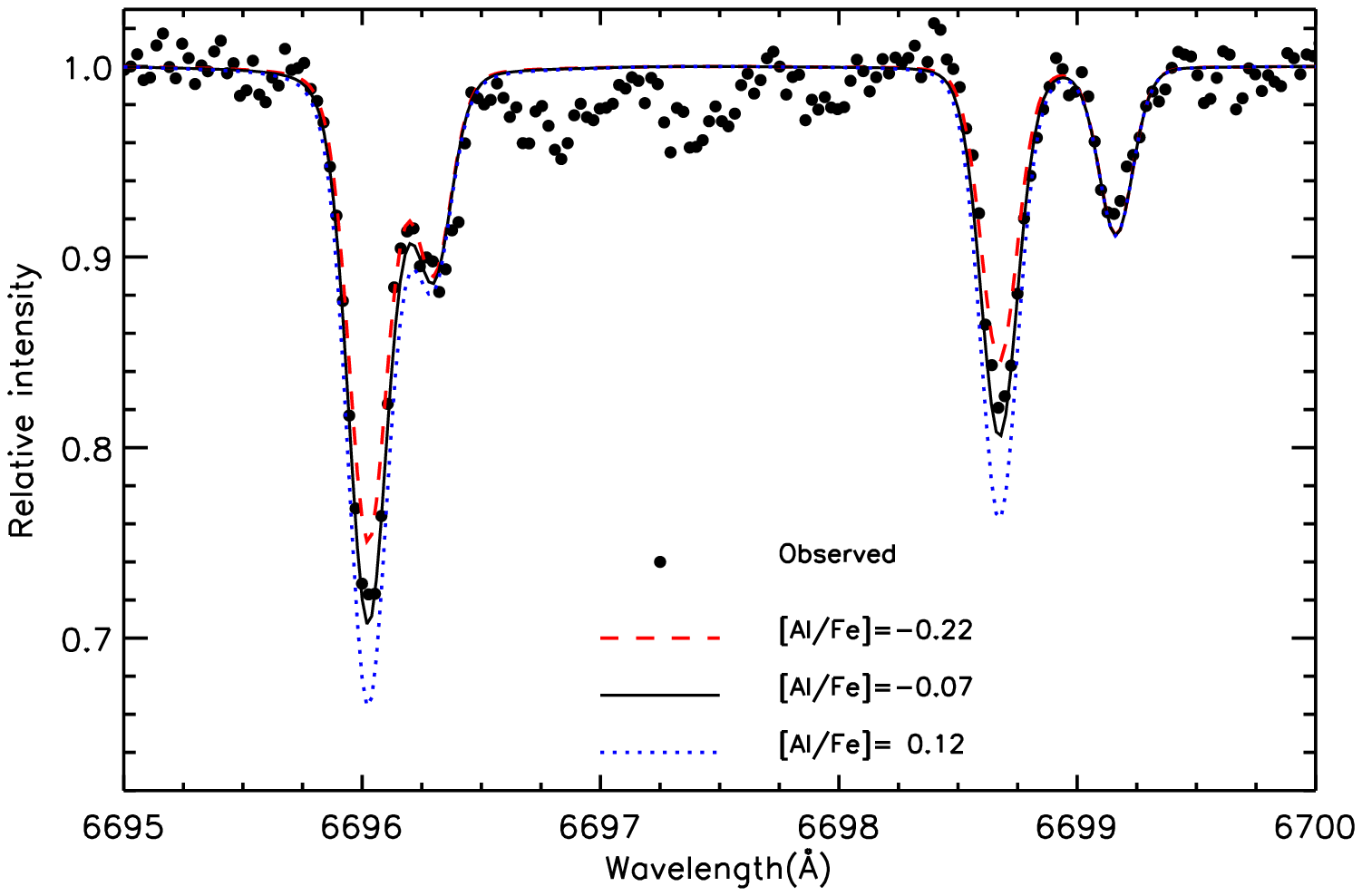} \\
\includegraphics[width=0.80\textwidth]{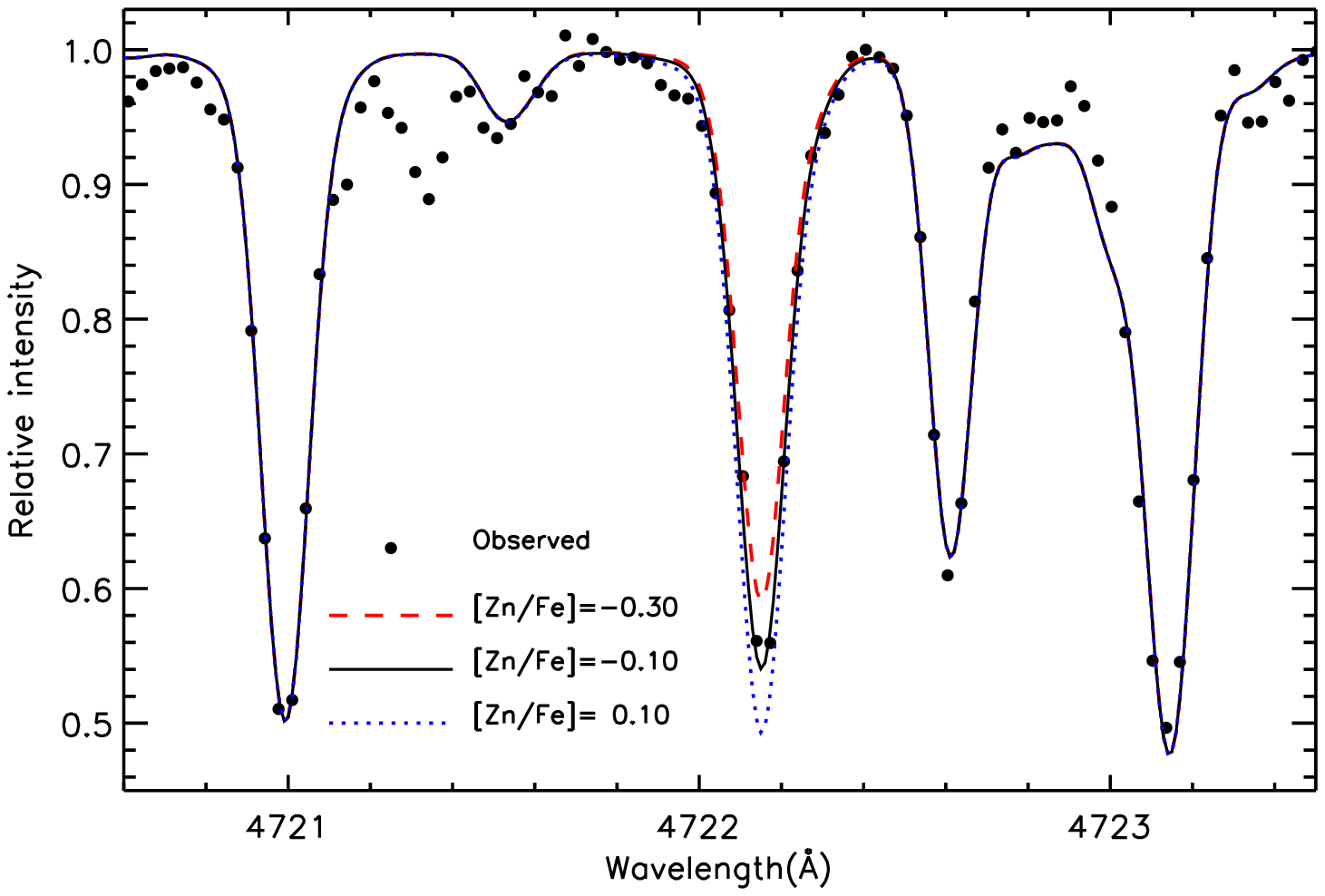} 
\end{array} $
\end{figure}

\clearpage

\begin{figure}
\plotone{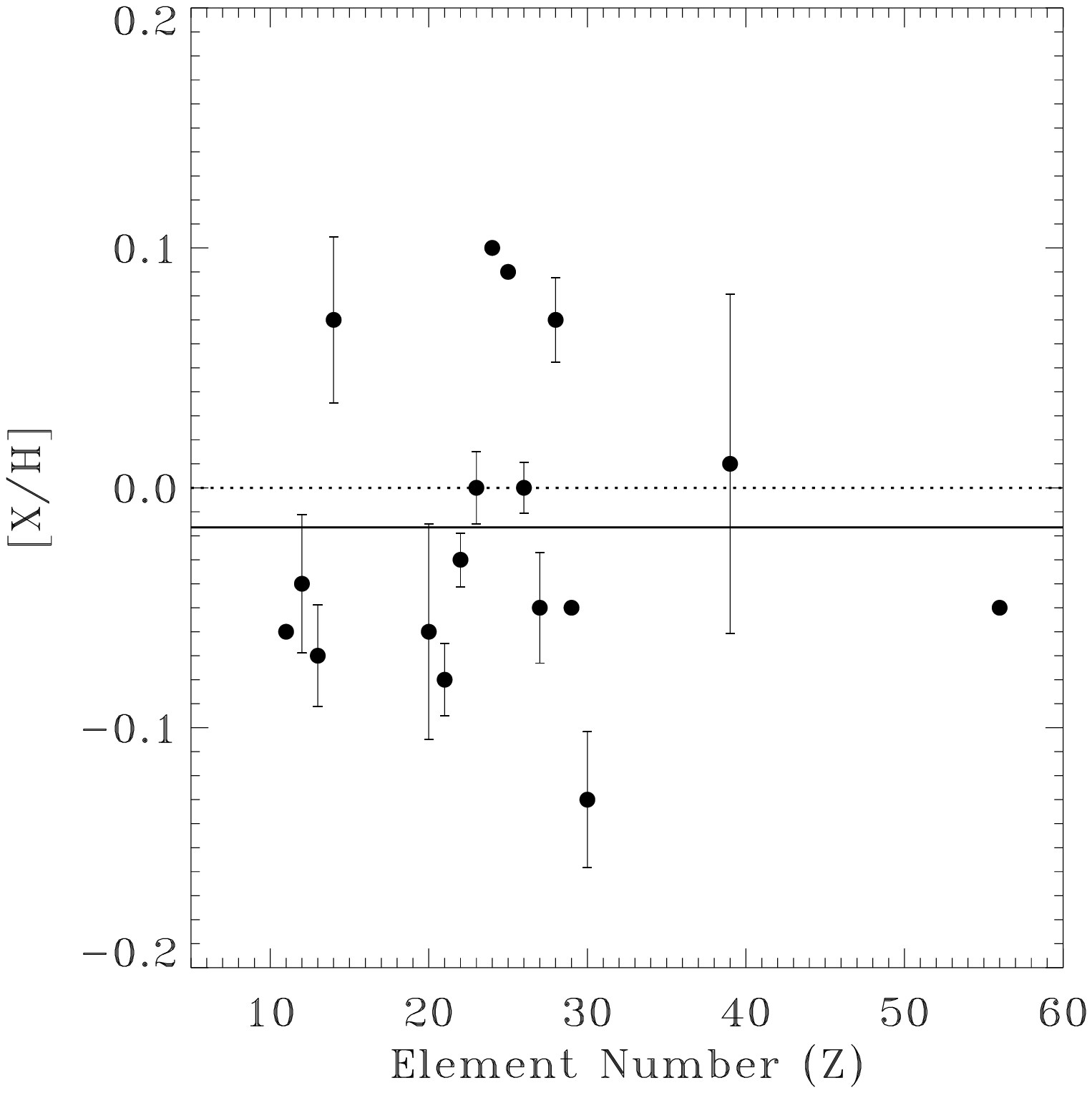}
\end{figure}

\clearpage

\begin{figure}
\plotone{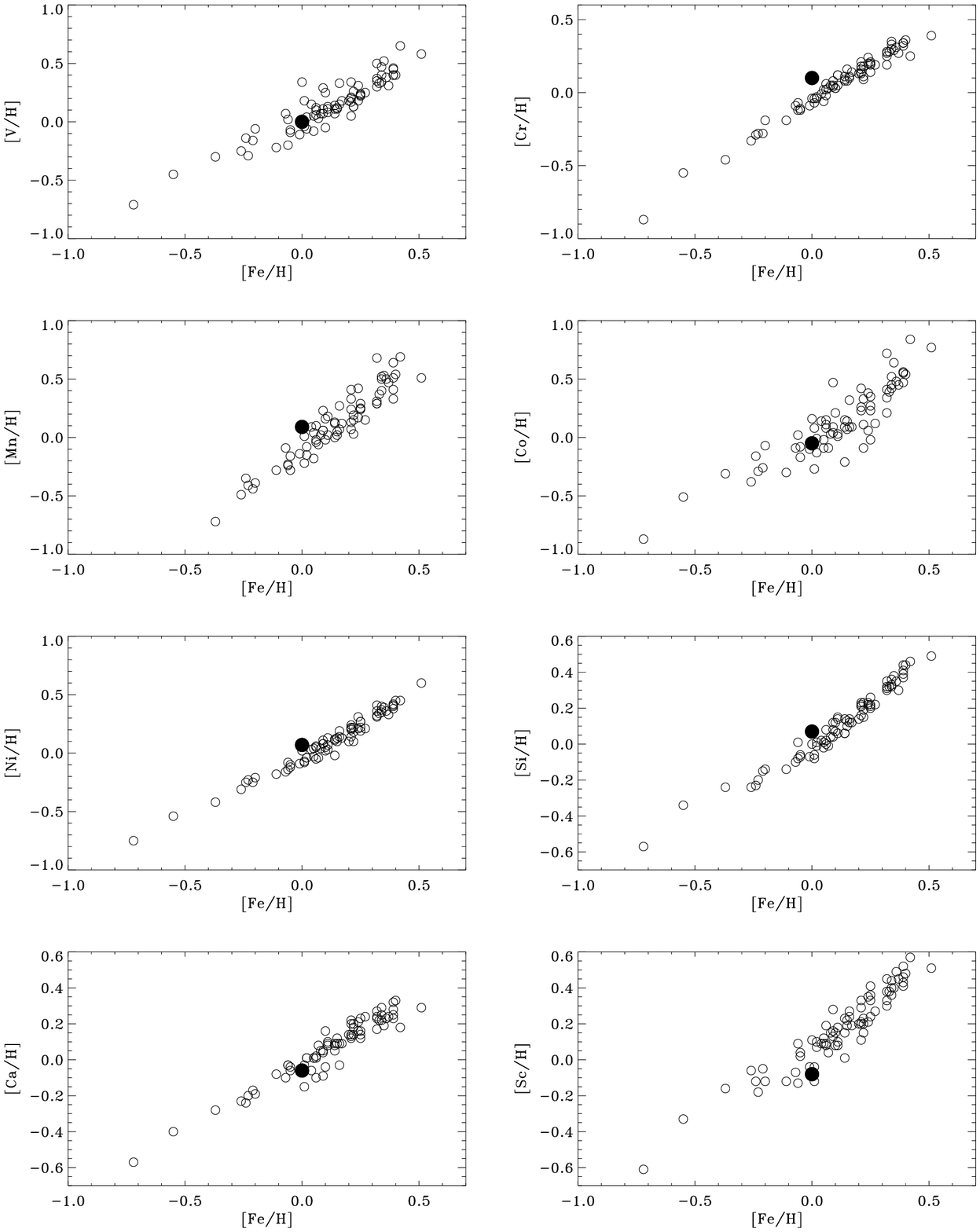}
\end{figure}

\clearpage

\begin{figure}
\plotone{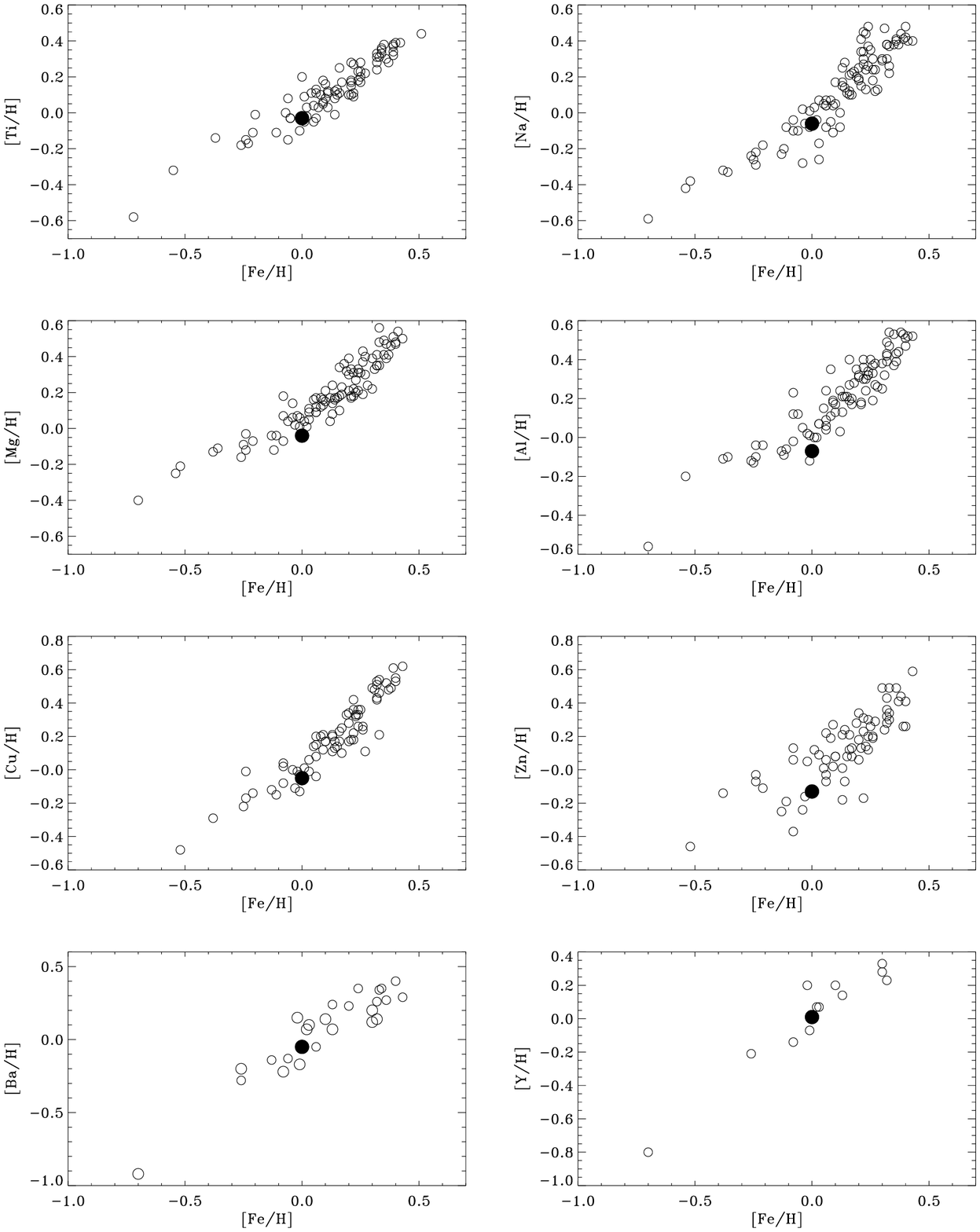}
\end{figure}

\clearpage

\begin{figure}
\plotone{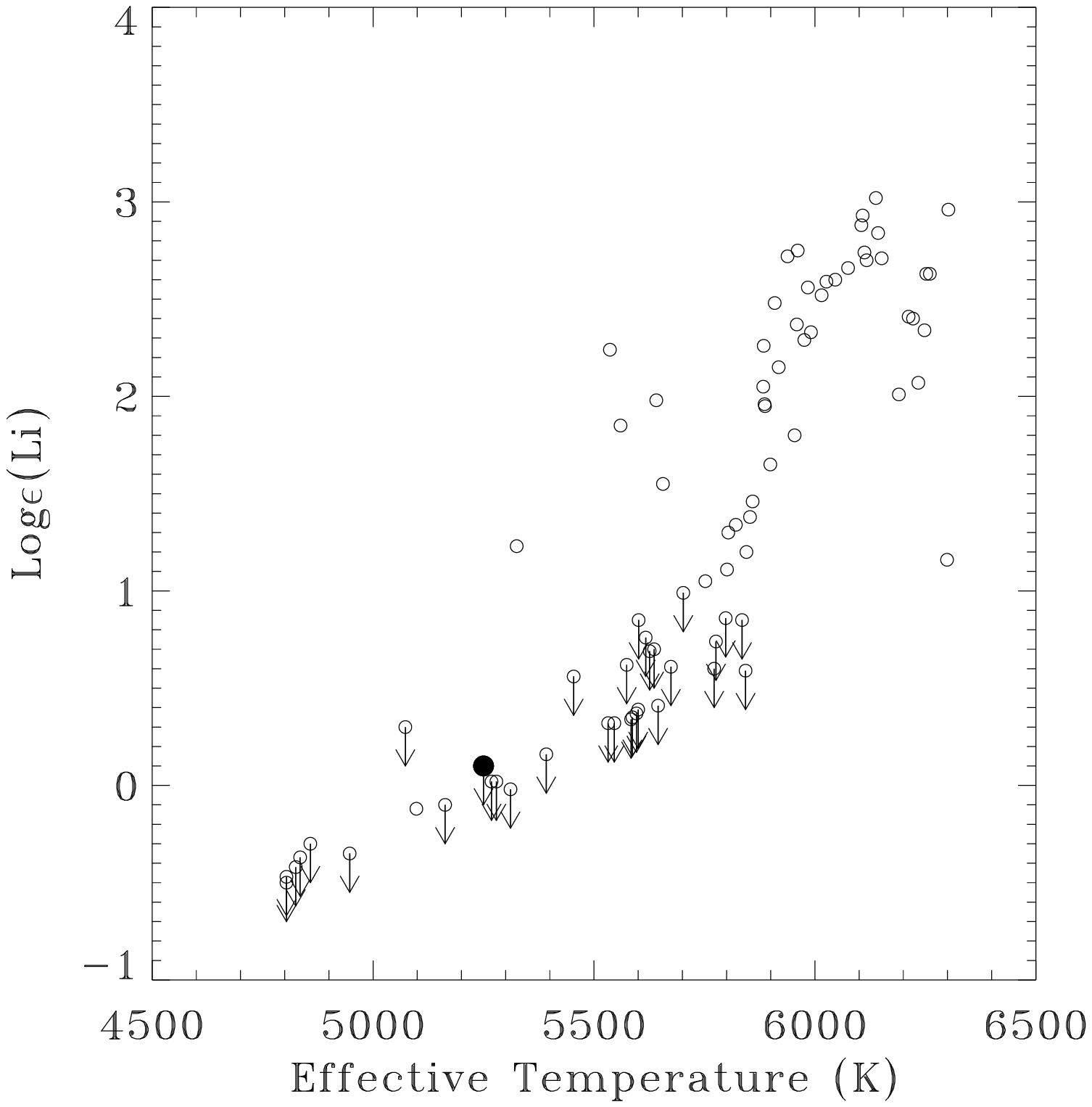}
\end{figure}

\clearpage

\begin{figure}
\plotone{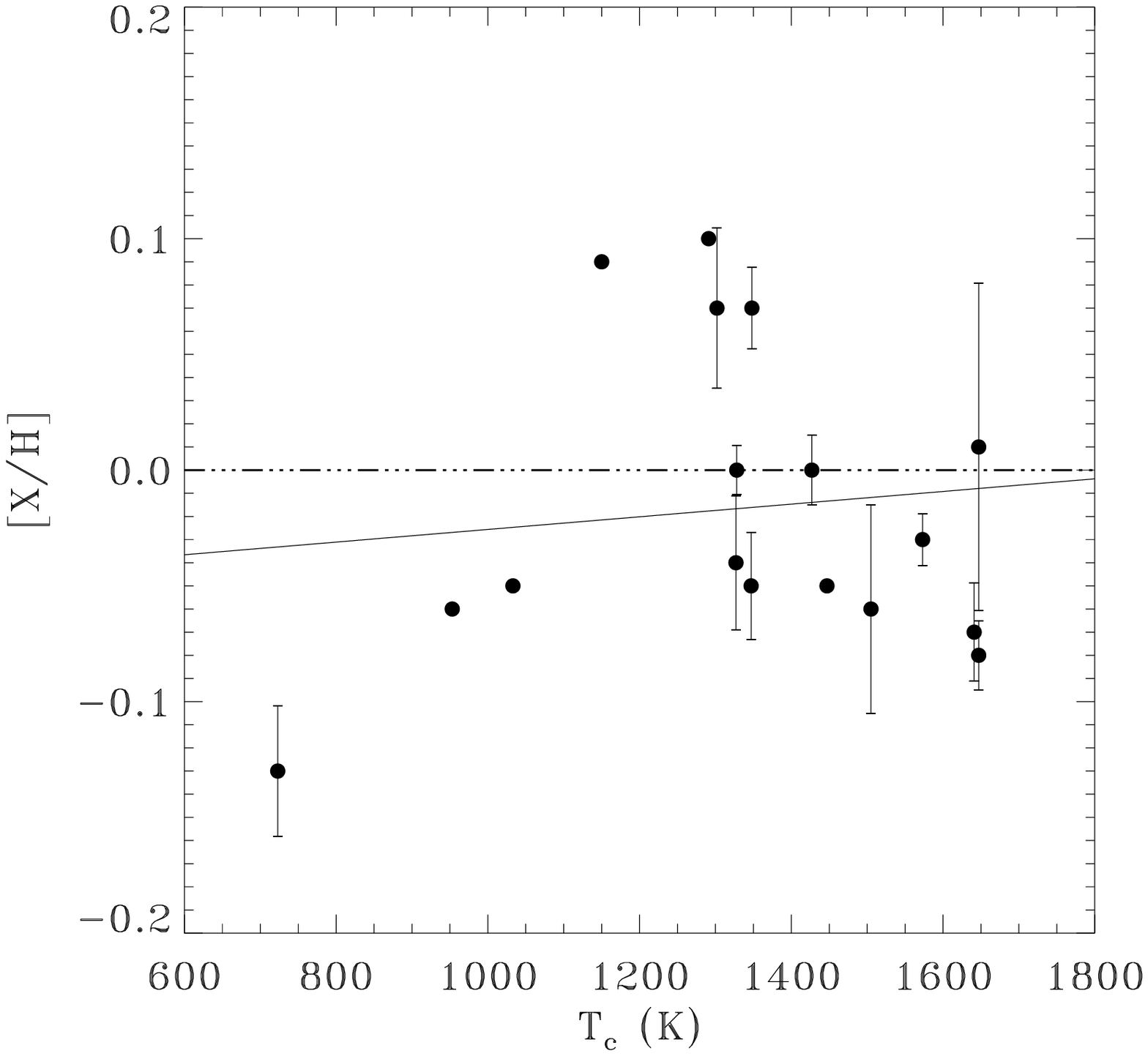}
\end{figure}

\clearpage
\begin{figure}
\plotone{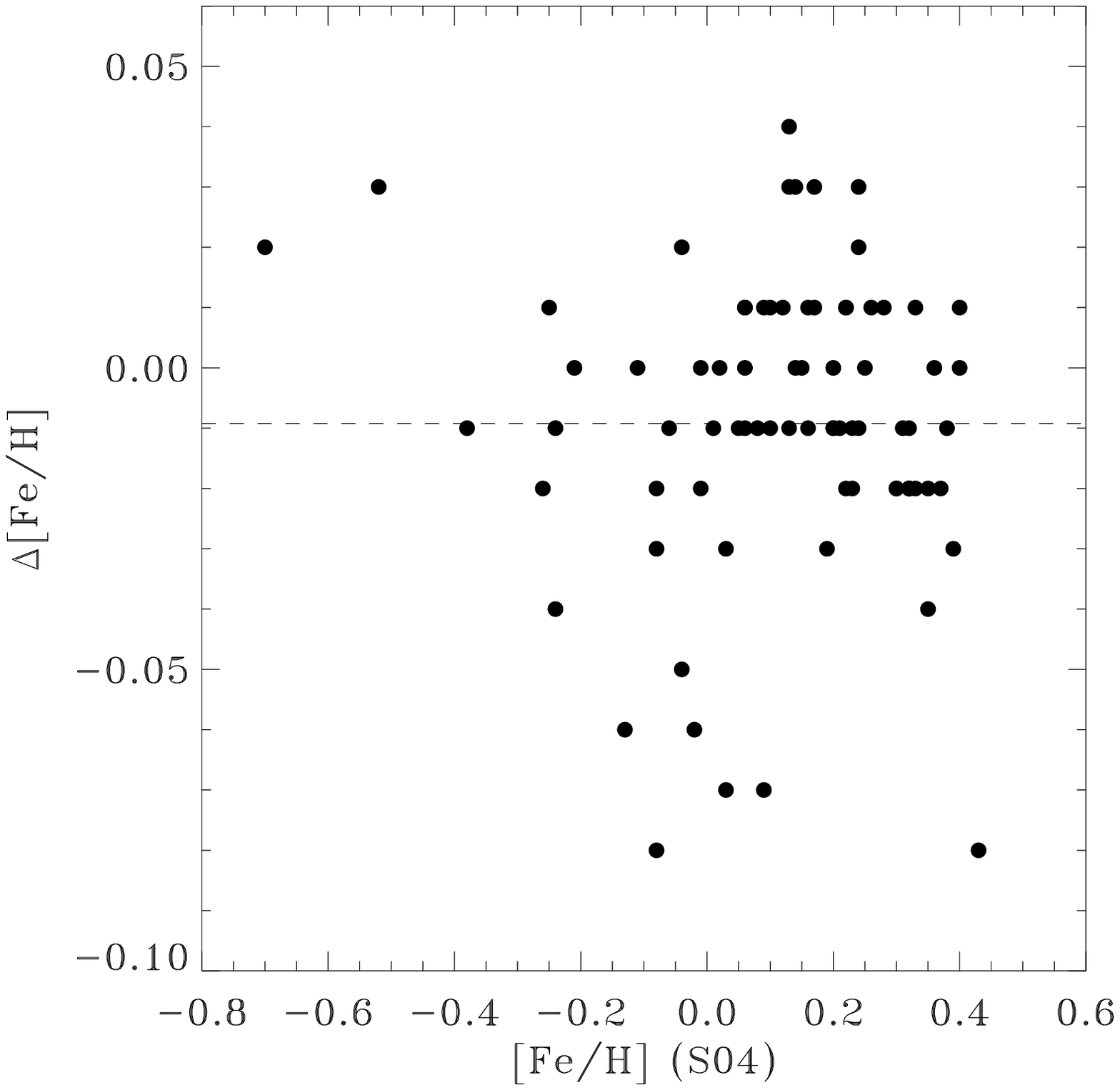}
\end{figure}

\clearpage

\begin{figure}
\plotone{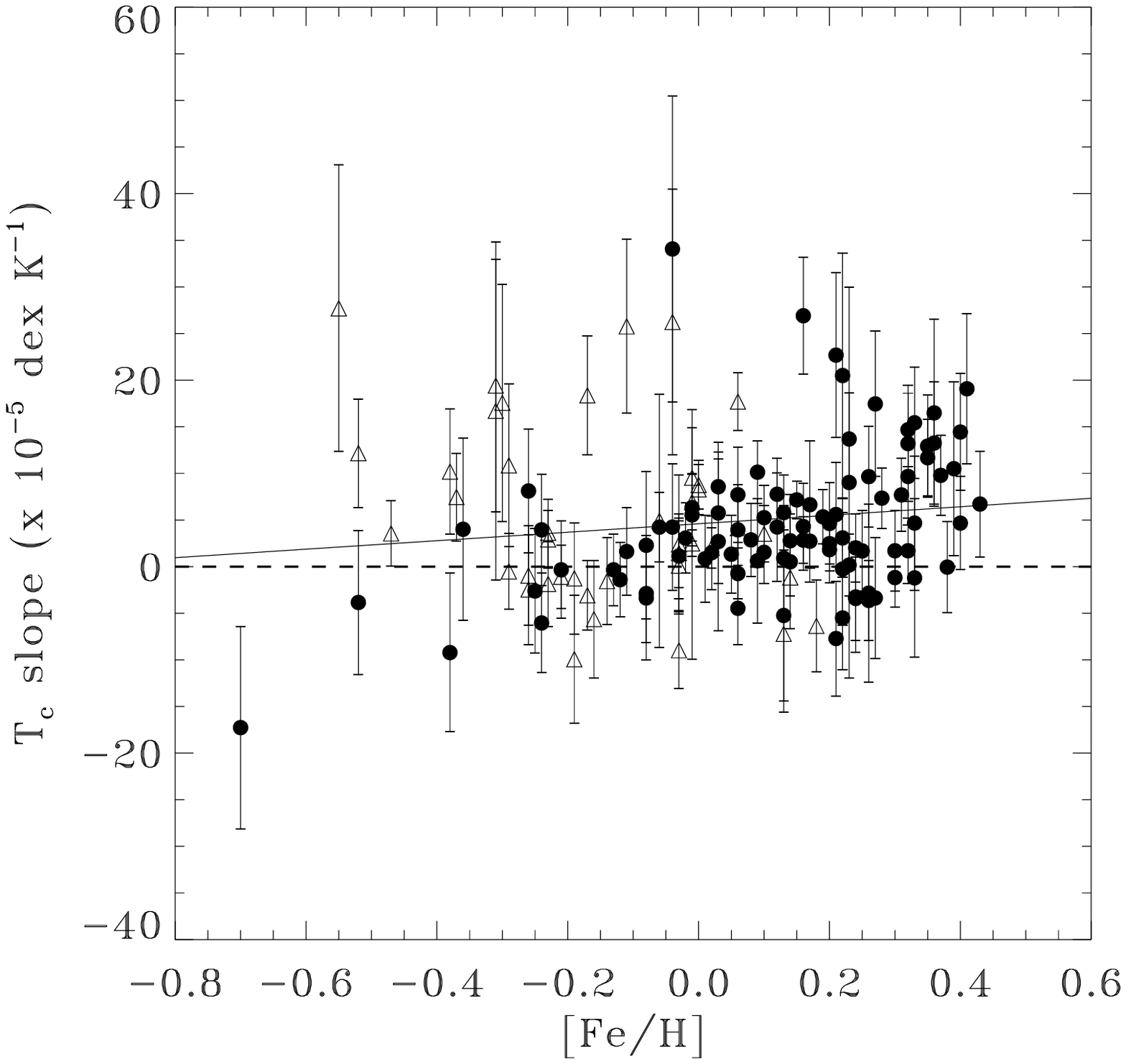}
\end{figure}

\begin{figure}
\plotone{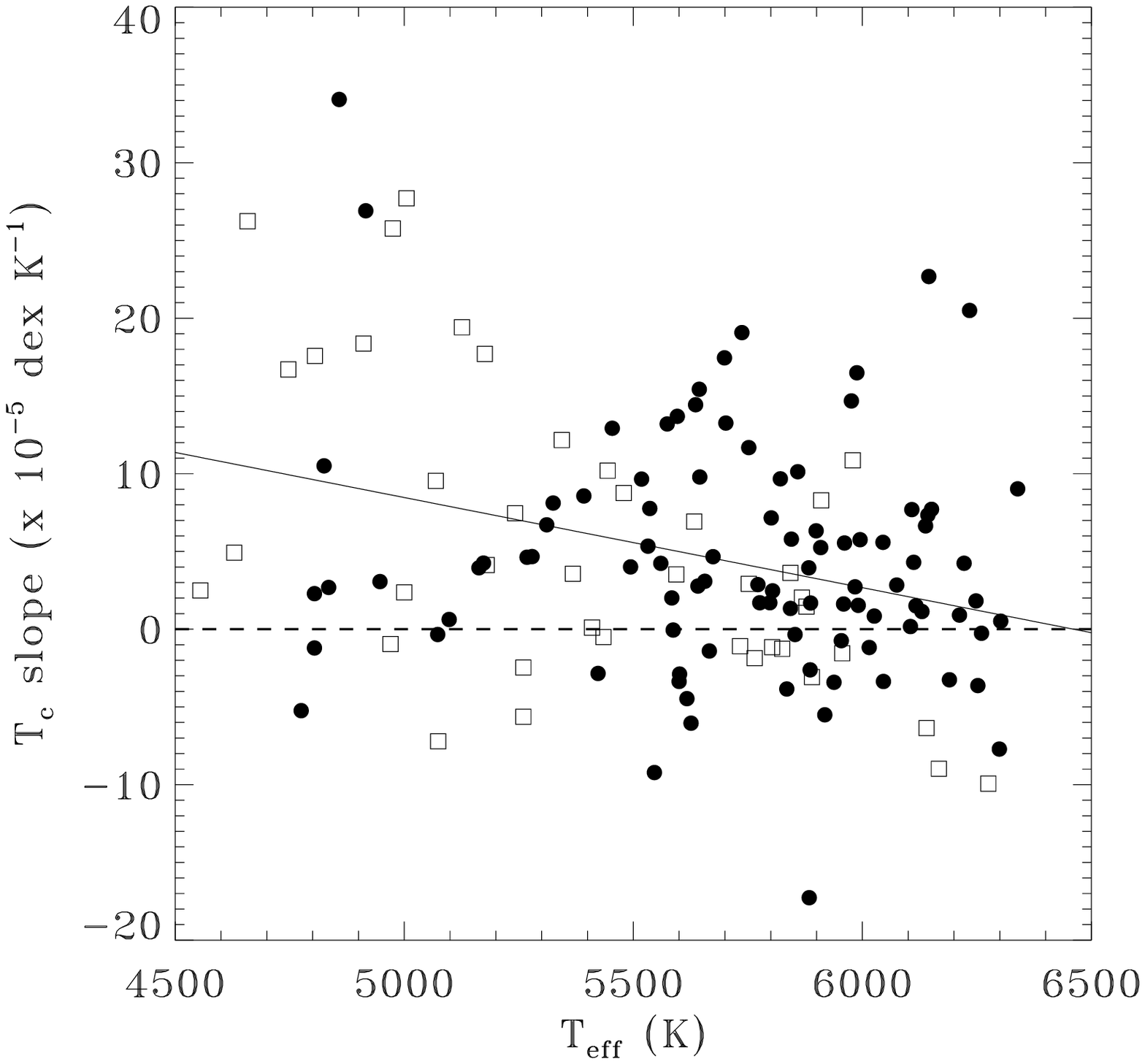}
\end{figure}

\clearpage

\begin{figure}
\centering
$\begin{array}{ll}
\includegraphics[width=0.52\textwidth]{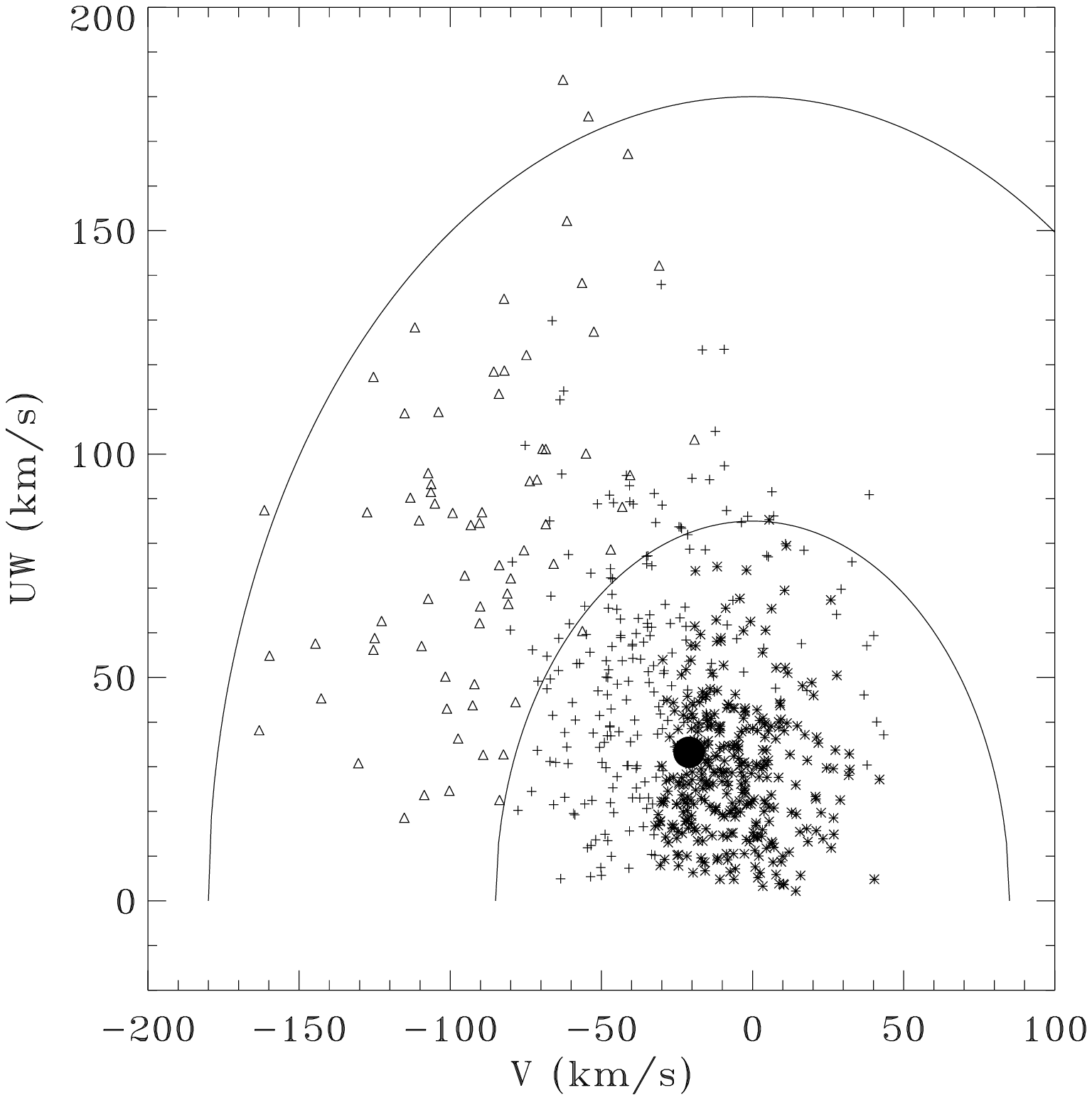} &
\includegraphics[width=0.51\textwidth]{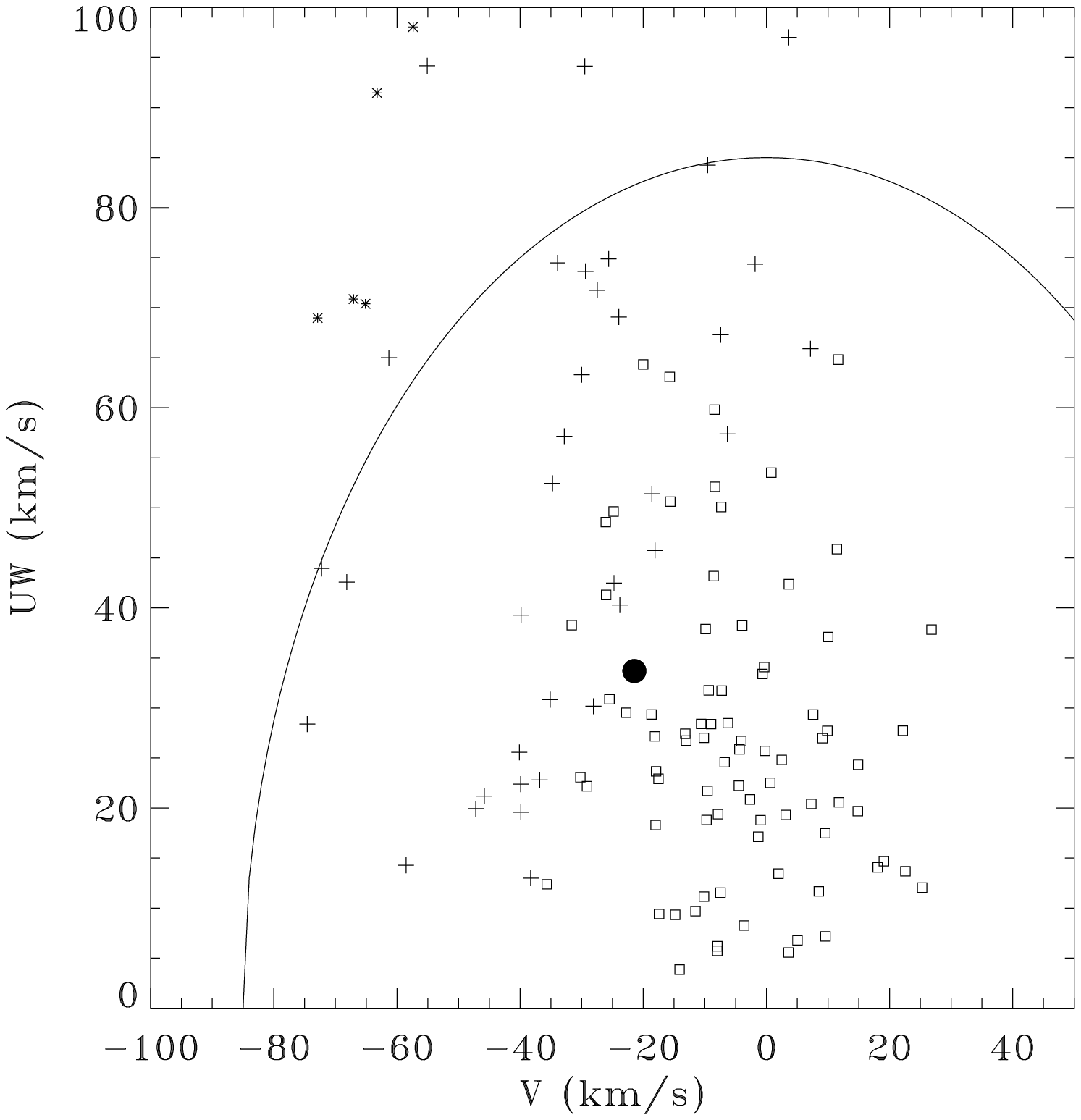} 
\end{array} $
\end{figure}

\clearpage

\begin{figure}
\plotone{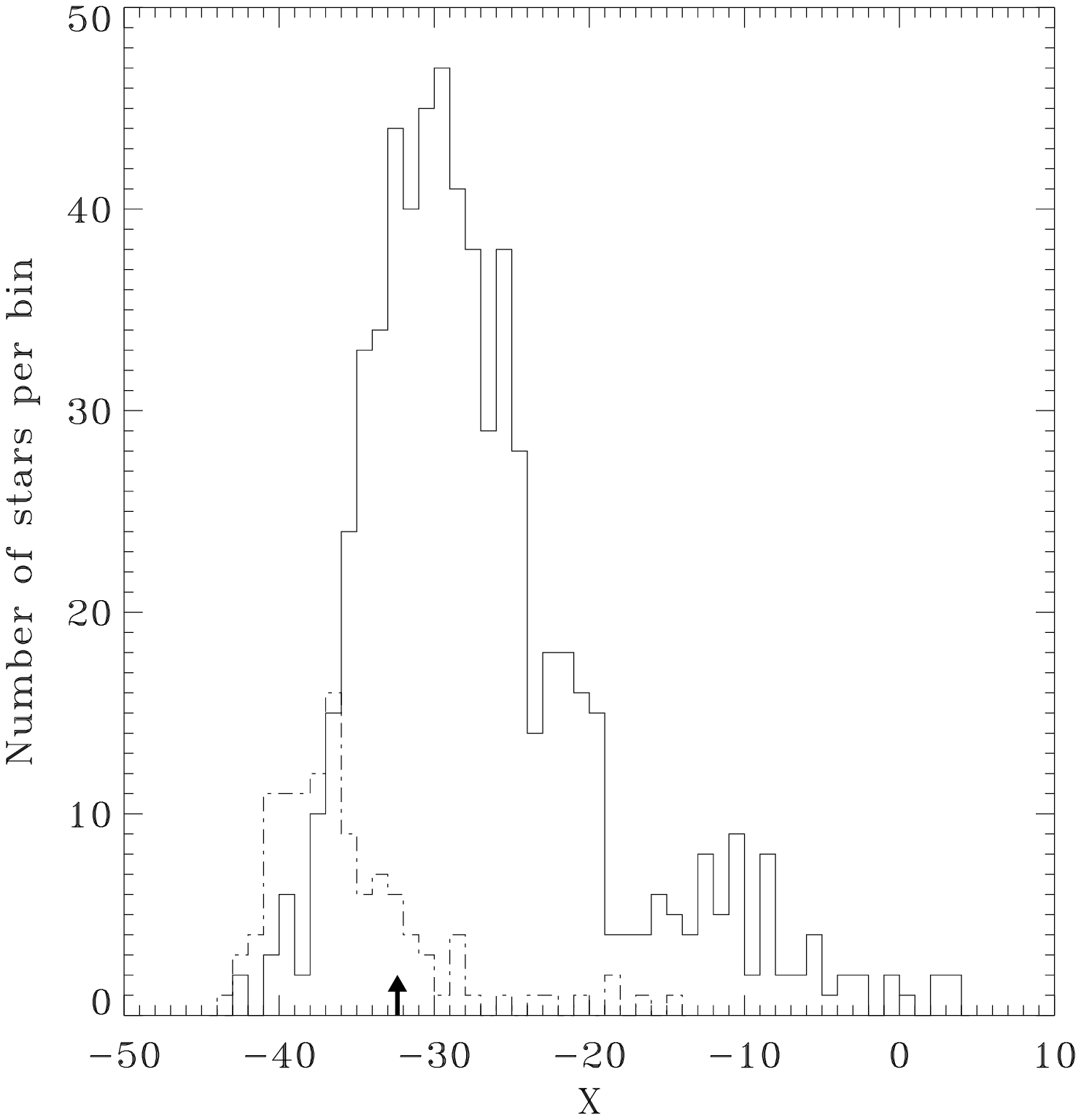}
\end{figure}

\clearpage

\begin{figure}
\epsscale{1.1}
$\begin{array}{ll}
\includegraphics[width=0.51\textwidth]{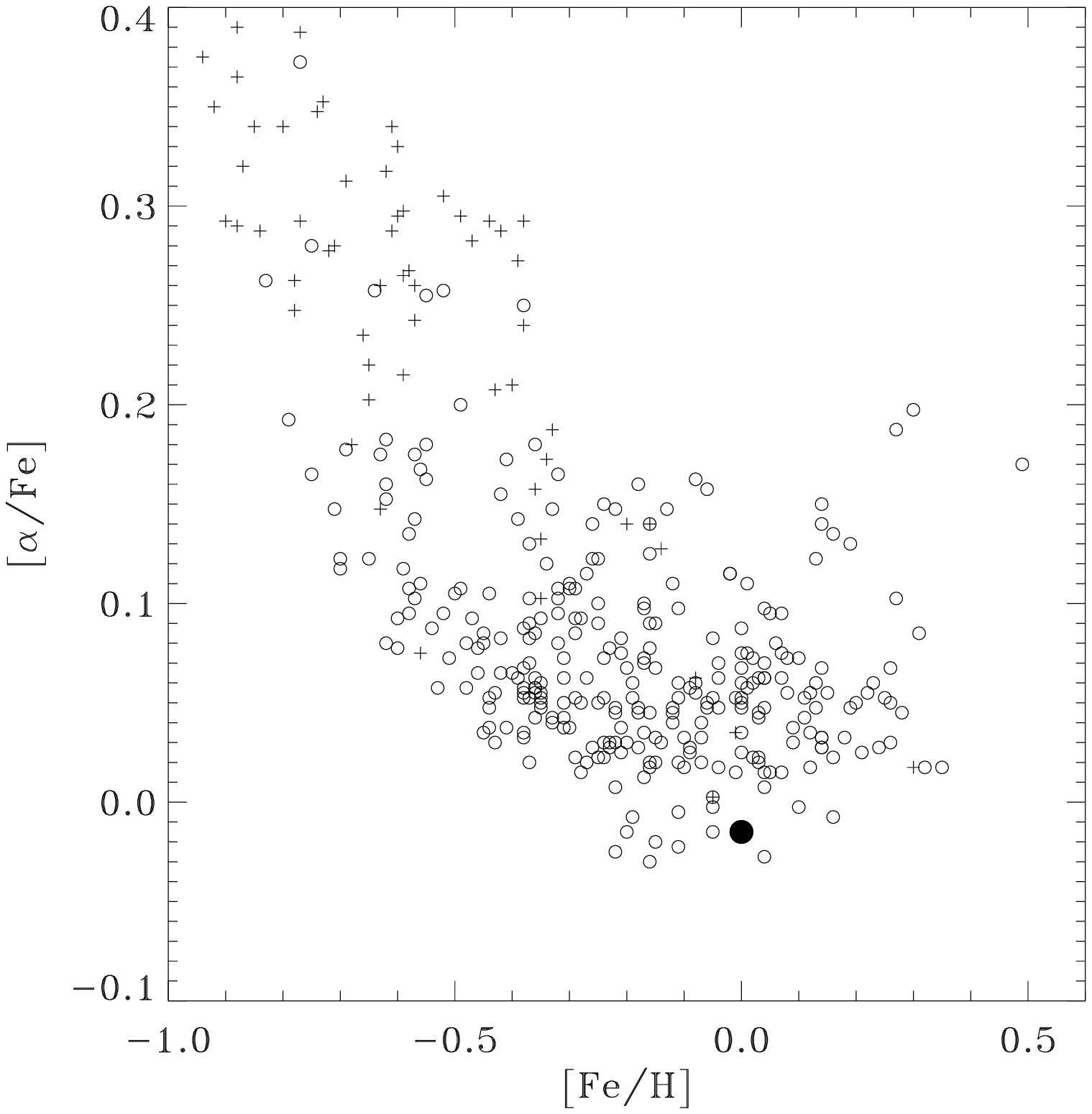} &
\includegraphics[width=0.51\textwidth]{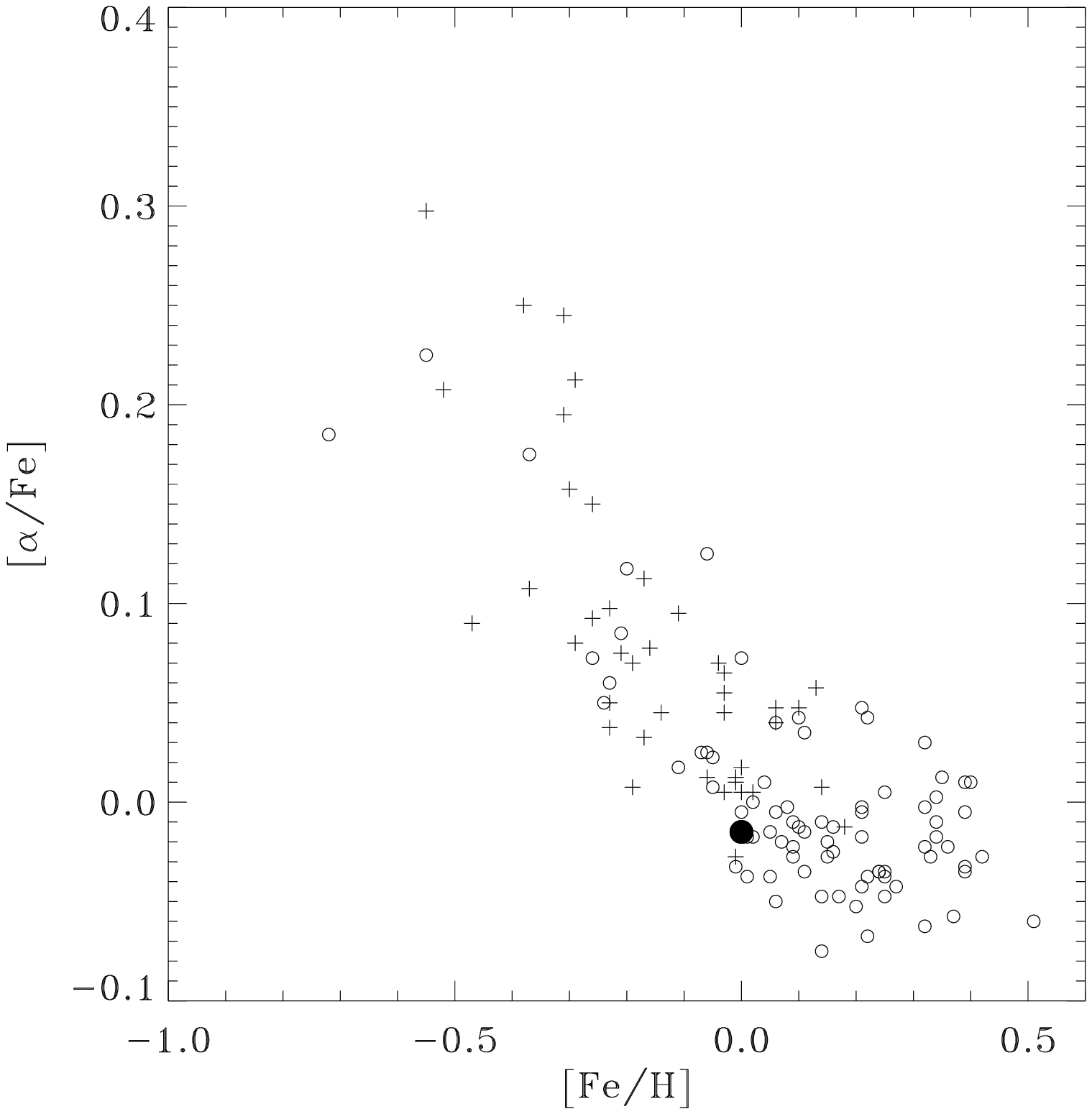} 
\end{array} $
\end{figure}

\clearpage

\begin{figure}
\epsscale{1.}
$\begin{array}{ll}
\includegraphics[width=0.52\textwidth]{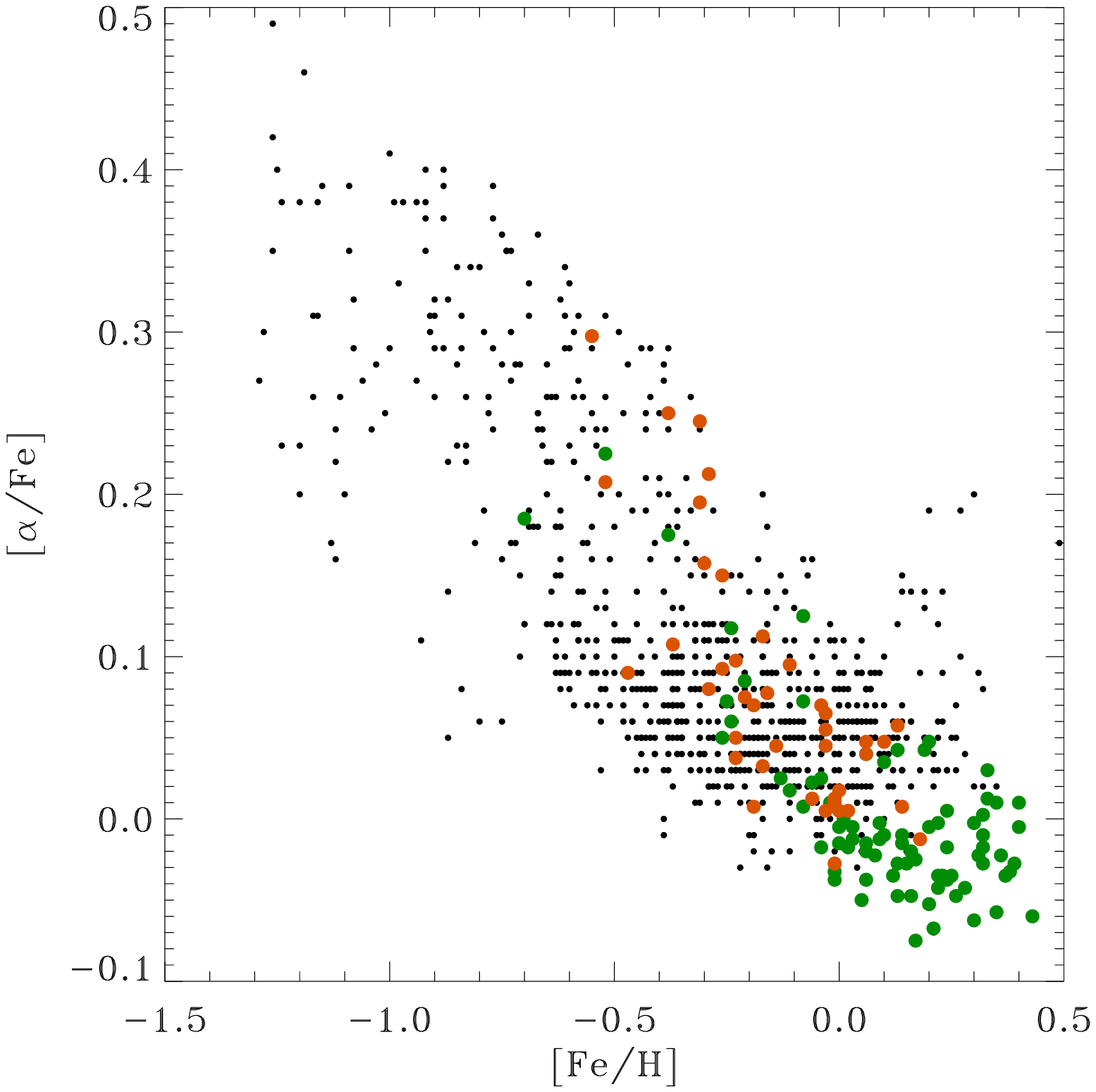} &
\includegraphics[width=0.52\textwidth]{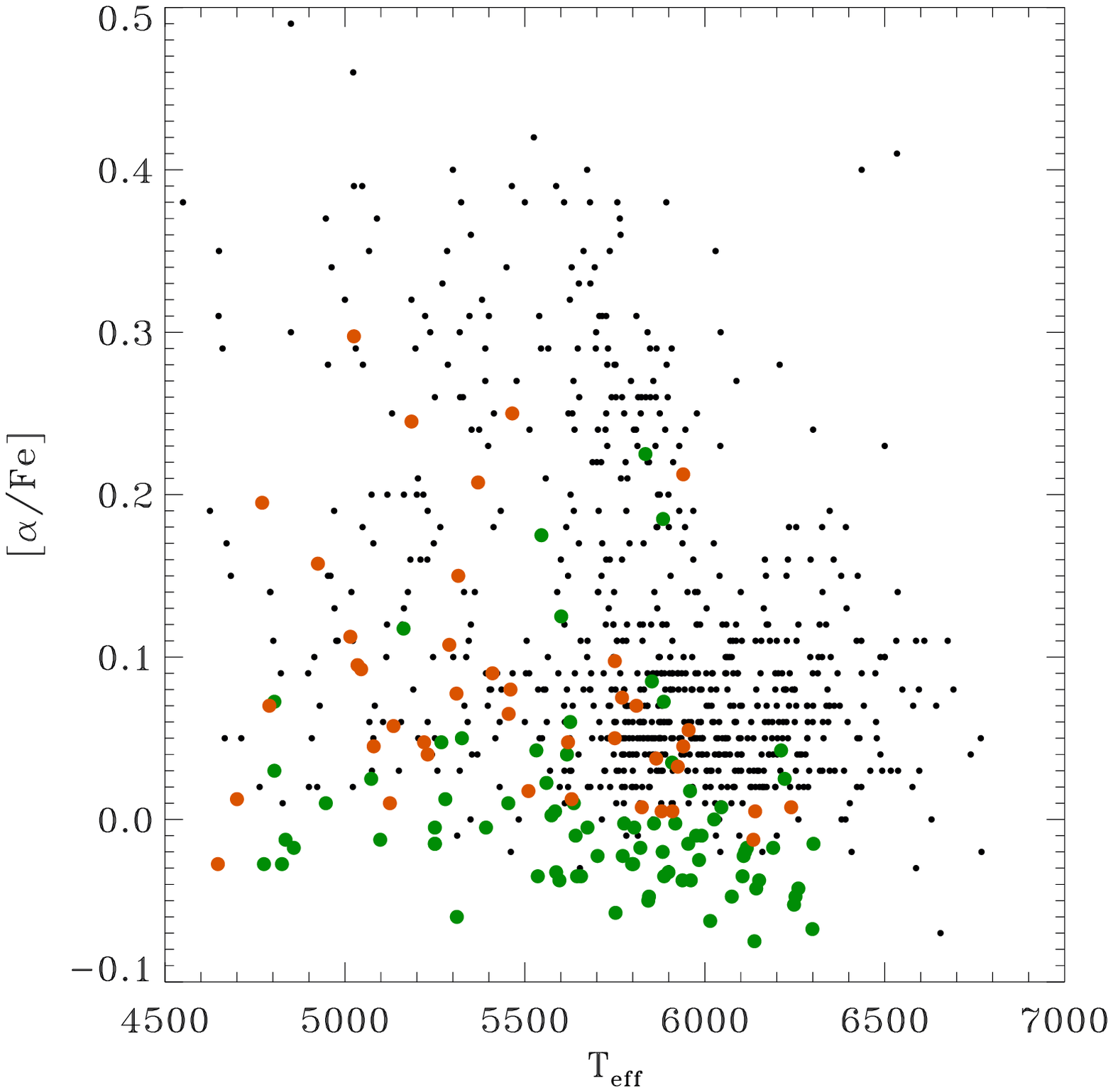} 
\end{array} $
\end{figure}

\clearpage

\begin{figure}
\epsscale{1.}
$\begin{array}{ll}
\includegraphics[width=0.51\textwidth]{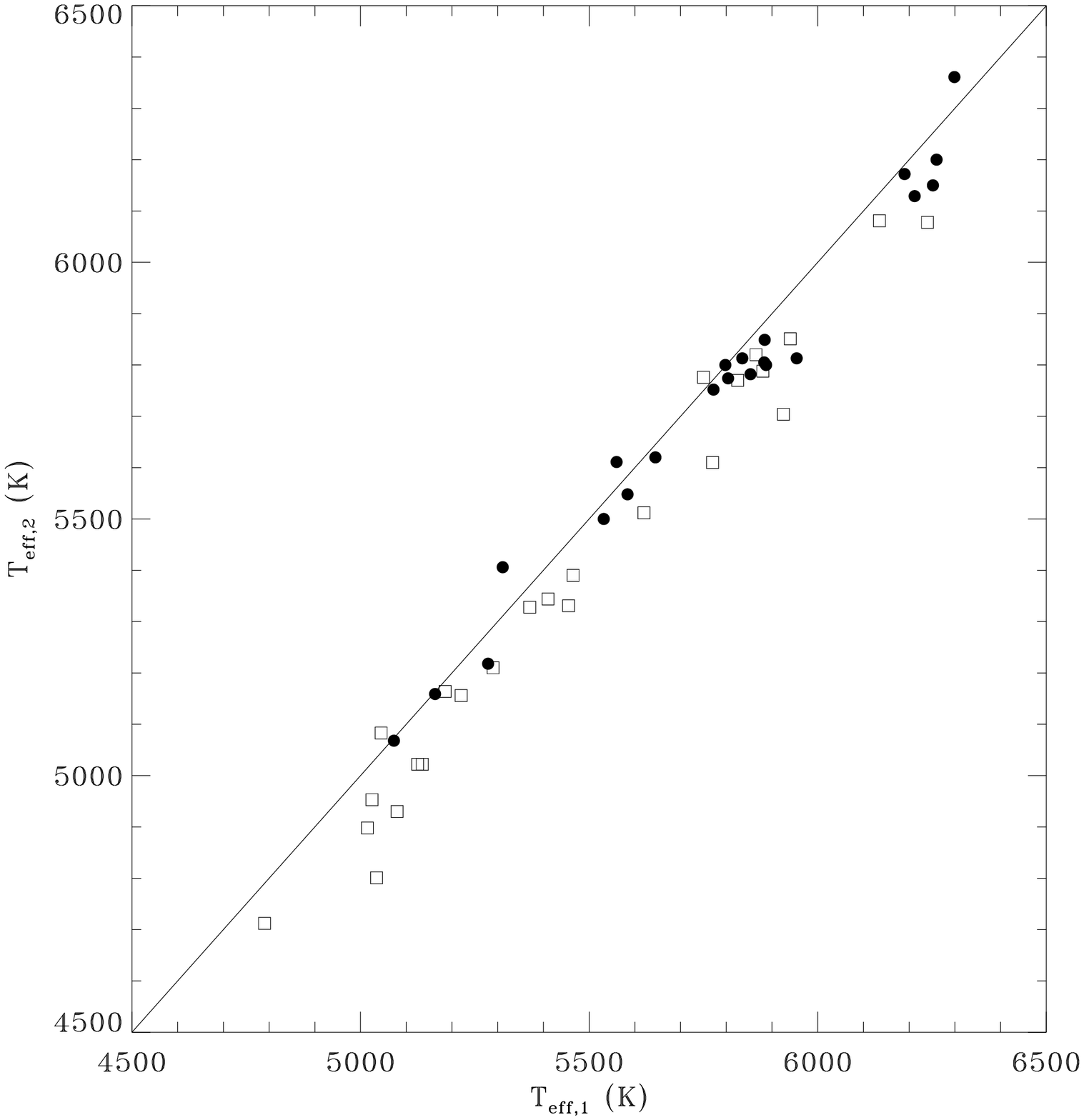} &
\includegraphics[width=0.49\textwidth]{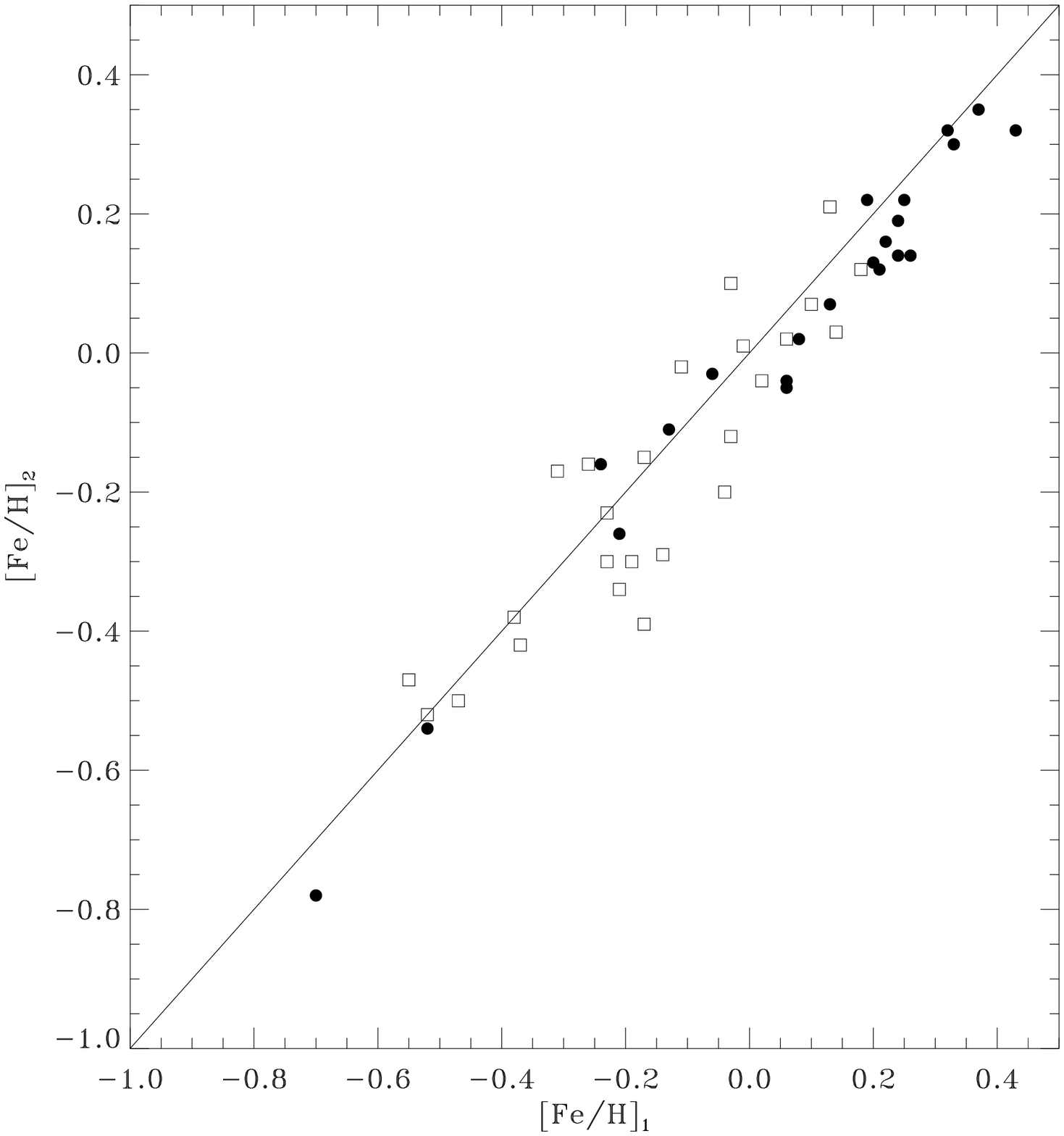} \\
\includegraphics[width=0.51\textwidth]{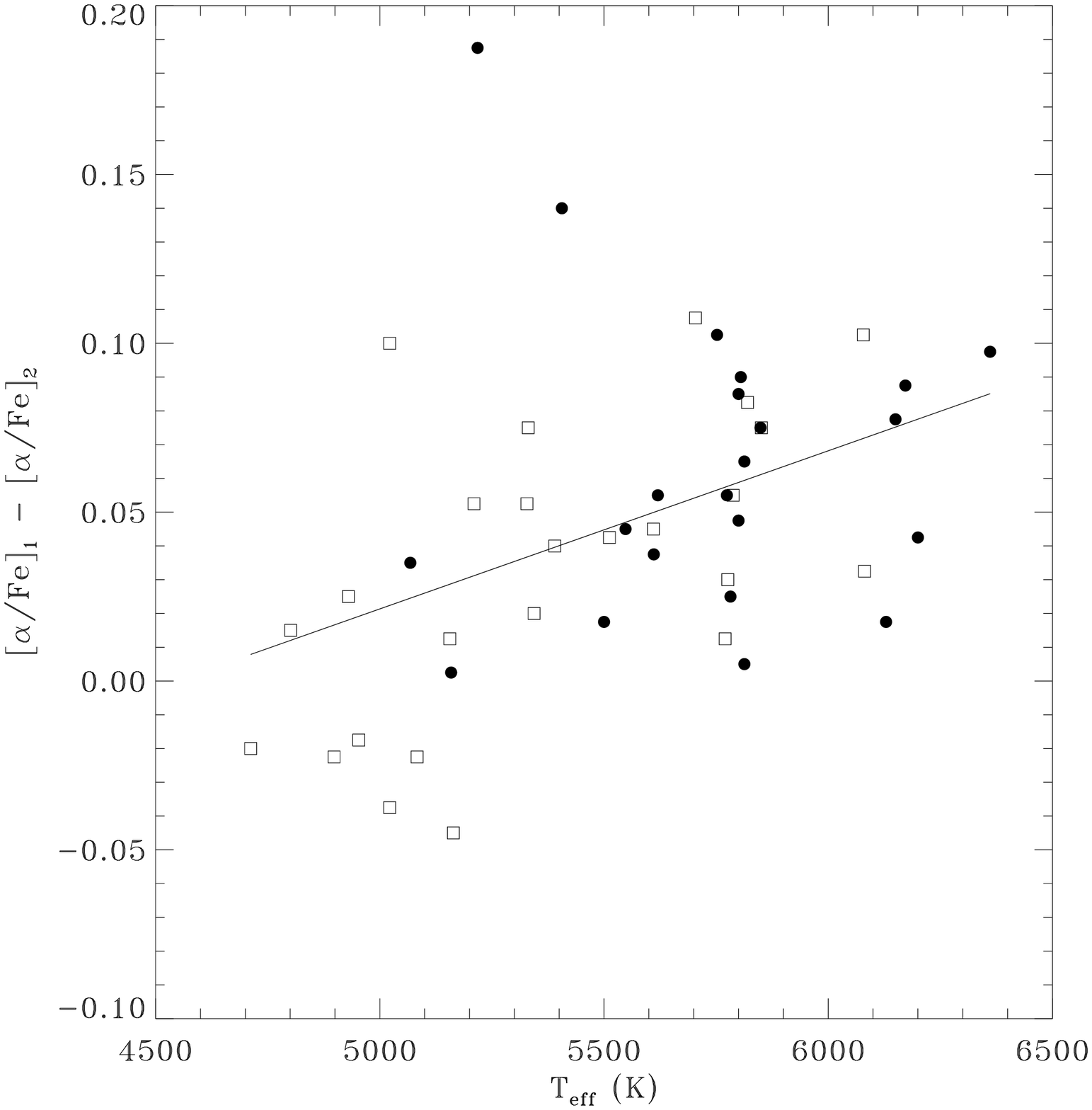} &
\includegraphics[width=0.49\textwidth]{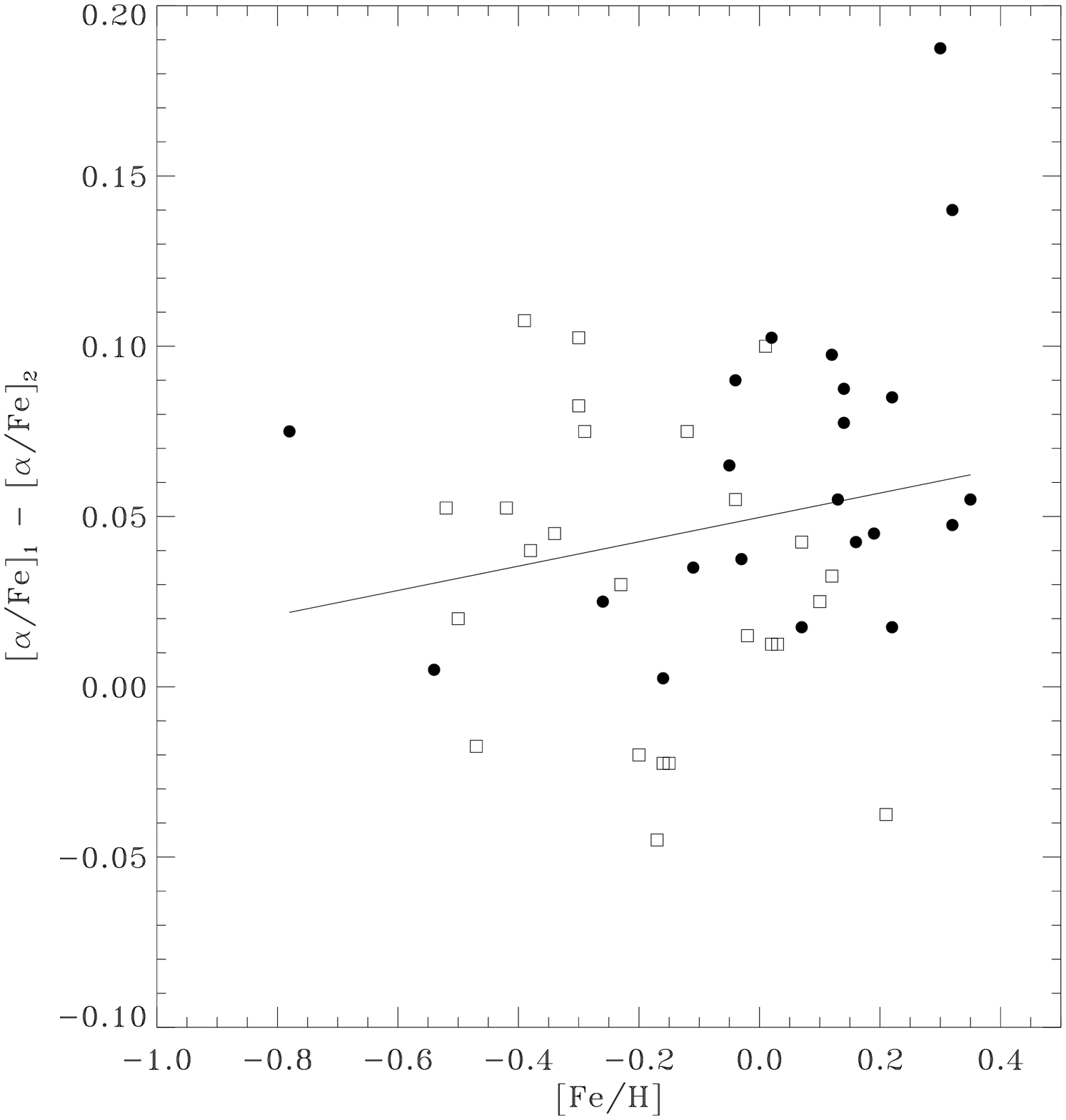} 
\end{array} $
\end{figure}

\clearpage

\begin{figure}
\epsscale{1.}
$\begin{array}{ll}
\includegraphics[width=0.52\textwidth]{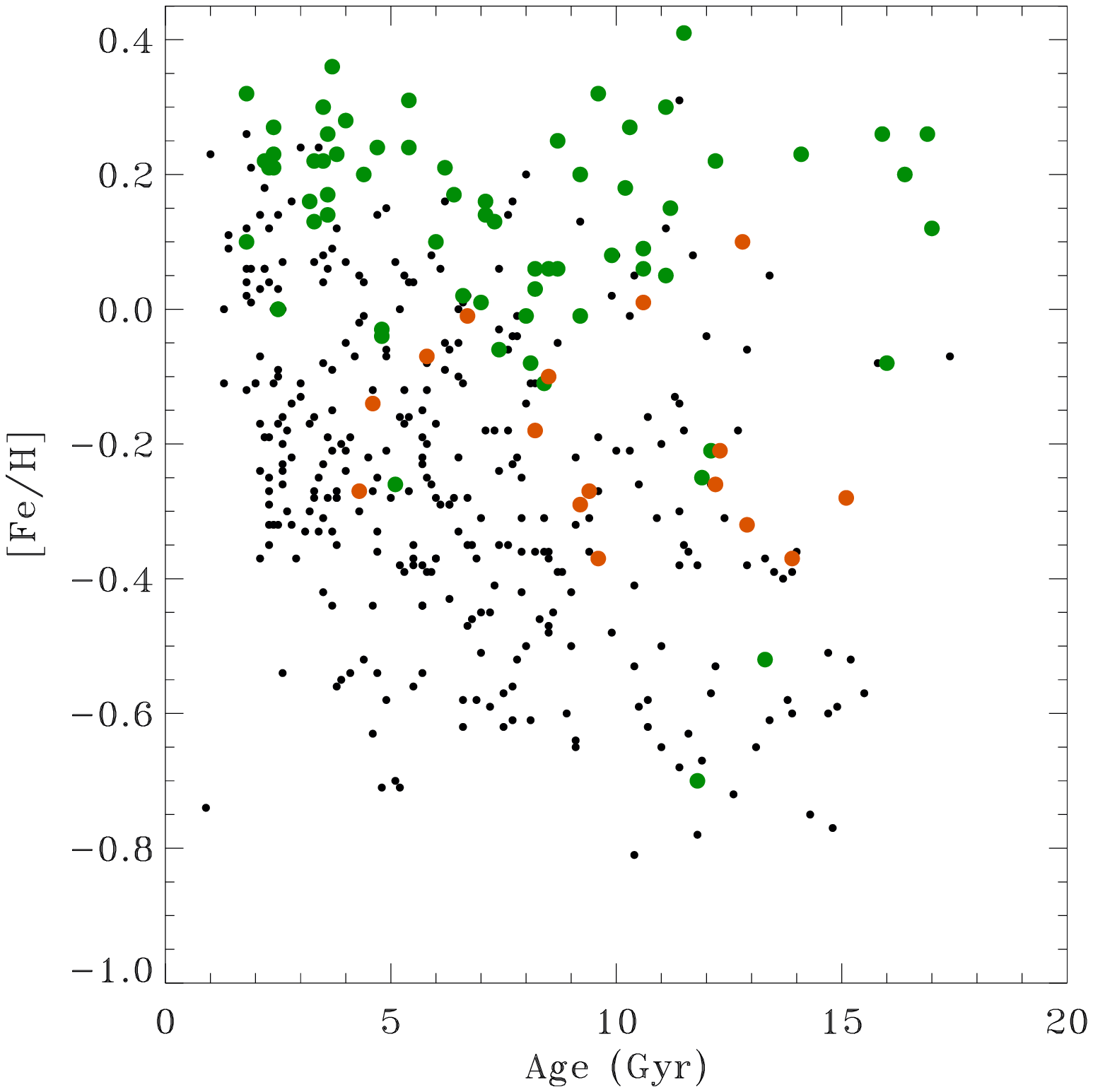} &
\includegraphics[width=0.53\textwidth]{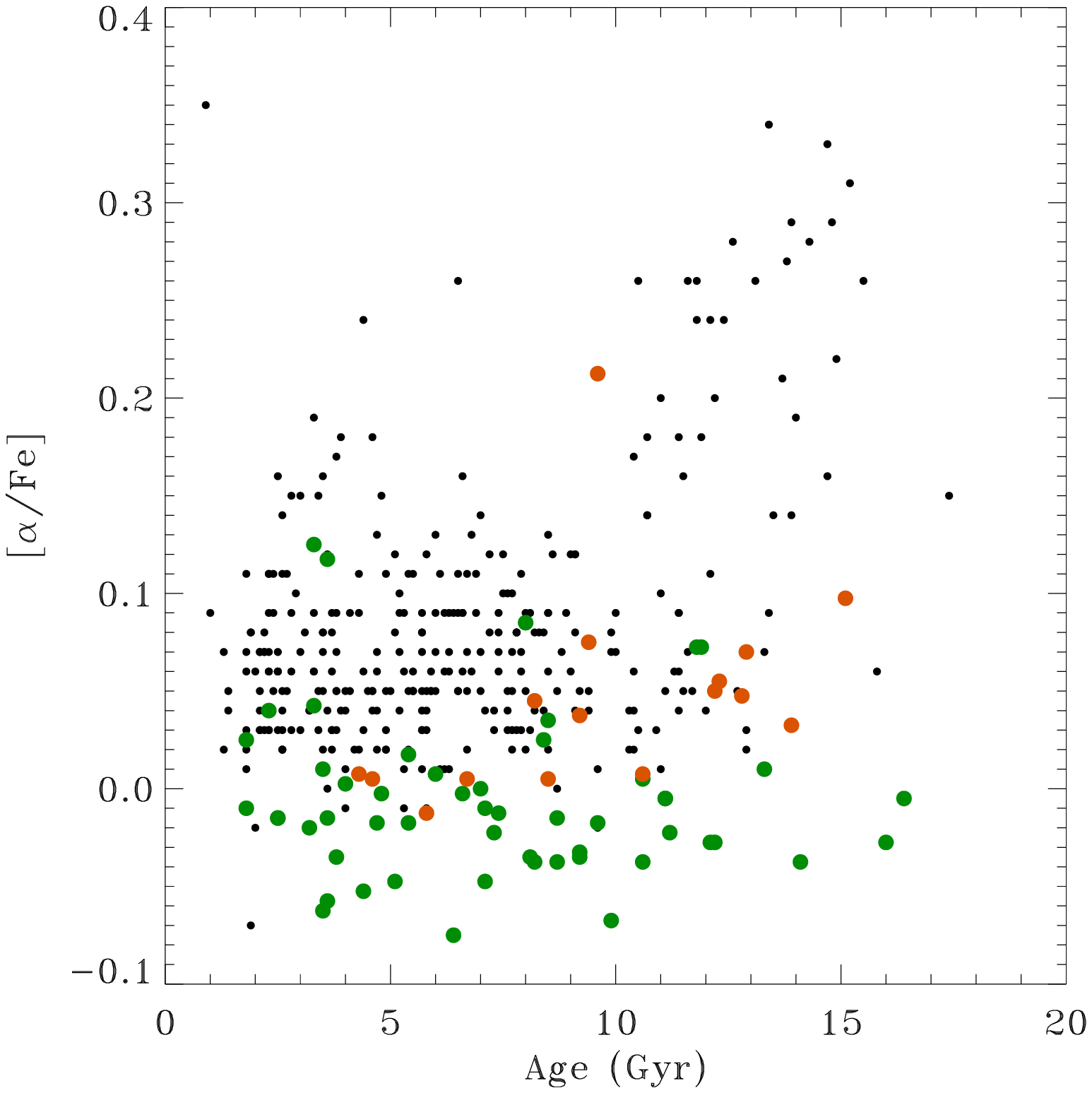} 
\end{array} $
\end{figure}

\clearpage

\end{document}